%% file: ms.tex
\documentclass[11pt]{article}
\input{header}
\usepackage{fullpage, parskip}

\title{VCBART: Bayesian trees for varying coefficients}
\author{Sameer K. Deshpande\thanks{Department of Statistics, University of Wisconsin--Madison. \nolinkurl{sameer.deshpande@wisc.edu}}, Ray Bai\thanks{Department of Statistics, University of South Carolina}, Cecilia Balocchi\thanks{School of Mathematics, University of Edinburgh}, Jennifer E. Starling\thanks{Mathematica Inc.}, Jordan Weiss\thanks{Stanford University}}

\newcommand{\revise}[1]{\textcolor{black}{#1}}

\begin{document}
\maketitle
\begin{abstract}
\input{abstract}
\end{abstract}

\newpage
\singlespacing 
\section{Introduction}
\label{sec:introduction}
\input{introduction}

\section{Background}
\label{sec:background}
\input{background}

\section{The VCBART procedure}
\label{sec:proposed_procedure}
\input{proposed_procedure}

\section{Simulation studies}
\label{sec:simulation_studies}
\input{synthetic_experiments}

\section{Real data examples}
\label{sec:real_data}

\subsection{HRS cognition data}
\label{sec:hrs_analysis}
\input{hrs_analysis}

\subsection{Philadelphia crime data}
\label{sec:philly_analysis}
\input{philly_analysis}

\section{Discussion}
\label{sec:discussion}
\input{discussion}

\section*{Acknowledgements}
Support for this work was provided by the University of Wisconsin--Madison Office of the Vice Chancellor for Research and Graduate Education with funding from the Wisconsin Alumni Research Foundation (S.K.D.); and National Science Foundation grant DMS-2015528 (R.B.).

{
\singlespacing
\small
\bibliographystyle{apalike}
\bibliography{vcbart_refs}
}

\newpage

\renewcommand{\theequation}{S\arabic{equation}}
\renewcommand{\thesection}{S\arabic{section}}  
\renewcommand{\thefigure}{S\arabic{figure}}  
\renewcommand{\thetable}{S\arabic{table}} 

\setcounter{equation}{0}
\setcounter{section}{0}
\setcounter{subsection}{0}
\setcounter{subsubsection}{0}

\onehalfspacing


{
\Large
\centering
\textbf{Supplementary Materials}
}

\input{supplement_outline}

\section{Proofs of asymptotic results}
\label{app:proofs}
\input{proofs}

\newpage
\section{Hyperparameter sensitivity}
\label{app:hyperparameter_sensitivity}
\input{hyperparameter_sensitivity}

\newpage
\section{Additional experimental results}
\label{app:additional_simulations}
\input{additional_sim_results}

\newpage
\section{Gibbs sampler derviation}
\label{app:gibbs_sampler}
\input{gibbs_sampler}

\newpage
\section{Additional figures}
\label{app:additional_figures}
\input{additional_figures}

\end{document}

%% file: header.tex
\usepackage{amsmath, amssymb, amsthm}
\usepackage{algorithm}
\usepackage{algpseudocode}
\usepackage{amsmath, amssymb, amsthm}
\usepackage{float, graphicx, multirow,subcaption, setspace}
\usepackage{comment}
\usepackage{url}
\usepackage{enumitem}
\usepackage{tikz}
\usepackage{bm, bbm}
\usetikzlibrary{shapes.geometric, positioning}
\usetikzlibrary{quotes, angles}
\usepackage{rotating}
\usepackage{xr, xr-hyper}
\usepackage{hyperref}
\usepackage{natbib}

\definecolor{SkyBlue}{RGB}{14, 118, 188}
\definecolor{BrightRed}{RGB}{223, 82, 78}
\definecolor{Green638}{RGB}{165,255,118} 
\definecolor{RevisedColor}{RGB}{25, 200, 150}

\hypersetup{pdfborder = {0 0 0.5 [3 3]}, colorlinks = true, linkcolor = BrightRed, citecolor = SkyBlue}

\newcommand{\Rlang}{\textsf{R}~}

\newcommand\numbereqn{\addtocounter{equation}{1}\tag{\theequation}}

\newcommand{\R}{\mathbb{R}} 
\newcommand{\E}{\mathbb{E}} 

\newcommand{\calC}{\mathcal{C}}

\newcommand{\calZ}{\mathcal{Z}}
\newcommand{\calE}{\mathcal{E}}
\newcommand{\bcalE}{\boldsymbol{\mathcal{E}}}

\newcommand{\cutset}{\calC}

\newcommand{\ind}[1]{\mathbbm{1}\left( #1 \right)} 

\newcommand{\normaldist}[2]{\mathcal{N}\left(#1,#2\right)} 
\newcommand{\mvnormaldist}[3]{\mathcal{N}_{#1}\left(#2,#3\right)} 

\newcommand{\bx}{\bm{x}}
\newcommand{\bz}{\bm{z}}

\newcommand{\br}{\bm{r}}

\newcommand{\Rcont}{R_{\text{cont}}}
\newcommand{\Rcat}{R_{\text{cat}}}

\newcommand{\bY}{\bm{Y}}
\newcommand{\bX}{\bm{X}}
\newcommand{\bZ}{\bm{Z}}
\newcommand{\bR}{\bm{R}}

\newcommand{\bEcal}{\boldsymbol{\mathcal{E}}}

\newcommand{\bbeta}{\boldsymbol{\beta}}

\newcommand{\btheta}{\boldsymbol{\theta}}
\newcommand{\bmu}{\boldsymbol{\mu}}

\newcommand{\bTheta}{\boldsymbol{\Theta}}
\newcommand{\bSigma}{\boldsymbol{\Sigma}}
\newcommand{\bOmega}{\boldsymbol{\Omega}}

\theoremstyle{plain}
\newtheorem{theorem}{Theorem}

\newtheorem{lemma}[theorem]{Lemma}

\theoremstyle{definition}

\newtheorem{ex}{Example}

\theoremstyle{plain}
\newtheorem{remark}{Remark}

%% file: abstract.tex
The linear varying coefficient models posits a linear relationship between an outcome and covariates in which the covariate effects are modeled as functions of additional effect modifiers.
Despite a long history of study and use in statistics and econometrics, state-of-the-art varying coefficient modeling methods cannot accommodate multivariate effect modifiers without imposing restrictive functional form assumptions or involving computationally intensive hyperparameter tuning.
In response, we introduce VCBART, which flexibly estimates the covariate effect in a varying coefficient model using Bayesian Additive Regression Trees.
With simple default settings, VCBART outperforms existing varying coefficient methods in terms of covariate effect estimation, uncertainty quantification, and outcome prediction.
We illustrate the utility of VCBART with two case studies: one examining how the association between later-life cognition and measures of socioeconomic position vary with respect to age and socio-demographics and another estimating how temporal trends in urban crime vary at the neighborhood level. 
An \textsf{R} package implementing VCBART is available at \url{https://github.com/skdeshpande91/VCBART}.

%% file: introduction.tex
\subsection{Motivation}
\label{sec:motivation}

The linear varying coefficient model specifies a linear relationship between an outcome $Y$ and $p$ covariates $X_{1}, \ldots, X_{p}$ that is allowed to change according to the values of $R$ \textit{effect modifiers} $Z_{1}, \ldots, Z_{R}$.
That is, the model asserts that
\begin{equation}
\label{eq:general_model}
Y= \beta_{0}(\bZ) + \beta_{1}(\bZ)X_{1} + \cdots + \beta_{p}(\bZ)X_{p} + \varepsilon,
\end{equation}
where the $\beta_{0}(\bZ), \ldots, \beta_{p}(\bZ)$ are \textit{functions} mapping $\R^{R}$ to $\R$ and the residual error $\varepsilon$ has mean zero.
In this paper, we use Bayesian Additive Regression Trees \citep[BART;][]{Chipman2010} to learn the covariate effect functions $\beta_{j}(\bZ),$ expressing each with an ensemble of binary regression trees.
Our proposed method, which we call VCBART, can produce extremely accurate estimates and well-calibrated uncertainty intervals of evaluations $\beta_{j}(\bz)$ and predictions $\E[Y \vert \bX = \bx, \bZ = \bz]$ without imposing rigid parametric assumptions or requiring computationally intensive tuning. 
VCBART further enjoys strong theoretical guarantees and scales gracefully to large datasets with tens of thousands of observations. 
The following applications motivate our work.

\textbf{Socioeconomic position and cognition.} 
A large body of evidence suggests that socioeconomic position (SEP) at different points in the life course is an important determinant of cognitive function in mid-life and older adulthood \citep{LuoWaite2005, LyuBurr2016, Marden2017, Greenfield2019, Zhang2020}. 
A critical open challenge in life course research involves estimating how the associations between cognition and various SEP measures evolve over time and with respect to sociodemographic characteristics.
Typically, authors related cognitive outcome to several SEP variables with linear models that included pre-specified interactions between the SEP covariates and characteristics like age, gender, and race.
For instance, both \citet{LyuBurr2016} and \citet{Marden2017} modeled an interaction between SEP and age while \citet{Aartsen2019} introduced an additional interaction with age-squared.
In doing so, these authors implicitly make strong parametric assumptions about how the associations of interest vary.
Unfortunately such functional form assumptions may be inadequate as many of these associations can weaken or level off as people age \citep[see][and references therein]{Dupre2007}.

In Section~\ref{sec:hrs_analysis}, we use VCBART to estimate whether and how the associations between total score on a battery of cognitive assessments and measures of SEP in childhood, early adulthood, and later-life can vary with respect to sociodemographic factors like age, race, and gender.
Our dataset comes from the Health and Retirement Study (HRS), which is a nationally representative longitudinal study of US-based adults, and contains $N = 67,988$ total observations of $n =10,812$ subjects.
Using VCBART, we found little temporal variation in the association between childhood SEP and later-life cognition.
Further, after adjusting for SEP in childhood and late adulthood, the association between early adulthood SEP and later-life cognition varied with respect to race and gender but did not vary substantially over time.

\textbf{Crime in Philadelphia}. Over multiple studies of neighborhood-level crime, Balocchi and colleagues have discovered that (i) partially pooling data across adjacent neighborhoods often improves crime forecasts but (ii) ignoring potential spatial discontinuities can yield highly biased forecasts \citep{Balocchi2019, Balocchi2021_resolution,Balocchi2022_crime}.
Spatial discontinuities not only occur along known geographic landmarks (e.g., highways, parks, or rivers) but can also coincide with less visible differences in neighborhood-level demographic and socioeconomic dimensions.
\citet{Balocchi2022_crime} fit linear models regressing a transformed crime density onto a time index in each of Philadelphia's 384 census tracts.
Concerned about potential spatial discontinuities in the tract-specific slopes and intercepts, they estimated two partitions of the tracts, one for the slopes and one for the intercepts, in which parameter values were similar within clusters but not across clusters. 
Working within a Bayesian framework, they identified several high posterior probability partition pairs with an ensemble optimization procedure that greedily searched over the space of pairs of partitions of census tracts.

Although they obtained promising predictive results, their analysis is limited by the assumption that the number of reported crimes within each census tract is monotonic in time.
An arguably more realistic model would (i) incorporate higher-order trends to better capture potential non-linearities and (ii) allow the corresponding model coefficients to vary across tracts.
That is, a more realistic model would allow for different non-monotonic crime trends in different parts of the city.
Unfortunately, extending \citet{Balocchi2022_crime}'s optimization procedure to cover higher-order models is computationally impractical, as it must search over a combinatorially vast product space of partitions. 
In Section~\ref{sec:philly_analysis}, we recast their model as a linear varying coefficient model with a single categorical effect modifier $Z$ recording the census tract.
Using VCBART to fit several tract-specific polynomial models, we find that a quartic model of crime provides a better fit to data than their linear model.

\subsection{Our contributions}

In both problems, we wish to make minimal assumptions about the functional form of the covariate effects $\beta_{j}(\bZ).$
Additionally, at least for the HRS data, we wish to identify which elements of $Z$ modify the effects of which elements of $X.$
Further, in both applications, calibrated and coherent uncertainty quantification is imperative.
For the HRS data, identifying disparities in the socioeconomic determinants of later-life cognition carries profound public health implications, especially in light of an aging population.
And for the crime dataset, accurate assessments of the uncertainty about crime forecasts can inform future public safety policy decisions.
Finally, we require a method that can scale to the size of the two motivating datasets, ideally without involving computationally intensive hyperparameter tuning.

Unfortunately, many existing state-of-the-art procedures for fitting varying coefficient models with $R > 1$ modifiers involve tuning several parameters with leave-one-out cross-validation or rigidly assume the $\beta_{j}(\bZ)$'s are additive in the $Z_{r}$'s.
Moreover, their default implementations often do not provide any out-of-sample uncertainty quantification.

We instead approximate each $\beta_{j}(\bZ)$ with a sum of regression trees.
On synthetic data, our proposed procedure VCBART exhibits superior covariate effect recovery compared to the current state-of-the-art \textit{without requiring any hand-tuning or strong structural assumptions}.
We additionally show that, under mild conditions, the VCBART posterior concentrates at a near-optimal rate, even with correlated residual errors.
To the best of our knowledge, our Theorem~\ref{thm:panel_concentration} is the first result demonstrating the theoretical near-optimality of Bayesian treed regression in settings with non-i.i.d. noise.
VCBART also generalizes \citet{Hahn2020}'s Bayesian Causal Forests (BCF) model in order to estimate heterogeneous treatment effects in settings with multiple binary or continuous treatments.

Here is an outline for the rest of the paper. 
We briefly review relevant background on VC modeling and BART in Section~\ref{sec:background}.
Then, in Section~\ref{sec:proposed_procedure}, we introduce VCBART, describe how to perform posterior inference, and state our asymptotic results.
In Section~\ref{sec:simulation_studies}, we demonstrate VCBART's excellent covariate effect recovery and predictive capabilities using synthetic data.
We apply VCBART to our motivating datasets in Section~\ref{sec:real_data} before outlining several avenues for future work in Section~\ref{sec:discussion}.

%% file: background.tex
\subsection{Varying coefficient models}
\label{sec:vc_models}

Since their introduction in \citet{Hastie1993}, varying coefficient models have been extensively studied and deployed in statistics and econometrics.
We give a very brief overview here; see \citet{FanZhang2008} and \citet{Franco-Villoria2019} for more comprehensive reviews.

When there is only a single modifier (i.e. $R = 1$), one popular approach to fitting the model in Equation~\eqref{eq:general_model} is to express each $\beta_{j}(\bZ)$ as a linear combination of pre-specified basis functions \citep[see, e.g.,][]{Hoover1998, Huang2002}.
Such a decomposition effectively reduces the functional regression problem to a high-dimensional linear regression problem for which numerous frequentist \citep[see, e.g.,][]{Wang2008, Wang2009, Wei2011} and Bayesian \citep{Bai2019b} regularization techniques have been proposed.
Kernel smoothing is another popular and theoretically supported alternative when $R = 1$ \citep{Wu2000}.

Varying coefficient models are commonly encountered in spatial statistics \citep[see, e.g.,][]{Gelfand2003, Finley2020}.
In these settings, there are usually $R = 2$ modifiers (space) or $R = 3$ modifiers (space and time) and the $\beta_{j}(\bZ)$'s are typically modeled with Gaussian processes.
When the number of observations is large, full Bayesian inference with these models can be computationally demanding, necessitating the use of clever approximations or advanced techniques like distributed computing \citep{Guhaniyogi2020}. 
Outside of spatial contexts, to accommodate $R > 1$ modifiers, \citet{Tibshirani2019} and \citet{Lee2018} respectively constrained the $\beta_{j}(\bZ)$'s to be linear and additive functions of the $Z_{r}$'s.
In contrast, \citet{Li2010} proposed a multivariate kernel smoothing estimator that imposes no rigid structural assumptions on the covariate effects.
However, the default implementation of their procedure tunes several bandwidth parameters with leave-one-out cross-validation. 
When $n, p, $ or $R$ are large, their procedure is computationally prohibitive. 
 
In the last decade, several authors have used regression trees to estimate the covariate effects.
\citet{Burgin2015}, for instance, modeled each $\beta_{j}(\bZ)$ with a single regression tree while \citet{WangHastie2012} and \citet{ZhouHooker2019} both constructed separate ensembles for each $\beta_{j}(\bZ)$ using boosting.
Unfortunately, existing tree-based procedures require substantial tuning and often do not automatically return any uncertainty estimates.

\revise{Before proceeding, we pause to highlight an important difference between VCBART and \citet{Bai2019b}.
That work is limited to settings with only one effect modifier (i.e., $R = 1$) and where interest lies primarily in determining which $X_{j}$'s affect $Y$ (i.e., whether $\beta_{j}(\bz) = 0$ for all $\bz$). 
In contrast, VCBART allows for an arbitrary number of effect modifiers, estimates individual evaluations $\beta_{j}(\bz),$ and determinines which $Z_{r}$'s drive variation in each function $\beta_{j}(\bZ).$}

\subsection{Bayesian Additive Regression Trees}
\label{sec:bart_background}

\citet{Chipman2010} introduced BART in the context of the standard nonparametric regression problem: given $n$ observations from the model $y \sim \normaldist{f(\bx)}{\sigma^{2}},$ they estimated $f$ with a sum of $M$ piecewise constant step functions, which they represented as binary regression trees.
Formally, they computed a posterior distribution over regression tree ensembles, which induces an approximate posterior distribution over regression function evaluations $f(\bx).$
Their regression tree prior, which we adopt and detail in Section~\ref{sec:regression_tree_prior}, introduces strong regularization and encourages trees to be ``weak learners'' in the sense that no one tree explains too much variation in the observed $y$'s.

The basic BART model has been extended successfully to survival analysis \citep{Sparapani2016}, multiple imputation \citep{Xu2016}, log-linear models \citep{Murray2019}, semi-continuous responses \citep{Linero2020}, and causal inference \citep{Hill2011, Hahn2020}.
With conceptually simple modifications, BART can also recover smooth \citep{LineroYang2018, Starling2019} and monotonic \citep{ Starling2020, Chipman2019} functions.
In each setting, BART-based methods often substantially outperform existing procedures in terms of function recovery, prediction, and ease-of-use.
Indeed, nearly every BART extension provides default hyperparameters that yield generally excellent performance ``off-the-shelf.''
Further, recent results in \citet{RockovaSaha2019} and \citet{Rockova2019} demonstrate BART's theoretical near-optimality under very mild assumptions. 
See \citet{Tan2019} and \citet{Hill2020} for more detailed reviews of BART and its many extensions.

%% file: proposed_procedure.tex
Without loss of generality, suppose that we have $\Rcont$ continuous modifiers, each of which has been re-scaled to lie in the unit interval $[0,1],$ and $\Rcat$ categorical or discrete modifiers. 
For $r = 1, \ldots, \Rcat,$ we will assume that the $r$-th categorical modifier can take $K_{r}$ distinct values contained in some discrete set $[K_{r}].$
We further concatenate our modifiers into a vector $\bZ$ whose first $\Rcont$ entries record the values of the continuous modifiers and whose last $\Rcat$ entries record the values of the categorical modifiers.
In other words, each observed modifier vector $\bz$ lies in the product space $\calZ = [0,1]^{\Rcont} \times [K_{1}] \times \cdots \times [K_{\Rcat}].$

Both motivating applications involve repeated measurements over time: in the HRS dataset, we have between four and eight observations per subject and in the crime dataset, we have yearly crime densities in each census tract.
For simplicity, we will refer to the observed units (i.e., HRS participants and Philadelphia census tracts) as ``subjects.''
For each subject $i = 1, \ldots, n,$ we observe $n_{i}$ triplets $(\bx_{it}, \bz_{it}, y_{it})$ of covariates $\bx$, modifiers $\bz$, and outcome $y$.
For all $i = 1, \ldots, n,$ and $t = 1, \ldots, n_{i},$ we model
\begin{equation}
\label{eq:panel_vc_model}
y_{it} = \beta_{0}(\bz_{it}) + \sum_{j = 1}^{p}{\beta_{j}(\bz_{it})x_{itj}} + \sigma \varepsilon_{it}. 
\end{equation}
We further assume $\boldsymbol{\varepsilon}_{i} = (\varepsilon_{i1}, \ldots, \varepsilon_{in_{i}})^{\top} \sim \mvnormaldist{n_{i}}{\mathbf{0}_{n_{i}}}{\bSigma_{i}(\rho)}$ where $\bSigma_{i}(\rho)$ is a correlation matrix with off-diagonal elements equal to $0 \leq \rho < 1.$
In other words, we assume an exchangeable or compound symmetry correlation structure for each subject's errors.
We assume that the noise vectors $\boldsymbol{\varepsilon}_{i}$'s are independent across subjects.
We do not assume that we observe each subject an equal number of times nor do we assume that the observation times are equally spaced. 

The key idea of VCBART is to approximate each function $\beta_{j}(\bZ)$ with its own regression tree ensemble. 
In Section~\ref{sec:regression_tree_prior} we describe the prior over the regression trees and in Section~\ref{sec:posterior_modifier_selection}, we outline strategies for posterior computation and modifier selection.

\subsection{Regression tree prior}
\label{sec:regression_tree_prior}

To set our notation, let $T$ be a binary decision tree that consists of a collection of internal nodes and a collection of terminal or \textit{leaf nodes}.
Each internal node of $T$ is associated with a decision rule $\{Z_{r} \in \calC\}.$ 
When $Z_{r}$ is a continuous variable, $\calC$ is a half-open interval of the form $[0, c)$ and when $Z_{r}$ is categorical, $\calC$ is a subset of $[K_{r - \Rcont}].$

\begin{figure}[h]
\centering
\begin{subfigure}[b]{0.48\textwidth}
\centering
\includegraphics[width = \textwidth]{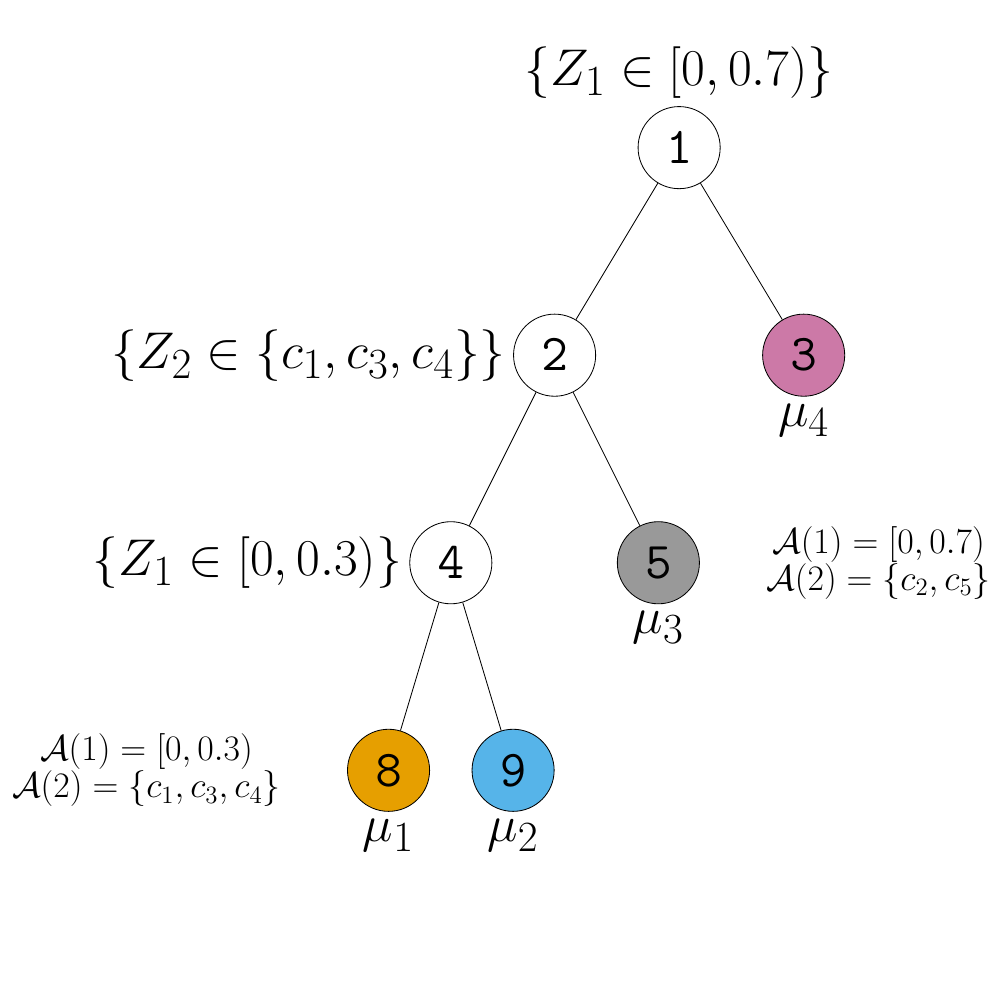}
\caption{}
\label{fig:tree_new}
\end{subfigure}
\begin{subfigure}[b]{0.48\textwidth}
\centering
\includegraphics[width = \textwidth]{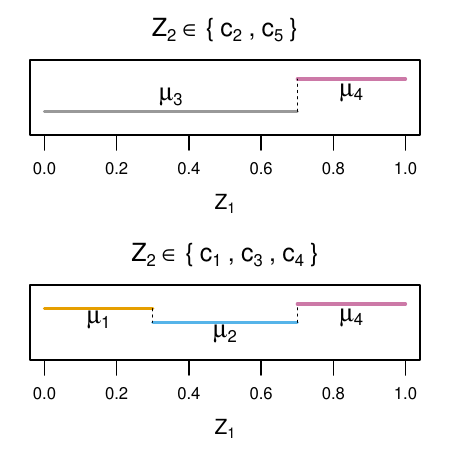}
\caption{}
\label{fig:step_new}
\end{subfigure}
\caption{A regression tree defined over $\calZ = [0,1] \times \{c_{1}, \cdots, c_{5}\}$ (a) and its step function representation (b). In (a), we also report the sets of available values of $Z_{1}$ and $Z_{2}$ at nodes \texttt{5} and \texttt{8}.}
\label{fig:tree}
\end{figure}

Given $T$ and any vector $\bz,$ we can imagine tracing a path from the root down the tree by following the decision rules.
Specifically, any time the path encounters the rule $\{Z_{r} \in\calC \},$ it proceeds to the left child if $z_{r} \in\calC.$
Otherwise, it proceeds to the right child.
Such decision-following paths continue until they reach a leaf node.
It is not difficult to verify that every such path terminates in a single leaf node, implying that $T$ induces a partition of $\calZ$ into disjoint regions, one for each leaf.
By associating each leaf node $\ell$ of $T$ with a scalar \textit{jump} $\mu_{\ell},$ the pair $(T, \bmu)$ represents a piecewise constant function over $\calZ,$ where $\bmu$ denotes the collections of jumps (see Figure~\ref{fig:tree} for an example).

For each $j = 0, \ldots, p,$ we introduce an ensemble $\calE^{(j)} = \{(T_{m}^{(j)}, \bmu_{m}^{(j)})\}_{m = 1}^{M}$ of $M$ regression trees to approximate the function $\beta_{j}(\bZ).$
We model each ensemble independently and place independent and identical priors on the individual regression trees within each ensemble.
Following \citet{Chipman2010}, we specify a scaled inverse-$\chi^{2}$ prior for the residual error variance $\sigma^{2} \sim \lambda \nu \chi^{-2}_{\nu}.$
Like that work, we take $\nu = 3$ and set $\lambda$ so that there is 90\% prior probability on the event that $\sigma$ is less than the standard deviation of the observed responses $Y.$
We complete our prior specification with a uniform prior on the autocorrelation $\rho.$

We specify the prior for an individual tree $(T_{m}^{(j)}, \bmu_{m}^{(j)})$ in the $j$-th ensemble compositionally with a marginal prior for the decision tree and a conditional prior for the jumps given the tree.
Just like in the original BART model, we model the jumps in $\bmu_{m}^{(j)}$ as \textit{a priori} independent $\normaldist{0}{\tau^{2}_{j}}$ random variables, where $\tau_{j}$ is a positive hyperparameter.
Thus, for each $\bz \in \mathcal{Z},$ the implied marginal prior of $\beta_{j}(\bz)$ is $\normaldist{0}{M\tau^{2}_{j}}.$
We have found taking $\tau_{j} = 0.5/\sqrt{M}$ to be an effective default choice; see Section S2 of the Supplementary Materials for a hyperparameter sensitivity analysis.

We specify the decision tree prior implicitly by describing how to sample from the prior. 
First, we draw the overall graphical structure with a branching process that starts from a root node, which is initially considered terminal.
Whenever a new terminal node is created at depth $d$, we attach two child nodes to it with probability $0.95(1 + d)^{-2}.$
The quadratic decay ensures that the process terminates at a finite depth.

Given the graphical structure, we draw decision rules at each non-terminal node in two steps.
First, we sample the splitting variable index $v \sim \text{Multinomial}(\theta_{j1}, \ldots, \theta_{jR})$ and $\theta_{jr}$ is the probability of selecting $Z_{r}.$
We place a symmetric $\textrm{Dirichlet}(\eta_{j}/R)$ prior on the vector of splitting probabilities and place a further Beta prior over $\eta_{j}/(\eta_{j} + R).$
The implied prior density of $\eta_{j}$ is proportional to $(R + \eta_{j})^{-(R+1)}.$
Although the $\eta_{j}$ prior favors sparsity, it has the capacity to include more variables as needed \citep[\S 3.3]{Linero2018}.
As evidenced by the simulation studies in \citet{Linero2018}, the prior hierarchy on the splitting variable works well in both the sparse setting, when the function depends on only a few variables, and in the non-sparse setting. 

Then, conditional on $r,$ we set $\cutset$ to be a random subset of $\mathcal{A}(r),$ the set of available $Z_{r}$ values at the current non-terminal node (see annotations in Figure~\ref{fig:tree_new}). 
When $Z_{r}$ is continuous, $\mathcal{A}(r)$ is an interval and we set $\cutset = [0,c)$ where $c$ is drawn uniformly from $\mathcal{A}(r).$
When $Z_{r}$ is categorical, $\mathcal{A}(r)$ is a discrete set.
How we draw the random subset $\cutset$ depends on whether $Z_{r}$ displays network structure like the census tracts in our crime data.
If not, we assign each element of $\mathcal{A}(r)$ to $\cutset$ with probability 0.5.
Otherwise, we form $\cutset$ by drawing a random network partition.
Briefly, we partition the vertices of the subgraph induced by $\mathcal{A}(r)$ by deleting a random edge from a random spanning tree of the subgraph; see Section~\ref{sec:philly_analysis} below and \citet{Deshpande2022_flexBART} for more details.

\subsection{Posterior computation and modifier selection}
\label{sec:posterior_modifier_selection}

\textbf{Conditional regression tree updates}.
Because the VCBART posterior is analytically intractable, we simulate posterior draws using a Gibbs sampler.
Briefly, we sequentially update each tree $(T_{m}^{(j)}, \bmu_{m}^{(j)})$ conditionally on the other $Mp-1$ regression trees using two steps.
First, we update the decision tree structure $T_{m}^{(j)}$ with a Metropolis-Hastings step and transition kernel that randomly grows or prunes the tree.
Like \citet{Chipman2010}'s original BART sampler, this update depends on the data and other regression trees through a particular partial residual.
Then, given $T_{m}^{(j)},$ we draw the jumps $\bmu^{(j)}_{m}$ from its conditional posterior distribution.
Essentially, this amounts to fitting an intercept-free conjugate Bayesian linear regression model in each leaf of the tree.
Whereas the jumps in the original BART model are conditionally independent, the correlated errors in Equation~\eqref{eq:panel_vc_model} render the jumps in VCBART conditionally dependent. 
After updating all trees in the ensemble for $\beta_{j},$ we perform a conjugate Dirichlet-Multinomial update for the vector of splitting probabilities $\btheta_{j}.$
After updating the trees in the $j^{\text{th}}$ ensemble and $\btheta_{j},$ we update $\eta_{j}$ using an independence Metropolis step.
After sweeping over all ensembles, we perform a conditionally conjugate Inverse Gamma update for $\sigma^{2}$ and sample a new $\rho$ using a random-walk Metropolis update on the unconstrained quantity $\log\left(\frac{\rho}{1-\rho}\right).$
See Section S4 of the Supplementary Materials for a detailed derivation of our Gibbs sampler.

\textbf{Modifier selection}. 
In our sampler, in order to update the vector of splitting index probabilities $\btheta_{j},$ we keep track of the number of times each modifier $Z_{r}$ is used in a decision rule in the ensemble $\mathcal{E}_{j}.$
Using these counts, we can estimate probability that each $Z_{r}$ is selected at least once in the ensemble $\mathcal{E}_{j}$ used to estimate $\beta_{j}(\bZ).$
These selection probabilities are analogous to the posterior inclusion probabilities encountered in Bayesian sparse linear regression.
Based on this interpretation, for each ensemble $\mathcal{E}_{j},$ we construct an analog of \citet{BarbieriBerger2004}'s median probability model by reporting those modifiers $Z_{r}$ whose selection probability exceeds 0.5.

\begin{remark}[Identifiability]
\label{remark:identifiability}
For each realization of the modifiers $\bz$, the vector of covariate function evaluations $\bbeta(\bz) = (\beta_{0}(\bz), \ldots, \beta_{p}(\bz))$ is identifiable as long as $\E[\bx\bx^{\top} \vert \bZ = \bz]$ is positive definite \citep{HuangShen2004}. 
The individual regression trees, however, are not identified. 
This is because we can switch the order of the trees without changing the predictions. 
Nevertheless, in our simulation studies, we did not experience any difficulties recovering the true covariate effects.
\end{remark}

In light of Remark~\ref{remark:identifiability}, we do not suggest diagnosing Markov chain convergence using the samples of individual trees.
We instead recommend tracking convergence using samples of $\sigma,$ the residual standard deviation.
In our experiments, we have found that although a large number of MCMC iterations (i.e., 20,000 or 50,000) may be needed for adequate mixing (e.g, Gelman-Rubin $\hat{R} < 1.1$), one can often obtain excellent predictions and well-calibrated uncertainty intervals using many fewer iterations (i.e., 2,000); see Section~\ref{sec:simulation_studies} and Section S2.1 of the Supplementary Materials.

\subsection{Posterior contraction of VCBART}
\label{sec:posterior_contraction}
 
With some minor modifications, VCBART enjoys essentially the same favorable theoretical properties as BART --- namely, the posterior concentrates at a nearly minimax optimal rate. 
To facilitate our theoretical analysis of the VCBART posterior, we assume that all modifiers lie in $[0,1]^{R}.$ In practice, we can always rescale continuous modifiers to lie in the unit interval, and with one-hot encoding, we can represent categorical modifiers with multiple binary indicators. 
However, our main theorem still holds as long when all modifiers in $\bZ$ are uniformly bounded.  
We additionally assume that the observed modifiers $\bz_{it} \in \mathbb{R}^{p}, i = 1, \ldots, n, t = 1, \ldots, n_i$, are fixed and follow a regular design (Definition 3.3 in \cite{Rockova2019}). 
We refer readers to \cite{Rockova2019} and \cite{RockovaSaha2019} for a technical definition. 
At a high level, design regularity implies that a \emph{k-d} tree partition \citep{Benley1975} of the dataset results in cells with similar diameters (i.e., similar maximal distance between two points in a cell). 

Like \citet{RockovaSaha2019}, we modify the decision tree prior so that the probability that the branching process continues growing at depth $d$ is $\gamma^{d}$ for some fixed $N^{-1} < \gamma < 1/2.$
We assume that for each $i = 1, \ldots, n$ and $t = 1, \ldots, n_i$, the ground truth varying coefficient model is
\begin{align} 
\label{eq:true_panel_model}
y_{it} = \beta_{0,0} (\bz_{it}) + \sum_{j=0}^{p} \beta_{0,j} (\bz_{it}) x_{itj} + \sigma_0 \varepsilon_{it},\end{align}
where each $\beta_{0,j}(\bZ), j = 0, \ldots, p$, is $\alpha_j$-H\"{o}lder continuous with $0 < \alpha_j \leq 1$, and $\bm{\varepsilon}_i \sim \mathcal{N}( \mathbf{0}_{n_i}, \bm{\Sigma}_{i} (\rho_0))$, where the $(j,k)^{\text{th}}$ entry of $\bm{\Sigma}_i(\rho_0)$ is $\mathbbm{1}(j=k) + \rho_0 \mathbbm{1}(j \neq k)$. 

Recall that $N = \sum_{i=1}^{n} n_i$ is the total number of observations for all $n$ subjects, $R$ is the number of effect modifiers, and $p$ is the number of covariates. Let $n_{\max} = \max \{ n_1, \ldots, n_n \}$ denote the maximum number of within-subject observations. 
For two nonnegative sequences $\{ a_n \}$ and $\{ b_n \}$, we write $a_n \asymp b_n$ to denote $0 < \lim \inf_{n \rightarrow \infty} a_n / b_n$ $\leq \lim sup_{n \rightarrow \infty} a_n / b_n < \infty$. Meanwhile, $a_n = O(b_n)$ means that for sufficiently large $n$, there exists a constant $C > 0$ independent of $n$ such that $a_n \leq C b_n$. 
We make the following assumptions.
\begin{itemize}
	\item [(A1)] The true varying coefficients satisfy $\max_{\bZ \in [0,1]^R} | \beta_{0,j} (\bZ) | < \infty$ for all $0 \leq j \leq p.$
	\item [(A2)] There is a constant $D > 1$ so that $\lvert x_{itj} \rvert \leq D$ for all $1 \leq i \leq n$, $1 \leq t \leq n_i$, $1 \leq j \leq p$.
	\item [(A3)] $R$, $p$, and $n_{\max}$ satisfy $R = O((\log N)^{1/2})$, $p = O(1)$, and $n_{\max} \asymp N / n$.
\end{itemize}
Assumption (A1) assumes that the true varying coefficients $\beta_j(\bZ)$'s are uniformly bounded. This assumption is also made in \citet{ZhouHooker2019} and is likely to be satisfied in practice since the $\bZ$'s are uniformly bounded. Assumption (A2) assumes that all of the covariates $\boldsymbol{X}$ are also uniformly bounded. Finally, Assumption (A3) specifies appropriate rates of growth for $R$, $p$, and $n_{\max}$.

Let $\bbeta$ and $\bbeta_{0}$ be $N \times (p+1)$ matrices whose respective $(i,j)$-th entries are $\beta_{j}(\bz_{it})$ and $\beta_{0,j}(\bz_{it}).$  Let $\lVert \bbeta - \bbeta_{0} \rVert_{N}^{2} = N^{-1}\sum_{i = i}^{n}{\sum_{t = 1}^{n_{i}}{\sum_{j = 0}^{p}{[\beta_{j}(\bz_{it}) - \beta_{0,j}(\bz_{it})]^{2}}}}$ be the squared empirical $\ell_2$ norm.
If we knew the true smoothness levels $\alpha_{j},$ the minimax rate for estimating $\bm{\beta}_0$ in $\lVert \cdot \rVert_N^2$ would be $\sum_{j=0}^{p} N^{-2 \alpha_j / (2 \alpha_j + R)}$ \citep{Rockova2019}. 
This is because the minimax estimation rate for each individual $\alpha_j$-H\"{o}lder smooth function $\beta_{0,j}(\bZ)$ is $N^{-2 \alpha_j / (2 \alpha_j + R)}$ \citep{Stone1982} and we do not assume that any of the $\beta_{0,j}(\bZ)$'s are zero functions.
In the absence of knowledge of the true $\alpha_j$'s, Theorem \ref{thm:panel_concentration} shows that VCBART can estimate the varying coefficients $\bm{\beta}_0$ at nearly this rate, sacrificing only a logarithmic factor.

\begin{theorem} \label{thm:panel_concentration}
	Under \eqref{eq:true_panel_model}, suppose we endow $(\bm{\beta}, \sigma^2, \rho)$ with the following VCBART prior: (a) For the BART priors on the varying coefficients $\beta_j$'s, the probability of splitting a tree node at depth $d$ is $q(d) = \gamma^{d}$ for some $1/N < \gamma < 1/2$ and the splitting variables are chosen uniformly at random; (b) $\sigma^2 \sim \lambda \nu \chi_{\nu}^{-2}$, where $\lambda, \nu > 0$; and (c) $\rho \sim \text{Uniform}(0,1)$. Assume that assumptions (A1)--(A3) hold. Then for some constant $\widetilde{C} > 0$ and $r_N^2 = \log N \times \sum_{j=0}^{p} N^{- 2 \alpha_j / (2 \alpha_j + R)}$,
		\begin{equation} \label{contraction_rate_panel_vc}
	\Pi \left( \bm{\beta}: \lVert \bm{\beta} - \bm{\beta}_0 \rVert_N > \widetilde{C} r_N | \bm{Y} \right) \rightarrow 0,
	\end{equation}
	in $\mathbb{P}_{\bm{\beta}_0}^{(N)}$-probability as $N, R \rightarrow \infty$.
\end{theorem}
\noindent The proof of Theorem \ref{thm:panel_concentration} is given in Section S1 of the Supplementary Materials.

We now pause to highlight the novelties of Theorem \ref{thm:panel_concentration}. 
First, most existing theoretical results for BART assume independent observations. 
Theorem \ref{thm:panel_concentration} appears to be the first to establish the near-optimality of BART under \emph{non}-independent errors. 
Secondly, within the broader literature on varying coefficient models, there are many theoretical results when there is only a \textit{single} modifier, i.e. $R=1$ \citep{HuangWuZHou2004, Wang2008, Wang2009, Wei2011, XueQu2012}. 
However, to our knowledge, \cite{ZhouHooker2019} are the only other authors who have derived asymptotic theory for varying coefficient models under \emph{multivariate} effect modifiers with $R > 1$. 
Our Theorem~\ref{thm:panel_concentration} strengthens the result of \cite{ZhouHooker2019} in two ways. 
First, \cite{ZhouHooker2019} only proved estimation consistency for their (frequentist) treed varying coefficient model. 
We sharpen their result by deriving a nearly-minimax posterior \emph{convergence rate}. 
Second, we allow the number of effect modifiers $R$ to \emph{diverge} with sample size in Assumption (A3), whereas the dimension $R$ is fixed in \cite{ZhouHooker2019}.

We also highlight a few limitations of Theorem~\ref{thm:panel_concentration}. 
First, the restriction that $0 < \alpha_j \leq 1$ is an inevitable consequence of using hard decision trees (i.e., step functions) to estimate the varying coefficient functions \citep{RockovaSaha2019, Rockova2019}. 
This fundamental limitation could be overcome if we replaced the hard decision rules with soft probabilistic decision rules \citep{LineroYang2018}. 
In this case, the restriction $0 < \alpha_j \leq 1$ could be relaxed to $\alpha_j > 0$ for all $j = 0, \ldots, p$. 
Such an extension of VCBART is the subject of ongoing work (see Section \ref{sec:discussion}).

We also assume that $O(p) = 1$ in Assumption (A3), which implies that the number of covariates $p$ is bounded. 
This assumption was also made in \cite{ZhouHooker2019} and is necessary to ensure that the posterior contraction rate $r_N^2 = \log N \times \sum_{j=0}^{p} N^{-2 \alpha_j / (2 \alpha_j + R)}$ decays to zero as $N \rightarrow \infty$. 
If we were to allow $p$ to diverge with $N$, then we would also need to assume sparsity in the varying coefficients (i.e., $\beta_{0,j}(\bZ) = 0$ for many $j \in \{ 0, 1, \ldots, p \}$). 
In this sparse set-up, let $S_0 \subset \{ 0, 1, \ldots, p \}$ denote the set of indices of the non-zero varying coefficients, with cardinality $\vert S_0 \rvert = s_0$. 
Under this sparse, high-dimensional varying coefficient model, the minimax estimation rate improves to $\widetilde{r}_N^2 = (s_0 \log p) / N + \sum_{j \in S_0} N^{-2 \alpha_j / (2 \alpha_j + R)}$ \citep{YangTokdar2015}. 
In this case, we could also allow $p$ and $s_0$ to diverge at the rates of $\log p = O(N^{\xi}), \xi \in (0, 1)$, and $s_0 = O(\log N)$ respectively \citep{YangTokdar2015}. 
VCBART does not perform \emph{selection} from the varying coefficient functions $\beta_j(\bZ)$'s, and we do \emph{not} assume sparsity in the $\beta_{0,j}(\bZ)$'s. 
The result in Theorem~\ref{thm:panel_concentration} thus applies only to the ``small $p$'' case. 
Nevertheless, given the scarce literature on varying coefficient models with \emph{multivariate} effect modifiers ($R>1$), Theorem~\ref{thm:panel_concentration} is still noteworthy and serves as an important first step towards further theoretical developments. 
We are currently working to extend VCBART to sparse, high-dimensional varying coefficient models where only a handful of the $\beta_j(\bZ)$'s are nonzero, and we will also investigate the posterior contraction rate in this context. 
This extension is briefly discussed in Section \ref{sec:discussion}.

Finally, in Theorem~\ref{thm:panel_concentration}, VCBART is only (nearly) minimax-optimal over the class of additive varying coefficient models where the individual functions have isotropic H\"{o}lder smoothness. 
This assumption may be restrictive. 
Recently, \citet{JeongRockova2023} and \citet{RockovaRousseau2023} have established the near-minimax optimality of Bayesian treed regression under even broader function classes which allow for heterogeneous smoothness, discontinuities, and spatial inhomogeneities. 
Investigating the asymptotic properties of VCBART under substantially relaxed smoothness assumptions is an interesting problem for future work.

%% file: synthetic_experiments.tex
\begin{figure}[h]
\centering
\includegraphics[width = \textwidth]{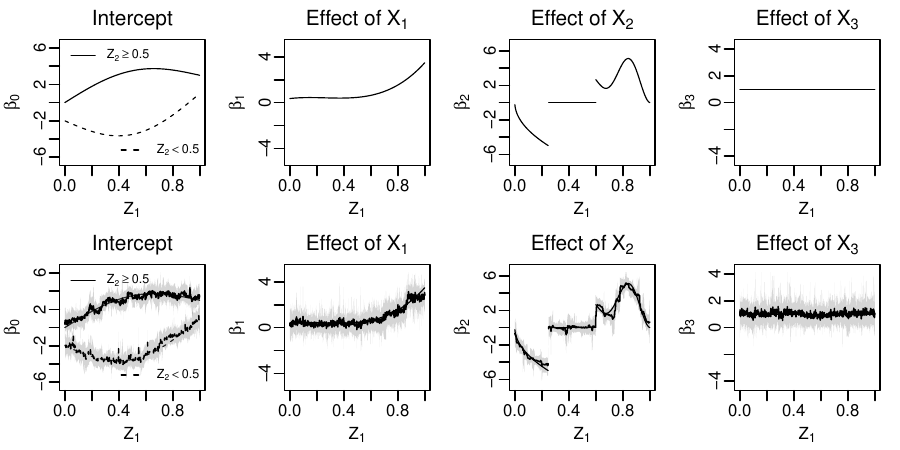}
\caption{ \textbf{Top row.} True functions $\beta_{0}, \ldots, \beta_{3}.$ \textbf{Bottom row.} VCBART's posterior mean (dark line) and point-wise 95\% credible interval (shaded) for each function}
\label{fig:p5R20_beta}
\end{figure}

We performed two simulation studies to understand VCBART's ability to (i) estimate covariate effect functions and predict future outcomes; (ii) identify which elements of $\bZ$ modify the effects of which elements of $X;$ and (iii) scale to large datasets.
We generated data for these experiments from the model in Equation~\eqref{eq:panel_vc_model} with $p = 5$ correlated covariates, $R = 20$ potential effect modifiers, $\sigma = 1,$ and independent within-subject errors.
The correlated covariates were drawn from a multivariate normal distribution with mean zero and a covariance matrix with entries $0.5^{\lvert i - j \rvert}.$
The modifiers were drawn uniformly from the interval $[0,1].$
To assess VCBART's ability to select relevant modifiers (i.e., identify the support of the $\beta_{j}(\bZ)$'s), we used the following covariate effect functions, which depend only five of the $R = 20$ modifiers, in all experiments:

\begin{align*}
\beta_{0}(\bz) &= 3z_{1} + (2 - 5\ind{z_{2} > 0.5})\sin\left(\pi z_{1}\right) - 2 \times \ind{z_{2} > 0.5} \\
\beta_{1}(\bz) &= \frac{\sin\left(2z_{1} + 0.5\right)}{4z_{1} + 1} + (2z_{1} - 0.5)^{4} \\
\beta_{2}(\bz) &= (3 - 3z_{1}^{2}\cos{(6\pi z_{1})}) \times \ind{z_{1} > 0.6} - 10\sqrt{z_{1}}\times \ind{z_{1} < 0.25} \\
\beta_{3}(\bz) &= 1 \\
\beta_{4}(\bz) &= 10\sin\left(\pi z_{1}z_{2}\right) + 20(z_{3} - 0.5)^{2} + 10z_{4} + 5z_{5} \\ 
\beta_{5}(\bz) &= \exp\left\{\sin\left(\left(0.9(z_{1} + 0.48)\right)^{10}\right)\right\} + z_{2}z_{3} + z_{4}. 
\end{align*}
The top row of Figure~\ref{fig:p5R20_beta} shows the functions $\beta_{0}, \ldots, \beta_{3}.$
The bottom row of Figure~\ref{fig:p5R20_beta} superimposes the VCBART posterior mean and 95\% credible intervals for $\beta_{0}, \ldots, \beta_{3}$ computed using a single dataset from our first experiment, which comprised four observations of $n = 250$ subjects for a total of $N = 1,000$ observed $(\bx_{it}, \bz_{it}, y_{it})$'s.
The posterior mean and uncertainty bands were computed after running four independent chains of VCBART for 2,000 iterations, discarding the first 1,000 samples of each as burn-in.
We approximated each $\beta_{j}(\bZ)$ with an ensemble of $M = 50$ trees and set each $\tau_{j} = 0.5/\sqrt{M}.$

Although the covariate effect functions have very different shapes, VCBART recovered each remarkably well: the posterior means of each $\beta_{j}(\bZ)$ closely tracked the shape of the true functions and, for the most part, the true function values were within the shaded point-wise 95\% credible intervals.
We observed similar behavior for the functions $\beta_{4}(\bZ)$ and $\beta_{5}(\bZ)$ that depend on multiple effect modifiers (see Figure S4 in the Supplementary Materials).
For these data, the Gelman-Rubin $\hat{R}$ value computed using the samples of $\sigma$ was around 1.2, which is somewhat higher than the heuristic convergence threshold of 1.1. 
Figure S1 of the Supplementary Materials shows a representative traceplots for $\sigma.$
While running each chain for many more iterations produced smaller $\hat{R}$ values and ostensibly better-mixing chains, we found that additional iterations did not meaningfully change the quality of VCBART's predictions and uncertainty intervals (Figure S2 of the Supplementary Materials).
Consequently, for the remainder of this section, we report the performance of VCBART based on running four chains for 2,000 iterations.

Repeating this experiment 25 times --- that is, fixing the $\bx_{it}$'s and $\bz_{it}$'s but simulating the $y_{it}$'s anew --- the mean coverage of the 95\% posterior credible intervals, averaged across all the $\beta_{j}$'s, exceeded 99\%. 
Further, averaging across all functions and replications, the mean sensitivity, specificity, precision and F1-score\footnote{Letting $\textrm{TP}, \textrm{TN}, \textrm{FP},$ and $\textrm{FN}$ be the numbers of true positive, true negative, false positive, and false negative identifications in the supports of the $\beta_{j}$'s, sensitivity is $\textrm{TP}/(\textrm{TP} + \textrm{FN}),$ specificity is $\textrm{TN}/(\textrm{TN} + \textrm{FP}),$ precision is $\textrm{TP}/(\textrm{TP}  + \textrm{FP})$ and the F1 score is $2\times \textrm{TP}/(2\times \textrm{TP} + \textrm{FP} + \textrm{FN})$} for modifier selection with our median probability model were 0.85, 0.99, 0.95, and 0.90.
Recalling that each $\beta_{j}(\bZ)$ depended on only a subset of the $R = 20$ modifiers, these results indicate excellent support recovery performance in the presence of several inactive variables.
In fact, VCBART retains its excellent ability to identify the relevant modifiers when $R$ is large and the effect modifiers are correlated and even in the presence of moderate-to-high correlation among the $X_{j}$'s; see Sections S3.4 and S3.5 of the Supplementary Materials.
Averaging across 25 replications, sequentially running four MCMC chains for 2,000 iterations each took under one minute on an M1 Mac.

We compared VCBART's ability to estimate covariate effects out-of-sample to those of the following methods: the standard linear model (\texttt{lm}); \citet{Li2010}'s kernel smoothing (\texttt{KS}; implemented in \citet{np_package}'s \textbf{np} package); \revise{\citet{Burgin2017}'s tree-based varying coefficient procedure (\texttt{TVC})}; and \citet{ZhouHooker2019}'s boosted tree procedure method (\texttt{BTVCM}).
\revise{\texttt{BTVCM} is not a boosted version of \texttt{TVC}: the latter grows a single deep tree for each $\beta_{j}(\bZ)$ and then prunes each tree using cross-validation while the former sets a fixed tree depth in each round of boosting.}
By default, \texttt{KS} respectively perform cross-validation to learn kernel bandwidth parameters.
\revise{We used the default cross-validation settings for \texttt{TVC} implemented in the \textsf{R} package \textbf{vcrpart} \citep{Burgin2017}.}
The default implementation of \texttt{BTVCM}, on the other hand, does not tune the number of trees or learning rate; in our experiments, we ran the method with 200 trees and with the learning rate used in \citet{ZhouHooker2019}'s experiments. 
Section S3.1 of the Supplementary Materials contains additional details about the implementation of VCBART and these competitors.

\begin{figure}[ht]
\centering
\begin{subfigure}[b]{0.24\textwidth}
\centering
\includegraphics[width = \textwidth]{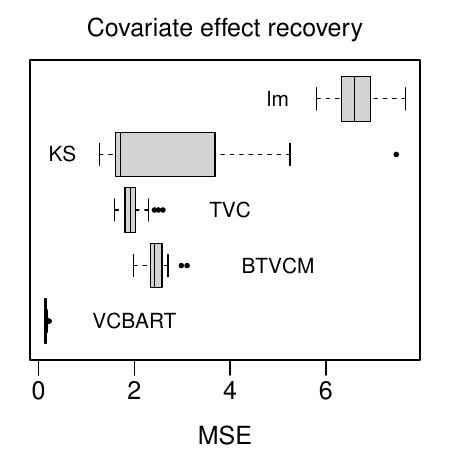}
\caption{}
\label{fig:p5R20_beta_mse}
\end{subfigure}
\begin{subfigure}[b]{0.24\textwidth}
\centering
\includegraphics[width = \textwidth]{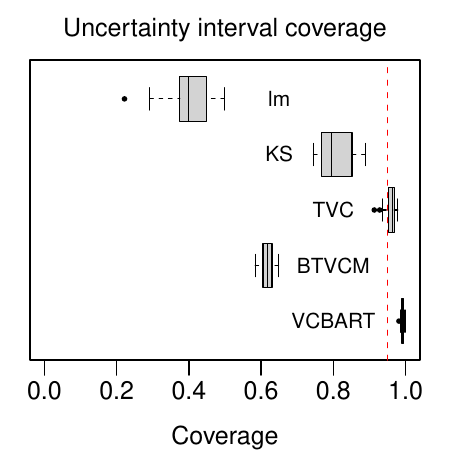}
\caption{}
\label{fig:p5R20_beta_cov}
\end{subfigure}
\begin{subfigure}[b]{0.24\textwidth}
\centering
\includegraphics[width = \textwidth]{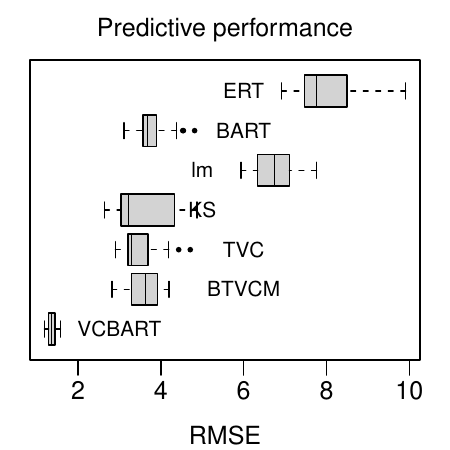}
\caption{}
\label{fig:p5R20_ystar_rmse}
\end{subfigure}
\begin{subfigure}[b]{0.25\textwidth}
\centering
\includegraphics[width = \textwidth]{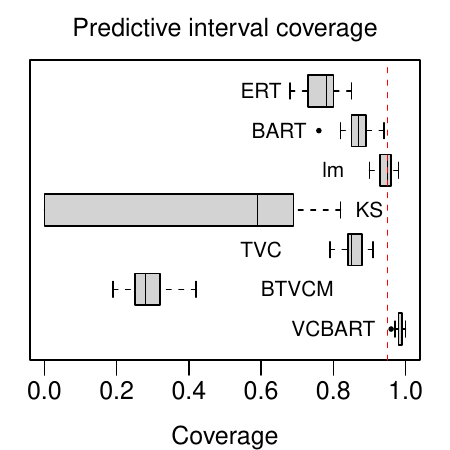}
\caption{}
\label{fig:p5R20_ystar_cov}
\end{subfigure}
\caption{(a) Average mean square error for estimating evaluations $\beta_{j}(\bz)$. (b) Average coverage of 95\% uncertainty intervals for evaluations $\beta_{j}(\bz).$ (c) Predictive root mean square error. (d) Coverage of 95\% prediction intervals. All measures reported over 25 testing datasets.}
\label{fig:p5R20_performance}
\end{figure}

We generated 25 different datasets comprising 250 training subjects and 25 testing subjects, each contributing four observations.
That is, we repeatedly trained each method using 1,000 total simulated observations and evaluated performance using an additional 100 observations.
As with Figure~\ref{fig:p5R20_beta}, we ran four chains VCBART for 2,000 iterations, discarding the first 1,000 as ``burn-in,'' with $M = 50$ trees per ensemble and setting each $\tau_{j} = 0.5/\sqrt{M}.$
Since the off-the-shelf implementations of \texttt{KS}, \revise{\texttt{TVC}}, and \texttt{BTVCM} do not return standard errors of the estimated covariate effects, we formed 95\% bootstrap confidence intervals for each $\beta_{j}(\bz)$ using 50 bootstrap resamples.
We ran these experiments in a shared high-throughput computing cluster \citep{chtc}.
Figures~\ref{fig:p5R20_beta_mse} and \ref{fig:p5R20_beta_cov} compare the out-of-sample mean square error for estimating $\beta_{j}(\bz)$ and uncertainty interval coverages, averaged across all six functions and over all 100 testing set observations.
We report function-by-function performance in Section S3.2 of the Supplementary Materials.

In terms of estimating out-of-sample evaluations $\beta_{j}(\bz),$ \texttt{lm}, which assumes that each covariate effect is constant with respect to $Z$, performed the worst while VCBART appeared the best.
VCBART additionally produced much more accurate predictions and better-calibrated uncertainty intervals than \texttt{BTVCM}, which also approximates each $\beta_{j}(\bZ)$ with a regression tree ensemble.
The higher coverage of VCBART's uncertainty intervals was not a by-product of exceedingly wide intervals.
In fact, VCBART's credible intervals for evaluations $\beta_{j}(\bz)$ tended to be shorter than the bootstrapped confidence intervals produced by \revise{\texttt{KS}, \texttt{BTVCM}, and \texttt{TVC}}.

Sequentially running four VCBART chains took, on average, about three minutes on our shared cluster.
Both \texttt{lm} and \texttt{BTVCM} were faster for these data, with \texttt{lm} running nearly instantaneously and \texttt{BTVCM} producing point estimates $\hat{\beta}_{j}(\bz)$ in about 90 seconds.
That said, both \texttt{lm} and \texttt{BTVCM} produced substantially worse point estimates than VCBART.
Although a carefully tuned \texttt{BTVCM} could yield better performance, we expect such tuning would take much longer.
We additionally found that obtaining point estimate $\hat{\beta}_{j}(\bz)$ with \revise{\texttt{TVC}} and \texttt{KS} \revise{respectively} about 1 hour and 9.6 hours, on average.
Sequentially running four chains of VCBART was substantially faster and produced better-calibrated uncertainty intervals than performing 50 bootstrap resamples for \texttt{BTVCM}, \texttt{KS}, \revise{and \texttt{TVC}}.
In summary, compared to existing methods for fitting varying coefficient models, VCBART was faster and produced more accurate predictions and better-calibrated uncertainty intervals off-the-shelf.

Intuitively, we should expect any method that reliably recovers individual out-of-sample evaluations $\beta_{j}(\bz)$ to yield highly accurate predictions of the outcome $Y.$
And indeed, Figure~\ref{fig:p5R20_ystar_rmse} shows that the out-of-sample predictive root mean square error (RMSE) of VCBART and the aforementioned competitors generally track with their estimation error.
Figure~\ref{fig:p5R20_ystar_rmse} additionally shows that VCBART achieved smaller predictive RMSE than two fully non-parametric methods, \texttt{BART} (implemented in \citet{bart_package}'s \textbf{BART} package) and extremely randomized trees (\texttt{ERT}; implemented in \citet{Wright2017}'s \textbf{ranger} package).
In our experiments, we allowed \texttt{BART} and \texttt{ERT} to use both $X$ and $Z$ to predict $Y.$ 
Although \texttt{BART} and \texttt{ERT} are facially more flexible than the other competitors, their predictive performance was worse than the correctly-specified varying coefficient methods.
Interestingly, however, \texttt{BART} and \texttt{ERT} produced predictive intervals with better coverage than the well-specified \texttt{BTVCM} and \texttt{KS} (Figure~\ref{fig:p5R20_ystar_cov}). 
That said, VCBART tended to return predictive intervals with better coverage than both \texttt{BART} and \texttt{ERT}.

To better understand how well VCBART scales to larger datasets, we conducted a second simulation study.
Similar to our first study, we generated 25 synthetic datasets with $n$ training subjects and 250 testing subjects, each contributing four observations.
We varied $n$ between 25 and 12,500, allowing us to see how VCBART performed when trained on as few as 100 and as many as 50,000 total observations. 
For brevity, we summarize the main findings of that study here and defer more detailed results to Section S3.3 of the Supplementary Materials.
Unsurprisingly, as $N$ increased, VCBART's estimation and prediction error decreased and the frequentist coverage of its credible intervals and posterior predictive intervals remained high.
In fact, VCBART produced uncertainty intervals with higher-than-nominal frequentist coverage when fit to as few as $N = 100$ total observations.
We additionally found that the median probability model's operating characteristics generally improved a $N$ increased.
For instance, the F1 score was 0.68 when $N = 200$ and exceeded 0.85 for $N > 1000.$
Averaging over 25 simulated datasets, sequentially running four VCBART chains for 2,000 iterations on an M1 Mac took around 55 seconds for $N = 1,000$ observations, 8.5 minutes for $N = 10,000$ observations, and around 48 minutes with $N = 50,000$ observations.

%% file: hrs_analysis.tex
To investigate how associations between later-life cognition and socioeconomic position at multiple life stages may vary over time and across sociodemographic characteristics, we analyzed publicly available data from the Health and Retirement Study (HRS).
The HRS is a nationally representative longitudinal survey of US adults over the age of 50 and their spouses of any age. 
Since 1992, the HRS has biennially assessed the economic, health, and social implications of aging through its core survey with response rates greater than 85\% in each wave.
The HRS is sponsored by the National Institute on Aging (NIA; U01AG009740) and is conducted by the University of Michigan. 
See \citet{Sonnega2014} for additional details about the HRS. 

In each wave, the HRS consistently administered a battery of cognitive tests that include tasks like listening to a series of 10 words and recalling as many possible immediately and several minutes later.
The HRS constructed a summary score that ranges from 0 to 35, with higher scores reflecting better cognitive functioning.
In our analysis, we use the total cognitive score as our outcome. 
Our covariate set includes measures of SEP at three distinct stages of life.
The first is a composite childhood SEP index, developed and validated by \citet{Vable2017}, which is based on the educational attainment and occupation of each subject's parents and their overall financial well-being in childhood.
Higher scores on this index correspond to higher SEP. 
We additionally included indicators of high school and college completion as proxies of SEP in early adulthood.
Finally, as measures of SEP in older adulthood, we used log-income and an indicator of labor force participation.
Our modifiers included age in months, gender, race, the U.S. Census division of subject's birthplace, and an indicator of whether the subject identified as Hispanic.

We restricted our analysis to HRS subjects who (i) had complete covariate, modifier, and outcome data in at least four HRS waves and (ii) were dementia-free in the first wave with complete data.
This resulted in analytic sample of $n = 10,812$ subjects who contributed a total of $N = 67,988$ person-years over the study period.
In total, we had $p = 5$ covariates and $R = 5$ potential modifiers.
Using the same hyperparameter setting as in Section~\ref{sec:simulation_studies} (i.e. $M = 50, \tau_{j} = 0.5/\sqrt{M}$), we simulated four Markov chains of 2,000 iterations each.
After discarding the first 1,000 iterations of each chain as ``burn-in'', we obtained a total of 4,000 MCMC samples from the VCBART posterior.
The $\hat{R}$ value based on the samples of $\sigma$ was 1.01.

Based on these samples, for each combination of $r$ and $j,$ we computed the posterior probability that the modifier $Z_{r}$ was used in at least one decision rule in the ensemble $\calE_{j}.$
We then computed the median probability model for each $\beta_{j}(\bz)$ by thresholding these probabilities at 0.5, as described in Section~\ref{sec:posterior_modifier_selection}. 
The median probability model included gender for all $\beta_{j}(\bZ)$ and age for all coefficient functions except for the one associated with college completion.
It also identified race as modifying the effects of childhood SEP, high school completion, and college completion but not later-life income or labor force participation.
Similarly, it identified birth place as modifying the effects of just childhood SEP and high school completion.
Taken together, these results suggest that, after adjusting for SEP in childhood and early-adulthood, difference in later-life cognitive score driven by differences in later-life SEP appear not to vary with respect to race and birthplace.

Further inspection revealed that the marginal posterior distributions of the estimated partial effects of later-life income for subjects in our dataset tended to concentrate within the interval [-0.3, 0.9].
For reference, the sample standard deviation of cognitive scores was about 4 points on a scale ranging rom 0 to 35 points.
This would suggest that, after adjusting for SEP in childhood and early adulthood, later-life income has a small predictive effect, if any, on later-life cognitive scores.

\begin{figure}[ht]

\begin{subfigure}[b]{0.32\textwidth}
\centering
\includegraphics[width = \textwidth]{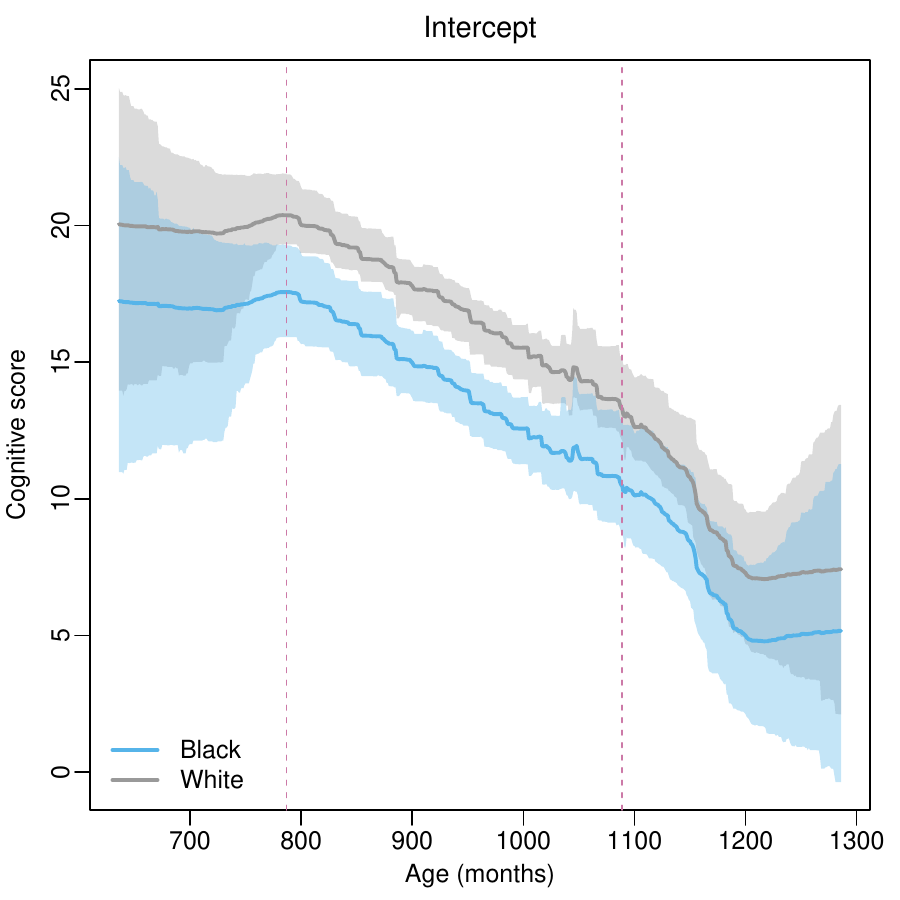}
\caption{}
\label{fig:hrs_intercept}
\end{subfigure}
\begin{subfigure}[b]{0.32\textwidth}
\centering
\includegraphics[width = \textwidth]{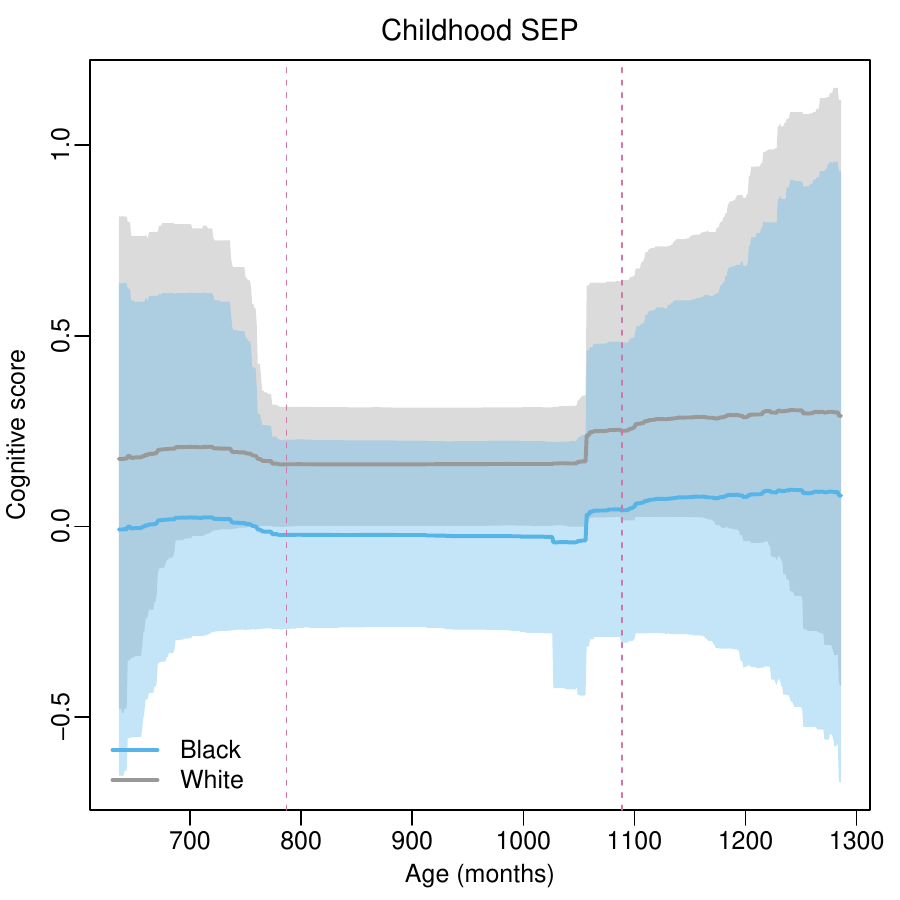}
\caption{}
\label{fig:hrs_cSEP}
\end{subfigure}
\begin{subfigure}[b]{0.32\textwidth}
\centering
\includegraphics[width = \textwidth]{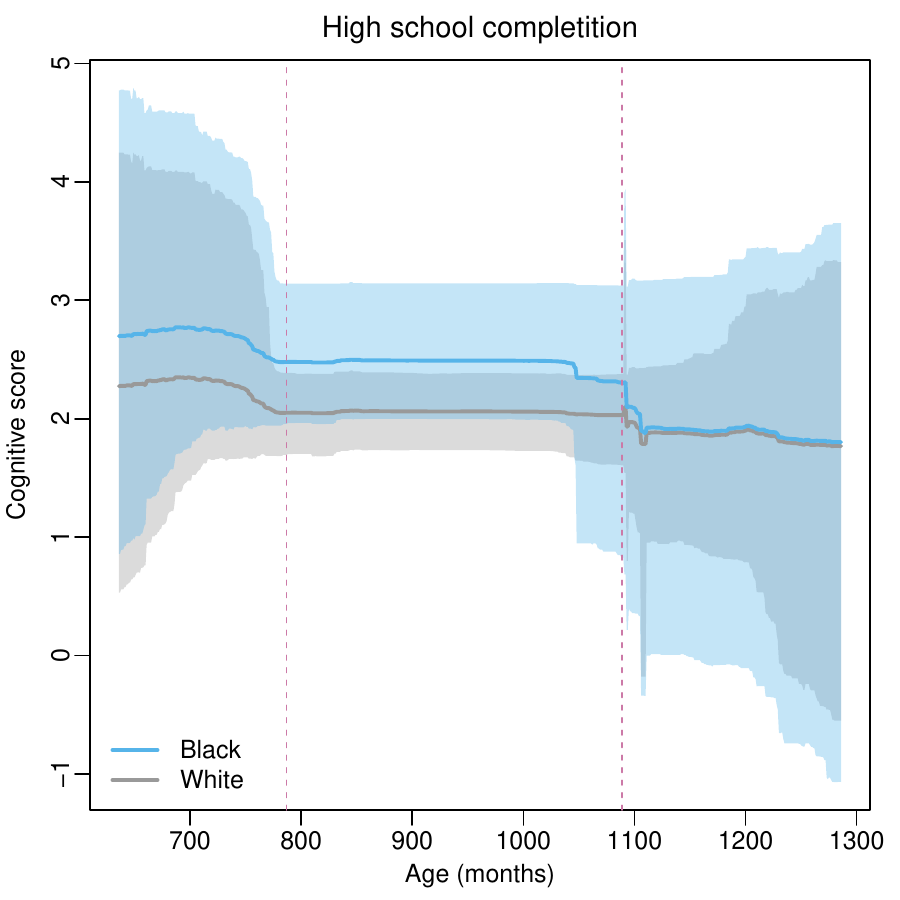}
\caption{}
\label{fig:hrs_hs}
\end{subfigure}
\caption{Posterior mean and pointwise 95\% credible intervals of the intercept (a) and partial effects of childhood SEP (b) and high school completion (c) on later-life cognitive score as functions of age. Dashed vertical lines delineate the 2.5\% and 97.5\% quantiles of the observed ages in our dataset.}
\label{fig:hrs_beta}
\end{figure}

Figure~\ref{fig:hrs_beta} shows the posterior mean and point-wise 95\% credible intervals of the predicted intercept and partial effects of childhood SEP and high school completion for two baseline individuals, one Black and one White.
Both individuals are females born in the East North Central Census Division of the U.S., which includes the states of Illinois, Indiana, Michigan, Ohio, and Wisconsin.
We can interpret the intercept as an average cognitive score for these baseline individuals if they (i) did not complete high school or college and (ii) did not report any earned income or labor force participation in older adulthood.

The dashed vertical lines in each panel show the 2.5\% and 97.5\% quantiles of the distribution of ages in our dataset. 
Unsurprisingly, there is much more uncertainty about the estimated effects outside of this age range. 
In Figure~\ref{fig:hrs_intercept}, we observe a steady decline in cognitive scores over time for both baseline individuals as well as a persistent gap between them, with the White individual having slightly higher baseline cognitive scores.
Figure~\ref{fig:hrs_cSEP} shows that the posterior distributions of the partial effect of childhood SEP for these individuals is concentrated on small positive and negative effect sizes.
In sharp contrast, however, we see in Figure~\ref{fig:hrs_hs} that high school completion is associated with higher cognitive scores for both baseline individuals.
Interestingly, we see that the estimated increase in cognitive score associated with high school completion is slightly higher for the Black baseline individual. 
Figure S12 in the Supplementary Materials plots the difference between these curves.

Recall that our median probability model detected temporal variation in the partial effect of high school completion.
At least for the two baseline individuals considered in Figure~\ref{fig:hrs_hs}, that temporal variation appears to manifest mostly in the lower tail of the distribution.
Between the ages of 65 and 90 years old, which covers approximately the middle 95\% of observed ages in our dataset, the partial effect appears, for all practical purposes, constant over time.

%% file: philly_analysis.tex
\begin{figure}[h]
\begin{subfigure}[b]{0.55\textwidth}
\centering
\includegraphics[width = \textwidth]{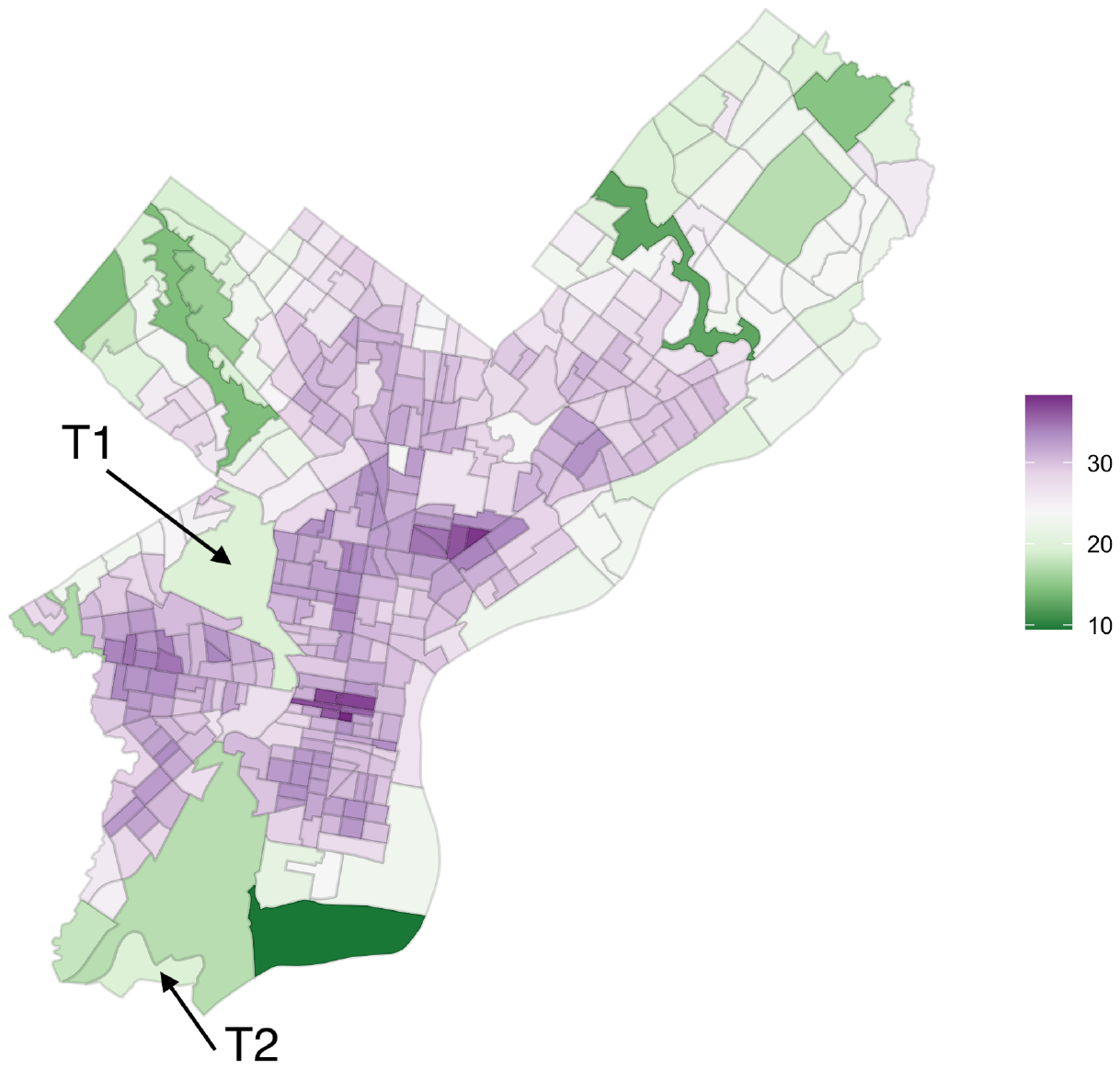}
\caption{}
\label{fig:philly_map}
\end{subfigure}
\begin{subfigure}[b]{0.44\textwidth}
\centering
\includegraphics[width = \textwidth]{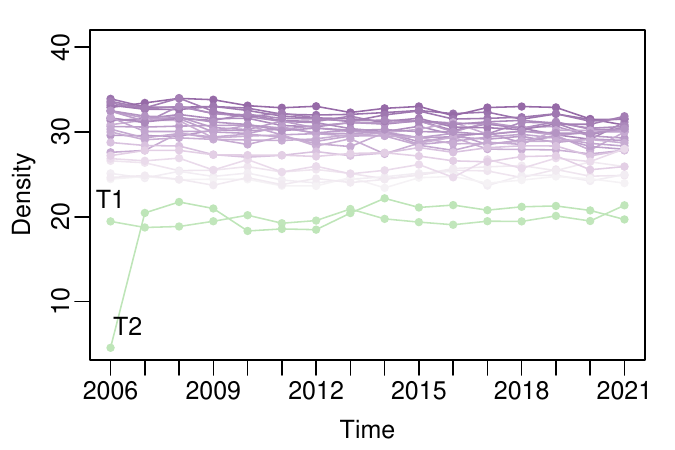}
\caption{}
\label{fig:philly_trace}
\end{subfigure}
\caption{Census tracts of Philadelphia colored according to the averaged transformed crime density (a). Trajectories of crime densities for tract labelled ``T1'' in (a), its neighbors, and the tract labelled ``T2'' (b). 
}
\label{fig:philly}
\end{figure}

The Philadelphia police department releases the latitude and longitude of every reported crime in the city at \url{opendataphilly.org}.
In each of the $n = 384$ census tracts in the city (Figure~\ref{fig:philly_map}), we computed the yearly crime density, defined as number of crimes per square mile, and applied an inverse hyperbolic sine transformation to counteract the considerable skewness. 
Let $y_{v,t}$ be the transformed crime density in census tract $v$ at time $t,$ with $t = 1$ corresponding to 2006 and $t = 16$ corresponding to 2021.
\citet{Balocchi2022_crime} modeled $y_{v,t} \sim \mathcal{N}(f^{(v)}(t), \sigma^{2})$ and computed first-order approximations of each $f^{(v)}.$
That is, they estimated tract-specific parameters $\beta_{0}^{(v)}$ and $\beta_{1}^{(v)}$ such that $f^{(v)}(t) \approx \beta_{0}^{(v)} + \beta_{1}^{(v)}\tilde{t},$ where $\tilde{t}$ is a centered and re-scaled version of the time index.
Rather than separately estimate each tract's parameters, they assumed that the intercepts and slopes were spatially clustered.
To estimate the latent clusterings, they ran several greedy searches across the space of pairs of census tract partitions.

Despite the reasonably accurate predictions obtained with their first-order approximation, a potential drawback of their analysis is the assumed monotonicity of crime over time.
An arguably more realistic model would allow for crime potential increases and decreases within each census tract.
Although it is conceptually easy to elaborate their model with higher-order terms, fitting such a model is computationally impractical.
Specifically, extending their approach to find an order-$d$ polynomial approximation involves searching over the product space of $d$ partitions.
The combinatorial vastness of this space renders efficient search all but infeasible. 

We can instead view their model and higher-order elaborations thereof as varying coefficient models in which the coefficients vary across the vertices of a network determined by census tract adjacency (Figure S13 in the Supplementary Materials).
Specifically, for a positive integer $d,$ we can approximate
\begin{equation}
\label{eq:philly_vc_model}
f^{(v)}(t) \approx \beta_{0}(v) + \beta_{1}(v)\tilde{t} + \beta_{2}(v)\tilde{t}^{2} + \cdots + \beta_{d}(v)\tilde{t}^{d},
\end{equation}
where we have suggestively expressed the $\beta_{d}$'s as functions of the census tract label $v.$

For a given approximation order $d,$ fitting the model in Equation~\eqref{eq:philly_vc_model} is relatively straightforward with VCBART.
To encourage spatial smoothness in the $\beta_{j}(v)$'s, we used a version of the BART prior that recursively partitions the tracts into spatially contiguous clusters.
This prior is identical to the one described in Section~\ref{sec:regression_tree_prior} except in how it draws the set $\cutset$ for decision rules involving the census tract label.
To draw such $\cutset$'s, we first compute the set $\mathcal{A}$ of available census tracts and then form a network whose vertices are the tracts in $\mathcal{A}$ and whose edges encode spatial adjacency relationships between those tracts.
Next, we sample a spanning tree of this network uniformly at random before deleting one edge at random, which partitions the network into two connected components.
These connected components correspond to spatially contiguous clusters of census tracts and we set $\cutset$ to be the tracts in one of these clusters.
See \citet[\S3.3]{Deshpande2022_flexBART} for more details about the prior.

We used VCBART to fit several versions of the model in Equation~\eqref{eq:philly_vc_model} with different approximation orders $d \in \{1, \ldots, 4\}$. 
We also consider two within-tract error correlation structure for each order $d$; the first structure assumes independence across observations while the second assumes that the errors are equally correlated across time points within each tract (i.e. compound symmetry). 
To compare the performance of each version, we formed 50 random training-testing splits and computed the in- and out-of-sample prediction error for each version (Table~\ref{tab:philly_loo}).
Each training split consisted of 90\% of the available data and the testing observations were held out uniformly at random.
We additionally fit a basic BART model using the implementation provided by \textbf{BART}.
\revise{
Running four MCMC chains for 2,000 iterations each for each VCBART model yielded Gelman-Rubin $\hat{R}$ values between 1.05 to 1.15.
Increasing the total number of iterations to 5,000 per chain yielded $\hat{R}$ values between 1.01 and 1.08.
}

\begin{table}[h]
\centering
\caption{In- and out-of-sample root mean square prediction error and run time (in hours) averaged over 50 replications in which 90\% of the data were used for training and 10\% for testing. Best performance in each column is bolded. \texttt{ind} stands for independent errors and \texttt{cs} stands for compound symmetry errors.}
\label{tab:philly_loo}
\begin{tabular}{lrrr} \hline
~ & RMSE (train) & RMSE (test) & Run time (hours) \\ \hline
$d = 1$, \texttt{ind} & 0.60 & 0.69 &  4.55\\
$d = 2$, \texttt{ind} & 0.53 & 0.65 &  7.61 \\
$d = 3$, \texttt{ind} & 0.51 & 0.64 &  10.44 \\
$d = 4$, \texttt{ind} & \textbf{0.49} & \textbf{0.63}  & 14 \\
$d = 1$, \texttt{cs} & 0.63 &0.71 & 4.68 \\
$d = 2$, \texttt{cs} & 0.56 & 0.67 & 8.35 \\
$d = 3$, \texttt{cs} & 0.54 & 0.66 & 11.01 \\
$d = 4$, \texttt{cs} & 0.51 &0.66 &  12.90 \\ 
\texttt{BART} & 0.68 & 0.75 & \textbf{2.4} \\ \hline
\end{tabular}
\end{table}

As we might expect, for both error structures, in-sample RMSE decreased with the approximation order $d.$
\revise{The out-of-sample RMSE also decreased as we increased $d.$} 
We further found that the compound symmetry model for the residual errors tended to produce worse in- and out-of-sample predictions.
Compared to BART, all of our VCBART models impose rather strong functional form assumptions about the temporal trend of crime rates.
On this view, we might regard BART as being much more flexible than VCBART since it does not make any such restrictions.
Nevertheless, we find that for these data, each of our VCBART models fits the data better than BART.
For all methods considered, the 95\% predictive intervals achieved coverages between 96\% and 98\% on both the training and testing data (results not shown).


\begin{figure}[ht]
\begin{subfigure}[b]{0.32\textwidth}
\centering
\includegraphics[width = \textwidth]{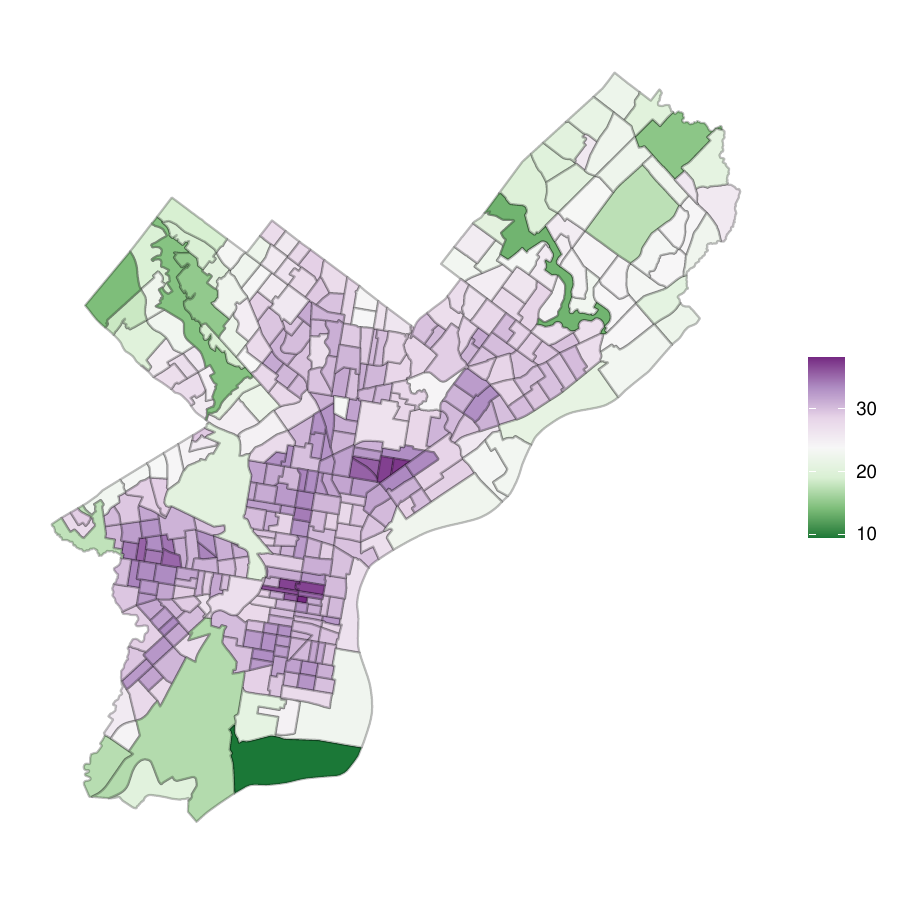}
\caption{}
\label{fig:philly_beta0}
\end{subfigure}
\begin{subfigure}[b]{0.32\textwidth}
\centering
\includegraphics[width = \textwidth]{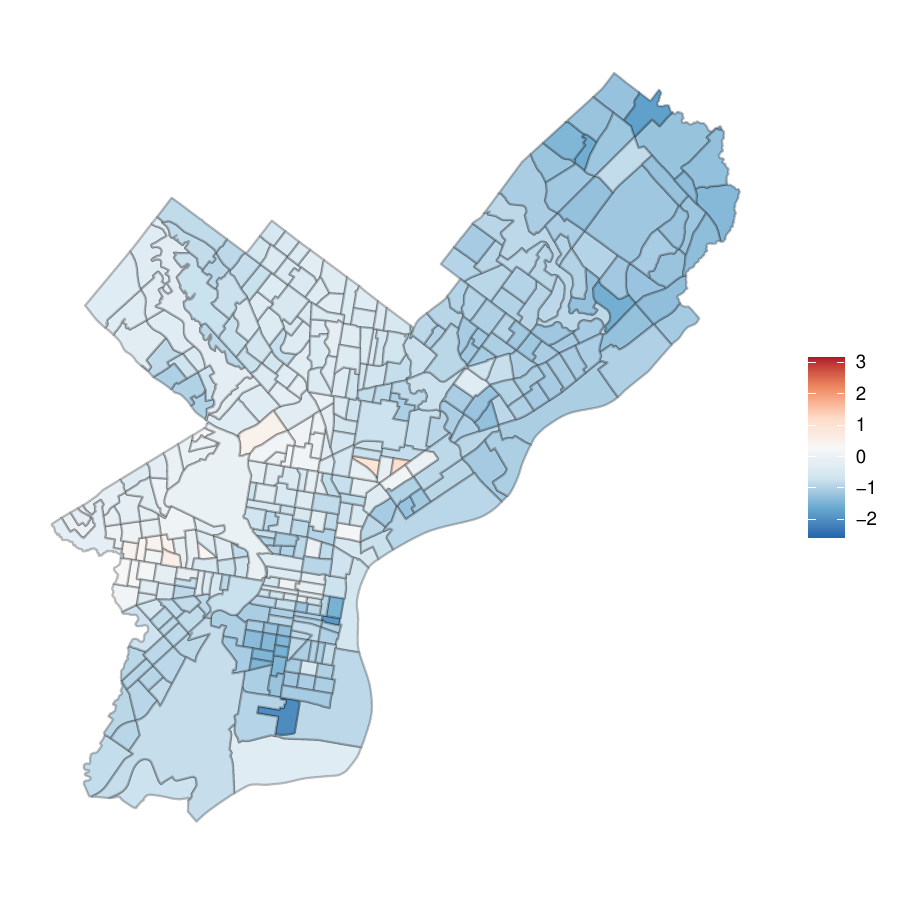}
\caption{}
\label{fig:philly_beta1}
\end{subfigure}
\begin{subfigure}[b]{0.32\textwidth}
\centering
\includegraphics[width = \textwidth]{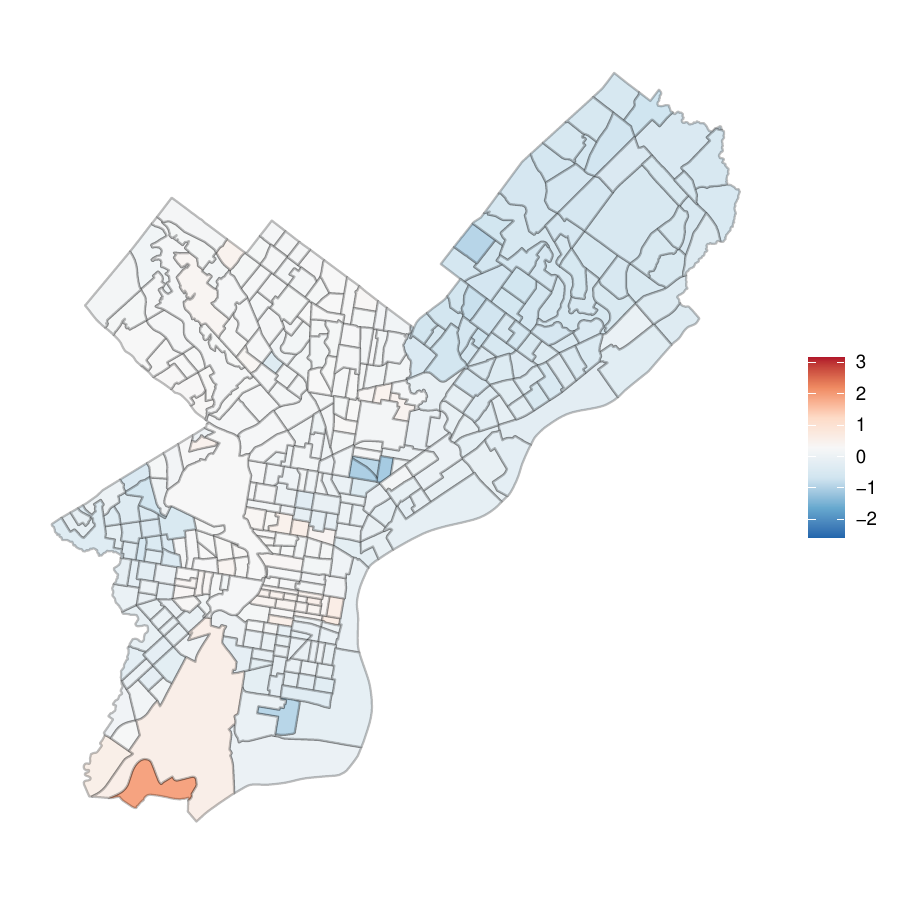}
\caption{}
\label{fig:philly_beta2}
\end{subfigure}

\begin{subfigure}[b]{0.32\textwidth}
\centering
\includegraphics[width = \textwidth]{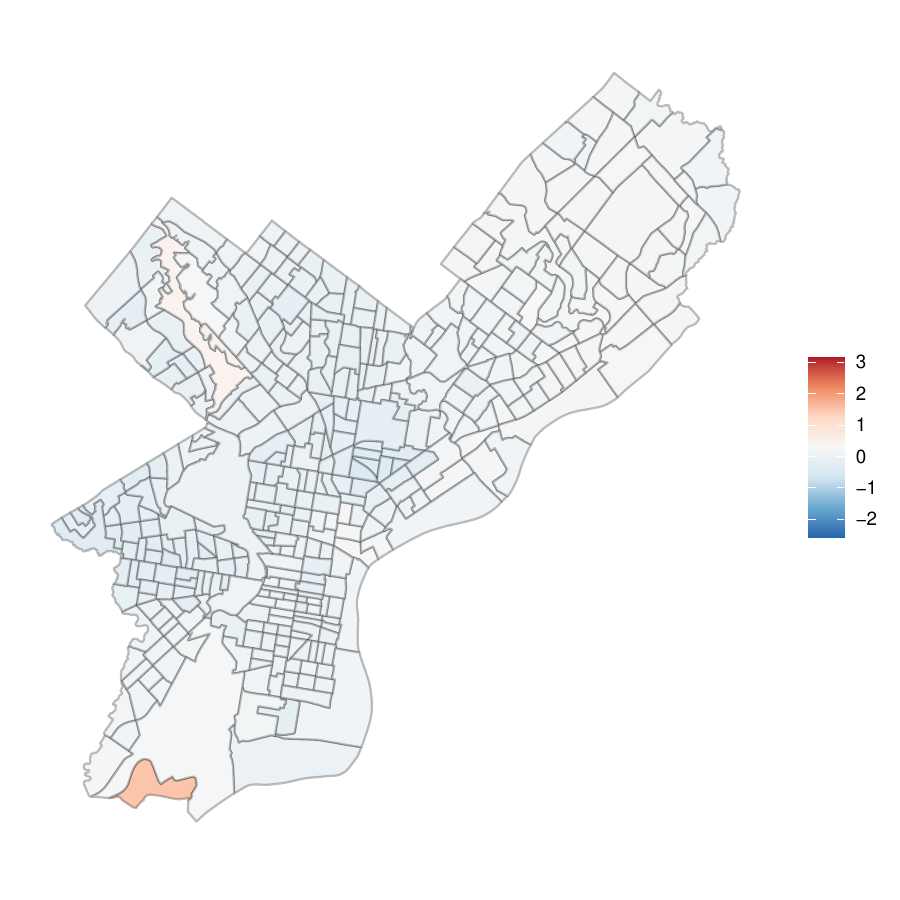}
\caption{}
\label{fig:philly_beta3}
\end{subfigure}
\begin{subfigure}[b]{0.32\textwidth}
\centering
\includegraphics[width = \textwidth]{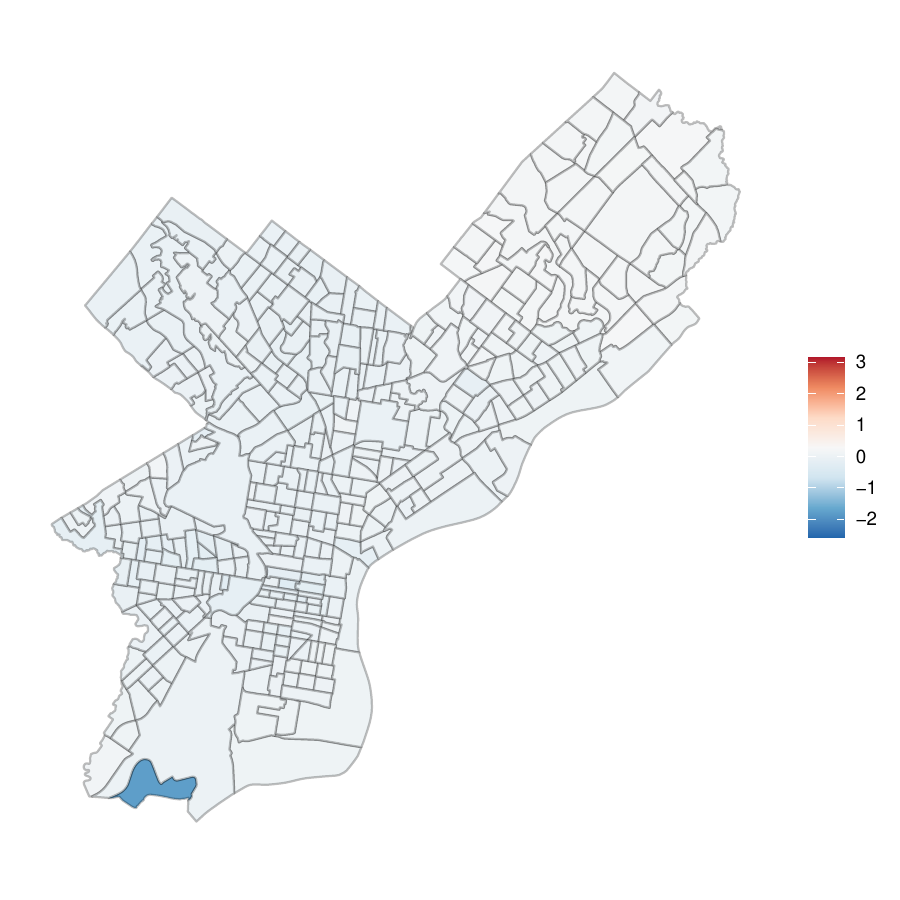}
\caption{}
\label{fig:philly_beta4}
\end{subfigure}
\caption{Posterior mean estimates of the varying coefficients in the best-fitting fourth-order approximation.}
\label{fig:philly_beta}
\end{figure}

Figure~\ref{fig:philly_beta} shows the posterior means of the coefficients from the best-fitting model, which was a fourth-order model.
The posterior mean of the first-order term $\beta_{1}$ was negative in 354 out of the 384 census tracts (92\%) and the posterior placed 95\% probability on negative $\beta_{1}$ values in 293 census tracts (76\%).
This finding is consistent with the general observation that crime fell city-wide in the early-half of the 2010's.
With one exception, the posterior distributions of $\beta_{3}$ and $\beta_{4}$ concentrated on very small values.
This finding is not entirely unexpected, in light of the relatively flat trace plots in Figure~\ref{fig:philly_trace} and for the other census tracts (not shown).
The one exception is a tract near the south of the city (labelled ``T2'' in Figure~\ref{fig:philly}), which experienced a dramatic increase in reported crimes from 2006 to 2007 and a somewhat oscillatory pattern over the next several years.
So while a quadratic polynomial was sufficient to capture the temporal variation in the rest of the city, a higher-order approximation was needed for this particular census tract.

%% file: discussion.tex
We have introduced VCBART for fitting linear varying coefficient models that retain the interpretability of linear models while displaying predictive accuracy on par with more flexible, blackbox regression models.
On simulated data, VCBART displayed covariate effect function recovery and predictive performance superior to existing state-of-the-art VC methods \textit{without intensive problem-specific tuning or imposing any structural assumptions about the functions $\beta_{j}(\bZ)$}.
It moreover returned coherent and generally well-calibrated uncertainty quantifications.
Theorem~\ref{thm:panel_concentration} shows that, up to a logarithmic factor, the VCBART posterior concentrates at minimax optimal rate.

In our analysis of the HRS data, we did not explicitly adjust for potential selection biases and confounding.
Indeed, it is possible that subjects with higher childhood SEP tended to be better educated and more well-off in later-life.
For a more rigorous study of causal effects of SEP on cognitive trajectories, we could use inverse-probability weights as in \citet{Marden2017}; indeed, incorporating observation-specific weights to the basic VCBART algorithm is relatively straightforward.
Relatedly, the Bayesian causal forest (BCF) models of \citet{Hahn2020} and \citet{Woody2020} are special cases of our more general VCBART model with single covariate. 
Studying VCBART's ability to estimate the simultaneous effects of multiple, possibly continuous, treatments under suitable identifying assumptions is an exciting avenue for future development.

In our motivating applications, the number of covariates $p$ was relatively small, compared to the number of total observation $N.$
One can easily imagine, however, settings in which $p$ is comparable to or larger than $N.$
In such settings, running VCBART is challenging as the coefficient functions are no longer identified without further assumptions like sparsity.
Under the assumption that only a handful of $\beta_{j}(\bZ)$'s are truly non-zero functions, we can estimate a sparse varying coefficient model by replacing the normal prior on tree jumps $\mu_{\ell}$ with spike-and-slab priors. 
Doing so involves introducing an ensemble-specific indicator and, conditionally on the relevant indicator, modeling all jumps in an ensemble as drawn from, for instance, a mean-zero Gaussian distribution with either a very small or very large variance. 
Such model elaboration inspires a natural Gibbs sampler that iterates between updating these indicators and updating trees within each ensemble.
Unfortunately, we would expect this Gibbs sampler to inherit the slow mixing of Gibbs samplers for spike-and-slab regression and for treed regression \citep{Ronen2022_mixing, KimRockova2023_mixing}.
A potentially more promising computational solution would combine existing fast mode-finding algorithms like EMVS \citep{RockovaGeorge2014_emvs} for sparse regression and the accelerated BART and BCF algorithms \citep{He2019, Krantsevich2021_xbcf} for treed regression.
Extending these accelerated algorithms to the general varying coefficient model setting is the subject of on-going work.

Because our implementation of VCBART utilizes hard decision trees, in which observations follow deterministic paths from the root to the leafs, it is unable to exactly recover smooth functions.
As demonstrated by \citet{Linero2022_softbart}, however, one can implement VCBART using soft decision trees where, for each $\bz$, the observations follow probabilistic paths from the root to the leafs \citep[see, e.g.,][and references therein]{LineroYang2018}.
Whereas a hard tree will output a single jump for a given $\bz$, a soft tree will output a weighted average of all jumps, thereby introducing additional smoothness. 
Extending our implementation of VCBART to support soft decision trees is the subject of ongoing work.

VCBART assumes homoscedastic, Gaussian residual errors and does not account for potential measurement errors of covariates. 
However, if these assumptions are violated, then the performance of VCBART may deteriorate. 
In the frequentist literature, there have been extensions to varying coefficient models that properly account for heteroscedasticity, skewed and/or heavy-tailed data, and measurement error \citep{ShenJMVA2014, XiongJAS2023, DongJoE2022}. 
VCBART can similarly be extended to these scenarios. 
For example, implementing a heteroskedastic version of VCBART in which the residual errors have variances of the form $\sigma^{2}(\bx, \bz)$ is relatively straightforward.
Indeed, following \citet{Pratola2019}, one could express log-variance $\log{\sigma^{2}(\bx,\bz)}$ with its own ensemble of regression trees.
Then in each MCMC iteration, one would update the trees in the variance ensemble conditionally on the ensembles for the mean function (i.e., the $\beta_{j}(\bZ)'s)$ and vice versa, just like the algorithm in \citet{Pratola2019}. 
Following \citet{Um2023SIM}, we could also implement a version of VCBART that is more robust to outliers by replacing the multivariate Gaussian residual error in \eqref{eq:panel_vc_model} with a multivariate skew-normal distribution.

While VCBART was shown to scale reasonably with sample size in Section~\ref{sec:simulation_studies}, further improvements in scalability are desirable. Recently, \citet{Luo2023_sharded} introduced a formal Bayesian procedure for dividing data into individual ``shards,'' fitting BART models to each shard, and combining predictions across shards.
We anticipate that a similarly sharded VCBART would be capable of scaling to even larger datasets than those considered here.
Finally, throughout this work we have assumed that the sets of covariates $X$ and effect modifiers $Z$ was fixed and known. 
While domain knowledge and theory often suggest natural choices of effect modifiers, we can readily envision scenarios in which it is not immediately clear whether a particular predictor should enter Equation~\eqref{eq:general_model} as a covariate, modifier, or both.
We leave automatic determination of the appropriate varying coefficient model specification to future work.

%% file: supplement_outline.tex
In Section~\ref{app:proofs}, we prove our main theoretical result (Theorem 3.1 in the main text).
We report the results of a sensitivity analysis to different model hyperparameter specifications and other algorithmic choices in Section~\ref{app:hyperparameter_sensitivity}.
We report additional simulation study results in Section~\ref{app:additional_simulations}.
We then derive our Gibbs sampling algorithm in Section~\ref{app:gibbs_sampler}.
Section~\ref{app:additional_figures} contains additional figures.

%% file: proofs.tex
\subsection{Proof overview and notation}

The proof of Theorem 3.1 (Section \ref{sec:main-thm-proof}) utilizes a technique recently developed by \cite{Ning2020} and \cite{Jeong2019}, which built upon the foundational theory pioneered by \citet{Ghosal2017}. 
In addition, we also leverage theory for Bayesian treed regression established by \cite{Rockova2019} and \cite{RockovaSaha2019}. The proof of Theorem 3.1 is broken up into two steps.

In the first step, we prove posterior contraction at the rate $r_N^2$ for VCBART with respect to the average R\'{e}nyi divergence of order 1/2 (defined below). 
We denote this divergence by $N^{-1} Q_{1/2}(f, f_0)$ where $f_0$ is the true density and $f$ is the VCBART density. 
This is proven by first showing that the VCBART prior places sufficient prior probability on small Kullback-Leibler neighborhoods around the true $f_0$. 
Then, we construct a sequence of approximating spaces (a sieve) $\mathcal{F}_N$ such that the prior probability \emph{outside} this sieve is exponentially small. 
Next, we construct a sequence of exponentially powerful statistical tests so that the probabilities of Type I and Type II error for testing $H_0: N^{-1} Q_{1/2}(f, f_0) = 0$ vs. $H_1: N^{-1 }Q_{1/2}(f,f_0) > C_1 r_N^2$ where $C_1 > 0$, are exponentially small for $f \in \mathcal{F}_N$. 
To construct these tests, we use the technique in \citet{Ning2020} and consider sequences of Neyman-Pearson likelihood ratio tests with alternatives in small subsets of the sieve. 
Finally, we show that the logarithm of the number of such subsets needed to cover the sieve can be asymptotically bounded above by $N r_N^2$.

In the second step, we convert the posterior contraction in average R\'{e}nyi divergence of order 1/2 to posterior contraction in terms of the squared empirical $\ell_2$ norm. 
In particular, because we assume Gaussian residual errors, the average R\'{e}nyi divergence of order 1/2 has a closed form which is particularly convenient to work with.

We use the following notation in our proofs. For two nonnegative sequences $\{ a_n \}$ and $\{ b_n \}$, we write $a_n \asymp b_n$ to denote $0 < \lim \inf_{n \rightarrow \infty} a_n/b_n \leq \lim \sup_{n \rightarrow \infty} a_n/b_n < \infty$. 
If $\lim_{n \rightarrow \infty} a_n/b_n = 0$, we write $a_n = o(b_n)$ or $a_n \prec b_n$.  
We use $a_n \lesssim b_n$ or $a_n = O(b_n)$ to denote that for sufficiently large $n$, there exists a constant $C >0$ independent of $n$ such that $a_n \leq Cb_n$. For a function $\beta$, $\lVert \beta \rVert_{\infty} = \max_{\bz \in [0,1]^{R}} | \beta (\bz)|$. 
For a vector $\boldsymbol{v}$, we denote its $\ell_2$ norm by $\lVert \boldsymbol{v} \rVert_2$. 
For a matrix $\boldsymbol{M}$, we denote its Frobenius norm by $\lVert \boldsymbol{M} \rVert_F$, its spectral norm by $\lVert \boldsymbol{M} \rVert_{sp}$, and its maximum entry in absolute value by $\lVert \boldsymbol{M} \rVert_{\max}$. 
For a square matrix $\bm{C}$, we let $\lambda_{\min} (\bm{C})$ and $\lambda_{\max} (\bm{C})$ denote its minimum and maximum eigenvalues and $\text{det}(\bm{C})$ denote its determinant.

For two densities $f$ and $g$, let $K(f,g) = \int f \log (f/g)$ and $V(f,g) = \int f | \log (f/g) - K(f/g) |^2$ denote the Kullback-Leibler (KL) divergence and KL variation respectively. Meanwhile, the R\'{e}nyi divergence of order 1/2 between $f$ and $g$ is denoted as $Q_{1/2}(f,g) = - \log \int f^{1/2} g^{1/2} d \nu$. 
The $\varepsilon$-covering number for a set $\Omega$ with semimetric $d$ is defined as the minimum number of $d$-balls of radius $\varepsilon$ needed to cover $\Omega$ and is denoted by $N(\varepsilon, \Omega, d)$. 
The metric entropy is the logarithm of the $\varepsilon$-covering number, i.e. $\log N (\varepsilon, \Omega, d)$. 
Finally, recall that $\bbeta$ and $\bbeta_{0}$ are $N \times (p+1)$ matrices whose respective $(i,j)$-th entries are $\beta_{j}(\bz_{it})$ and $\beta_{0,j}(\bz_{it})$, and $\lVert \bbeta - \bbeta_{0} \rVert_{N}^{2} = N^{-1}\sum_{i = i}^{n}{\sum_{t = 1}^{n_{i}}{\sum_{j = 0}^{p}{[\beta_{j}(\bz_{it}) - \beta_{0,j}(\bz_{it})]^{2}}}}$ is the squared empirical $\ell_2$ norm.

\subsection{Proof of Theorem 3.1} \label{sec:main-thm-proof}

Before proving Theorem 3.1, we first prove a preliminary lemma.

\begin{lemma} \label{Lemma:1}
	Let $\bm{\Sigma}_i(\rho)$ be the $n_i \times n_i$ covariance matrix for the $i$th subject whose $(j,k)$th entry is $\mathbbm{1}(j=k) + \rho~\mathbbm{1}(j \neq k), 1 \leq j, k \leq n_i$, with $0 \leq \rho < 1$. Recall that $n_{\max} = \max \{ n_1, \ldots, n_n \}$. Under the assumptions of Theorem 3.1, we have for any $(\sigma_1^2, \rho_1), (\sigma_2^2, \rho_2) \in (0, \infty) \times [0,1)$ and $\rho_0 \in [0,1)$, 
	\begin{enumerate}
		\item $\max_{1 \leq i \leq n} \lVert \sigma_1^2 \bm{\Sigma}_i (\rho_1) - \sigma_2^2 \bm{\Sigma}_i (\rho_2) \rVert_F^2 \lesssim n_{\max}^2 (\sigma_1^2 - \sigma_2^2) + n_{\max}^2 \sigma_2^4 | \rho_1 - \rho_2 |^2$.
		\item $1 \lesssim \min_{1 \leq i \leq n} \lambda_{\min} ( \bm{\Sigma}_i (\rho_0) ) \leq \max_{1 \leq i \leq n} \lambda_{\max} (\bm{\Sigma}_i (\rho_0)) \lesssim 1$.
	\end{enumerate}
\end{lemma}

\begin{proof}
	According to the proof of Theorem 9 in \cite{Jeong2019}, we have
	\begin{align*}
		\max_{1 \leq i \leq n} \lVert \sigma_1^2 \bm{\Sigma}_i(\rho_1) - \sigma_2^2 \bm{\Sigma}_i (\rho_2) \rVert_F^2 & \leq \frac{1}{n} \sum_{i=1}^{n} \lVert \sigma_1^2 \bm{\Sigma}_i (\rho_1) - \sigma_2^2 \bm{\Sigma}_i (\rho_2) \rVert_F^2 \\
		& \leq \frac{1}{n} \sum_{i=1}^{n} \left\{ (\sigma_1^2 - \sigma_2^2)^2 \lVert \bm{\Sigma}_i (\rho_1) \rVert_F^2 + \sigma_2^4 \lVert \bm{\Sigma}_i (\rho_1) - \bm{\Sigma}_i (\rho_2) \rVert_F^2 \right\} \\
		& \leq n_{\max}^2 (\sigma_1^2 - \sigma_2^2)^2 + \sigma_2^4 n_{\max}^2 | \rho_1 - \rho_2 |^2,
	\end{align*}
	where we used the fact that $n_i \leq n_{\max}$ for all $i = 1, \ldots, n$,  $\lVert \bm{\Sigma}_i (\rho_1) \rVert_{\max} = 1$, and $\lVert \bm{\Sigma}_i (\rho_1) - \bm{\Sigma}_i (\rho_2 ) \rVert_{\max} = | \rho_1 - \rho_2 |$ in the last line of the display. 
	This proves the first assertion of the lemma.
	
Also by the proof of Theorem 9 in \cite{Jeong2019}, for every $i = 1, \ldots, n$,
\begin{align*}
		1 - \rho_0 = \lambda_{\min} (\bm{\Sigma}_i(\rho_0)) \leq \lambda_{\max} (\bm{\Sigma}_i (\rho_0)) = 1 + (n_i - 1 )\rho_0.
\end{align*}
With $n_i \leq n_{\max}$ and $N = \sum_{i=1}^{n} n_i \leq n_{\max} n \asymp N$, it must be that $n_{\max} \asymp 1$ under Assumption (A3). 
The second assertion of the lemma now follows.
\end{proof}

\begin{proof}[Proof of Theorem 3.1]
Throughout this proof, we let $\xi_{it} = \beta_0(\bz_{it}) + \sum_{j=1}^{p} \beta_j (\bz_{it}) x_{itj}$, and let $\bm{\xi}_i = (\xi_{i1}, \ldots, \xi_{in_i})^{\top}$ denote the $n_i \times 1$ vector corresponding to the $i$th subject. Let $\bm{\xi} = (\bm{\xi}_1^{\top}, \ldots, \bm{\xi}_n^{\top})^{\top}$ be the $N \times 1$ vector of all the $\xi_{it}$'s. 
Let $\xi_{0,it} = \beta_{0,0} (\bz_{it}) + \sum_{j=1}^{p} \beta_{0,j} (\bz_{it}) x_{itj}$, and define $\bm{\xi}_{0,i} = (\xi_{0,i1}, \ldots, \xi_{0,in_i})^{\top}$ and $\bm{\xi}_0 = (\bm{\xi}_{0,1}^{\top}, \ldots, \bm{\xi}_{0,n}^{\top})^{\top}$. 
Let $\mathbb{P}_{\bm{\beta}_0}^{(N)}$ denote the probability measure underlying the true model,
\begin{equation} \label{true_panel_model}
	y_{it} = \beta_{0,0} (\bz_{it}) + \sum_{j=0}^{p} \beta_{0,j} (\bz_{it}) x_{itj} + \sigma_0 \varepsilon_{it}, \hspace{.3cm} 1 \leq i \leq n, 1 \leq t \leq n_i,
\end{equation}
where $\bm{\varepsilon}_i = (\varepsilon_{i1}, \ldots, \varepsilon_{in_i} )' \overset{ind}{\sim} \mathcal{N}_{n_i} (\bm{0}_{n_i}, \bm{\Sigma}_i(\rho_0)), \rho_0 \in [0,1)$. 

Recall $N = \sum_{i=1}^{n} n_i$ is the total number of observations. 
For the $i$th observation, the response $\bm{y}_i = (y_{i1}, \ldots, y_{in_i})^{\top}$ has the density $\bm{y}_i \sim \mathcal{N}_{n_i} (\bm{\xi}_i, \sigma^2 \bm{\Sigma}_i (\rho_0 ) )$. 
Let $f_i$ denote the density for $\bm{y}_i$ and $f = \prod_{i=1}^{n} f_i$. 
Analogously, we let $f_{0,i}$ be the density for $\bm{y}_i \sim \mathcal{N}_{n_i}(\bm{\xi}_{0,i}, \sigma^2 \bm{\Sigma}_i (\rho_0 ) )$, and let $f_0 = \prod_{i=1}^{n} f_{0,i}$. 
For shorthand, we denote $\bm{\Sigma}_i = \bm{\Sigma}_i(\rho)$ and $\bm{\Sigma}_{0,i} = \bm{\Sigma}_i(\rho_0)$, while $\bm{\Sigma} = \text{bdiag}(\bm{\Sigma}_1, \ldots, \bm{\Sigma}_n)$ and $\bm{\Sigma}_0 = \text{bdiag}(\bm{\Sigma}_{0,1}, \ldots, \bm{\Sigma}_{0,n})$ are $N \times N$ block diagonal matrices with respective diagonal blocks $\bm{\Sigma}_i$ and $\bm{\Sigma}_{0,i}$, $i =1, \ldots, n$.
\vspace{.05cm}

\noindent \textbf{Step 1: posterior contraction in R\'{e}nyi divergence of order 1/2.} We first prove that under the conditions of Theorem 3.1 and $r_N^2 = \log N \times \sum_{j=0}^{p} N^{-2 \alpha_j / (2 \alpha_j + R)}$,
\begin{equation} \label{renyi-contraction}
	\Pi \left( \frac{1}{N} Q_{1/2}(f, f_0) \geq C_1 r_N^2 \mid \bY \right) \rightarrow 0,
\end{equation}
in $\mathbb{P}_{f_0}^{(N)}$ probability as $N, R \rightarrow \infty$ for some $C_1 > 0$. As established in \cite{Ning2020}, \eqref{renyi-contraction} will be proven if we can show that for some $C_2, C_3 > 0$,
\begin{equation} \label{prior-concentration-condition-2}
\Pi \left( K(f_0, f) \leq N r_N^2, V(f_0, f) \leq N r_N^2 \right) \gtrsim \exp (- C_2 N r_N^2),
\end{equation}
and there exists a sieve $\{ \mathcal{F}_N \}_{N=1}^{\infty}$ such that
\begin{equation} \label{sieve-condition-2}
\Pi ( \mathcal{F}_N^c) \leq \exp(- C_3 N r_N^2),
\end{equation}
and a test function $\varphi_N$ such that
\begin{equation} \label{testing-conditions-2}
\begin{array}{l}
	\mathbb{E}_{f_0} \varphi_N \leq e^{-C_1 N r_N^2}, \\
	\displaystyle \sup_{f \in \mathcal{F}_N : \frac{1}{N} Q_{1/2}(f, f_0) > C_1 r_N^2} \mathbb{E}_f (1- \varphi_N) \lesssim e^{-C_1 N r_N^2 / 16}.
\end{array}
\end{equation} 
We first prove \eqref{prior-concentration-condition-2}. For $i = 1, \ldots, n$, we denote $\bm{\Sigma}_i^{\star} = (\sigma^2 / \sigma_0^2) \bm{\Sigma}_{0,i}^{-1/2} \bm{\Sigma}_i \bm{\Sigma}_{0,i}^{-1/2}$. Denote the ordered eigenvalues of $\bm{\Sigma}_i^{\star}$ by $\lambda_{it}, 1 \leq t \leq n_i$, and let $\bm{\Sigma}^{\star} = \textrm{bdiag} (\bm{\Sigma}_1^{\star}, \ldots, \bm{\Sigma}_n^{\star} )$ be the $N \times N$ block diagonal matrix with the $\bm{\Sigma}_i^{\star}$'s as the diagonal blocks. Noting that the $n$ subjects are independent, we have
\begin{align*}
& K(f_0, f) = \frac{1}{2} \left\{ \sum_{i=1}^{n} \sum_{t=1}^{n_i} (\lambda_{it} - 1 - \log \lambda_{it} ) + \frac{ \lVert \bm{\Sigma}^{-1/2} (\bm{\xi} - \bm{\xi}_0 ) \rVert_2^2}{\sigma^2} \right\}, \\
& V(f_0, f) = \left[ \sum_{i=1}^{n} \sum_{t=1}^{n_i} \frac{(1-\lambda_{it})^2}{2} \right] + \frac{\sigma_0^2}{(\sigma^2)^2} \lVert \bm{\Sigma}_0^{1/2} \bm{\Sigma}^{-1} (\bm{\xi} - \bm{\xi}_0 ) \rVert_2^2.  
\end{align*}
Define the sets,
\begin{align*}
& \mathcal{A}_1 = \left\{ ( \sigma^2, \rho) : \sum_{i=1}^{n} \sum_{t=1}^{n_i} ( \lambda_{it} - 1 - \log \lambda_{it} ) \leq N r_N^2,~\sum_{i=1}^{n} \sum_{t=1}^{n_i} (1 - \lambda_{it})^2 \leq N r_N^2 \right\}, \\
& \mathcal{A}_2 = \left\{ (\bm{\beta}, \sigma^2, \rho) : \frac{ \lVert \bm{\Sigma}^{-1/2} (\bm{\xi} - \bm{\xi}_0 ) \rVert_2^2}{\sigma^2} \leq N r_N^2,~\frac{\sigma_0^2}{(\sigma^2)^2} \lVert \bm{\Sigma}_0^{1/2} \bm{\Sigma}^{-1} ( \bm{\xi} - \bm{\xi}_0 ) \rVert_2^2 \leq \frac{N r_N^2}{2}  \right\}.
\end{align*}
Then $\Pi ( K(f_0, f) \leq N r_N^2, V(f_0, f) \leq N r_N^2 ) = \Pi ( \mathcal{A}_2 \mid \mathcal{A}_1 ) \Pi ( \mathcal{A}_1)$, and so we can consider $\Pi (\mathcal{A}_2 \mid \mathcal{A}_1)$ and $\Pi(\mathcal{A}_1)$ separately.

We first focus on lower-bounding $\Pi(\mathcal{A}_1)$. Arguing as in the proof of Lemma 1 of \cite{Jeong2019}, we can expand $\log \lambda_{it}$ in the powers of $(1-\lambda_{it})$ to get $\lambda_{it} - 1 - \log \lambda_{it}\sim (1-\lambda_{it})^2/2$. By the proof of Lemma 1 in \cite{Jeong2019}, we also have that $\sum_{t=1}^{n_i} (1- \lambda_{it})^2 \lesssim \lVert \sigma^2 \bm{\Sigma}_i - \sigma_0^2 \bm{\Sigma}_{0,i} \rVert_F^2$. Thus, for some constants $b_1, b_2 > 0$, we have as a lower bound for $\Pi(\mathcal{A}_1)$,
	\begin{align*} \label{Pi-A-lower-bound-1}
		\Pi(\mathcal{A}_1) & \geq \Pi \left( (\sigma^2, \rho) : \sum_{i=1}^{n} \lVert \sigma^2 \bm{\Sigma}_i - \sigma_0^2 \bm{\Sigma}_{0,i} \rVert_F^2 \leq b_1^2 N r_N^2  \right) \\
		& \gtrsim \Pi \bigg( (\sigma^2, \rho) : n_{\max}^2 (\sigma^2 - \sigma_0^2)^2 + \sigma_0^4 n_{\max}^2 | \rho - \rho_0 |^2 \leq b_2^2 n_{\max} r_N^2 \bigg) \\
		& \geq \Pi \left( \sigma^2 : | \sigma^2 - \sigma_0^2 | \leq \frac{b_2 r_N}{(2 n_{\max})^{1/2}} \right) \Pi \left( \rho: | \rho - \rho | \leq \frac{b_2 r_N}{ \sigma_0^2 (2 n_{\max})^{1/2}} \right), \numbereqn  
	\end{align*}
	where we used Lemma \ref{Lemma:1} and Assumption (A3) that $n_{\max} \asymp N/n$ in the second line. Note that since $\sigma^2 \sim \lambda \nu \chi_2^{-2}$, we also have $\sigma^2 \sim \text{InverseGamma}( \nu /2, \nu \lambda / 2 )$, and thus,
	\begin{align*} \label{Pi-A-lower-bound-2}
		\Pi \left( \sigma^2: | \sigma^2 - \sigma_0^2 | \leq \frac{b_2 r_N}{(2 n_{\max})^{1/2}} \right) & \gtrsim \int_{\sigma_0^2}^{\sigma_0^2 + \frac{b_2 r_N}{(2 n_{\max})^{1/2}}} (\sigma^2)^{-\nu/2 - 1} \exp \left( - \frac{\nu \lambda}{2 \sigma^2} \right) d \sigma^2 \\
		& \gtrsim \left( \sigma_0^2 + \frac{b_2 r_N}{(2 n_{\max})^{1/2}} \right)^{-\nu / 2 - 1} \\
		& \geq \exp \left( -\frac{C_2 N r_N^2}{4} \right), \numbereqn
	\end{align*}
	where we used Assumption (A3) in the last line, and $C_2$ is the constant in \eqref{prior-concentration-condition-2}. Since $\rho \sim \text{Uniform}(0,1)$, we also have 
	\begin{align*} \label{Pi-A-lower-bound-3}
		\Pi\left( \rho : | \rho - \rho_0 | \leq \frac{b_2 r_N}{\sigma_0^2 (2 n_{\max})^{1/2}} \right) & \geq \Pi \left( \rho : \rho_0 \leq \rho \leq \rho_0 + \frac{b_2 r_N}{\sigma_0^2 (2 n_{\max})^{1/2}} \right)	\\
		& =  \frac{b_2 r_N}{\sigma_0^2 (2 n_{\max})^{1/2}} \\
		& \geq \exp \left( - \frac{C_2 N r_N^2}{4} \right). \numbereqn 	
		\end{align*}
Combining \eqref{Pi-A-lower-bound-1}-\eqref{Pi-A-lower-bound-3} gives that
\begin{equation} \label{Pi-A-lower-bound-4}
	\Pi (\mathcal{A}_1) \gtrsim \exp \left( - \frac{C_2 N r_N^2}{2} \right), 
\end{equation}
where $C_2$ is the constant in \eqref{prior-concentration-condition-2}.

 Next we focus on bounding $\Pi ( \mathcal{A}_2 \mid \mathcal{A}_1 )$ from below. 
 Arguing as in Lemma 5.1 of \cite{Ning2020}, we have $\mathcal{A}_1 \supset \{ \lVert \sigma^{-2} \bm{\Sigma}^{-1} - \sigma_0^{-2} \bm{\Sigma}_0 \rVert_F \leq \epsilon_n / b_3 \}$ for some constant $b_2 > 0$ and sufficiently large $n$, which implies that $\lVert \sigma^{-2} \bm{\Sigma}^{-1} \rVert_{sp} \lesssim 1$ and $\lVert \bm{\Sigma}^{\star -1} \rVert_{sp} \lesssim 1$, where $\bm{\Sigma}^{\star}$ was defined in the paragraph below \eqref{testing-conditions-2}. Therefore, conditioning on $\mathcal{A}_1$,
 \begin{equation*}
 		\frac{\lVert \bm{\Sigma}^{-1/2} (\bm{\xi} - \bm{\xi}_0) \rVert_2^2}{\sigma^2} \lesssim \frac{\lVert \bm{\xi} - \bm{\xi}_0 \rVert_2^2}{\sigma^2},
 	\end{equation*}
 	and
 	\begin{equation*}
 		\frac{\sigma_0^2}{(\sigma^2)^2} \lVert \bm{\Sigma}_0^{1/2} \bm{\Sigma}^{-1} (\bm{\xi} - \bm{\xi}_0) \rVert_2^2 \lesssim \frac{\lVert \bm{\xi} - \bm{\xi}_0 \rVert_2^2}{\sigma^2}. 
 	\end{equation*}
 By Lemma \ref{Lemma:1} and using the fact that the eigenvalues of a block-diagonal matrix are the eigenvalues of the individual blocks, we also have that $\lVert \bm{\Sigma} \rVert_{sp} \lesssim 1$. Therefore, $ \sigma^{-2} \lVert \bm{\Sigma}^{-1} \rVert_{sp} = \lVert \sigma^{-2} \bm{\Sigma}^{-1} \rVert_{sp} \lesssim 1$ implies that $\sigma^{-2} \lesssim 1$ on event $\mathcal{A}_1$. It thus follows that conditional on $\mathcal{A}_1$, both expressions on the left-hand side of the inequalities in the set $\mathcal{A}_2$ can be bounded above by a constant multiple of $\lVert \bm{\xi} - \bm{\xi}_0 \rVert^2$.
  Thus, for some constant $b_4 > 0$, we have as a lower bound for $\Pi ( \mathcal{A}_2 \mid \mathcal{A}_1 )$,
\begin{align*} \label{lower-bound-A2condA1-panel}
& \Pi ( \mathcal{A}_2 \mid \mathcal{A}_1 ) \geq \Pi \left( \bm{\beta}: \lVert \bm{\xi} - \bm{\xi}_0 \rVert_2^2 \leq \frac{N r_N^2}{2 b_4}  \right) \\
& ~ = \Pi \left( \bm{\beta}: \sum_{i=1}^{n} \sum_{t=1}^{n_i} \left[\left\{ \beta_0 (\bz_{it}) - \beta_{0,0}(\bz_{it}) \right\} + \sum_{j=1}^{p} \left\{ \beta_j(\bz_{it}) - \beta_{0,j}(\bz_{it}) \right\} x_{itj} \right]^2 \leq \frac{N r_N^2}{2 b_4} \right) \\ 
& ~ \geq \Pi \left( \bm{\beta} : \sum_{i=1}^{n} \sum_{t=1}^{n_i} \sum_{j=0}^{p} [ \beta_j(\bz_{it}) - \beta_{0,j}(\bz_{it}) ]^2 \leq \frac{N r_N^2}{2 b_4 D^2 (p+1)} \right) \\
& ~ \geq \displaystyle \prod_{j=0}^{p} \Pi \left( \beta_j: \sum_{i=1}^{n} \sum_{t=1}^{n_i} [ \beta_j(\bz_{it}) - \beta_{0,j}(\bz_{it}) ]^2 \leq \frac{N (r_N^{j})^2}{2 b_4 D^2 (p+1)} \right) \\
& ~ \geq \prod_{j=0}^{p} \Pi \left( \beta_j : \lVert \beta_j - \beta_{0,j} \rVert_N \leq \frac{r_N^j}{D [2 b_4 (p+1)]^{1/2}} \right), \numbereqn
\end{align*} 
where $r_N^j =  (\log N)^{1/2} N^{-\alpha_j / (2 \alpha_j + R)}$ for $j=0, \ldots, p$, and we used Assumption (A2) that $|x_{itj} | \leq D, D > 1$, for all $i = 1, \ldots, n, t = 1, \ldots, n_i, j = 1, \ldots, p$, and an application of the Cauchy-Schwarz inequality in the third line of the display. Since we want to show that $\Pi(\mathcal{A}_2 \mid \mathcal{A}_1 ) \gtrsim \exp (- C_2 N r_N^2 / 2)$, it suffices to show (based on the last line of \eqref{lower-bound-A2condA1-panel}) that for each varying coefficient function $\beta_j, j=0, \ldots, p$,
\begin{equation} \label{lower-bound-functionals-panel}
\Pi \left( \beta_j : \lVert \beta_j - \beta_{0,j} \rVert_N \leq \frac{r_N^j}{D [2 b_4 (p+1)]^{1/2}} \right) \gtrsim e^{-C_2 N (r_N^j)^2 / 2},
\end{equation}
where $C_2 > 0$ is the constant in \eqref{prior-concentration-condition-2}. The lower bound in \eqref{lower-bound-functionals-panel} holds due to Assumption (A1) that $\lVert \beta_{0,j} \rVert_{\infty} < \infty$ for all $0 \leq j \leq p$, Assumption (A3) that $R=O((\log N)^{1/2})$ and $p=O(1)$, and Equation (3) in the proof of Theorem 7.1 of \citet{RockovaSaha2019}. Thus, combining \eqref{lower-bound-A2condA1-panel}-\eqref{lower-bound-functionals-panel}, we have that
\begin{equation} \label{Pi-A2A1-lower-bound}
	\Pi(\mathcal{A}_2 \mid \mathcal{A}_1 ) \gtrsim \exp \left( - \frac{C_2 N r_N^2}{2} \right).
\end{equation}
It now follows from \eqref{Pi-A-lower-bound-4} and \eqref{Pi-A2A1-lower-bound} that
\begin{equation*}
	\Pi( K(f_0, f) \leq N r_N^2, V(f_0, f) \leq N r_N^2) = \Pi (\mathcal{A}_2 \mid \mathcal{A}_1) \Pi (\mathcal{A}_1) \gtrsim \exp \left( - C_2 N r_N^2 \right),
\end{equation*}
which proves the prior mass condition \eqref{prior-concentration-condition-2}.

We now construct our sieve $\mathcal{F}_N$ so that the condition in \eqref{sieve-condition-2} holds. Specifically, for some constant $C_4 > 0$, we consider the sieve,
\begin{small}
	
	\begin{align} \label{sieve}
	\mathcal{F}_N = \left\{ (\bm{\beta}, \sigma^2, \rho): 0 < \sigma^2 < e^{2 C_4 N r_N^2 / \nu}, e^{-C_4 N r_N^2} < \rho < 1 - e^{-C_4 N r_N^2}, \beta_j \in \mathcal{B}_N^{j}, j=0, \ldots, p \right\},
	\end{align}
	\end{small}
	\noindent where for each $j=0, \ldots, p$, $\mathcal{B}_N^{j}$ is defined as the sieve in the proof of Theorem 7.1 of \citet{RockovaSaha2019}, i.e., the union of all functions belonging to the class of functions $\mathcal{F}_{\varepsilon}$ defined in Equation (5.5) of \citet{Rockova2019}) that are supported on a valid ensemble $\mathcal{V} \mathcal{E}$ (Definition 5.3 of \cite{Rockova2019}), where \textit{each} tree in the $j$th ensemble $\mathcal{E}_j$ has at most $l_N^{j}$ leaves, and $l_N^{j}$ is chosen as $l_N^{j} = \lfloor C_5 N (r_N^j)^2 / \log N \rfloor \asymp N^{R / (2 \alpha_j + R)}$, for sufficiently large constant $C_5 > 0$. With this choice of sieve, we have 
	\begin{equation} \label{upper-bound-sieve-complement-1}
	\Pi (\mathcal{F}_N^c) \leq \Pi ( \sigma^2 \geq e^{2 C_4 N r_N^2 / \nu} ) + \Pi( \rho \leq e^{-C_4 N r_N^2} ) + \Pi (\rho \geq 1 - e^{-C_4 N r_N^2}) + \displaystyle \sum_{j=0}^{p} \Pi (\beta_j \notin \mathcal{B}_n^{j}).
	\end{equation}
	Since the prior for $\sigma^2$ is $\sigma^2 \sim \lambda \nu \chi_2^{-2}$, we have
	\begin{align*} \label{upper-bound-sieve-complement-2}
		\Pi( \sigma^2 \geq e^{2 C_4 N r_N^2 / \nu} ) & \asymp \int_{e^{2 C_4 N r_N^2 / \nu}}^{\infty} (\sigma^2)^{-\nu/2-1} \exp \left( - \frac{\nu \lambda}{2 \sigma^2} \right) d \sigma^2 \\
		& \lesssim \int_{e^{2 C_4 N r_N^2 / \nu}}^{\infty} (\sigma^2)^{-\nu/2-1} \\
		& \asymp e^{- C_4 N r_N^2}. \numbereqn
	\end{align*}
	 Since $\rho \sim \text{Uniform}(0,1)$, it is also clear that
	 \begin{equation} \label{upper-bound-sieve-complement-3}
	 	\Pi(\rho \leq e^{-C_4 N r_N^2}) = e^{-C_4 N r_N^2}~~\text{ and }~~\Pi(\rho \geq 1 - e^{C_4 N r_N^2}) = e^{-C_4 N r_N^2}.
	 \end{equation}
We now focus on bounding the final term in \eqref{upper-bound-sieve-complement-1} from above. Let $L_{m}^{j}$ denote the number of terminal leaf notes in the $m$th tree of the $j$th tree ensemble $\mathcal{E}_j$ corresponding to the varying coefficient $\beta_j(\bz)$. Noting that all the BART priors have the same number of trees $M$, we have that the second term in \eqref{upper-bound-sieve-complement-1} can be bounded above as
\begin{align*} \label{upper-bound-sieve-complement-4}
\displaystyle \sum_{j=0}^{p} \Pi ( \beta_j \notin \mathcal{B}_n^j ) & \leq \displaystyle \sum_{j=0}^{p} \Pi \left( \displaystyle \bigcup_{m=1}^{M} \{ L_m^j > l_N^{j} \} \right) \\
& \leq \displaystyle \sum_{j=0}^{p} \sum_{m=1}^{M} \Pi ( L_m^j > l_N^j ) \\
& \lesssim M \sum_{j=0}^{p} e^{-\widetilde{D}_j l_N^j \log l_N^j } \\
& \lesssim e^{-C_6 N r_N^2},  \numbereqn
\end{align*}
where each $\widetilde{D}_j > 0$ is a constant corresponding to the $j$th ensemble, and $C_6 > 0$ is some constant. 
In the above display, we used Corollary 5.2 of \citet{RockovaSaha2019} in the third line. 
In the last line, we used Assumption (A3) that $p = O(1)$ and the fact that $l_N^{j} = \lfloor C_5 N (r_N^j)^2 / \log N \rfloor$, so each individual term in the summation in the third line can be bounded above by a constant multiple of $\exp(-C_6 N r_N^2 - \log (p+1))$.
Combining \eqref{upper-bound-sieve-complement-1}-\eqref{upper-bound-sieve-complement-4} gives that the $\Pi(\mathcal{F}_N^c) \leq e^{-C_3 N r_N^2}$ for some $C_3 > 0$, i.e. the first condition in \eqref{sieve-condition-2} holds for the sieve $\mathcal{F}_N$ constructed in \eqref{sieve}.

Finally, we establish the existence of a test function $\varphi_N$ so that \eqref{testing-conditions-2} holds. 
Let $\bm{\Sigma}_i = \bm{\Sigma}_i(\rho)$, $\bm{\Sigma}_{1,i} = \bm{\Sigma}_i(\rho_1)$, and let $\bm{\Sigma} = \text{bdiag}(\bm{\Sigma}_1, \ldots, \bm{\Sigma}_n)$ and $\bm{\Sigma}_1 = \text{bdiag}(\bm{\Sigma}_{1,1}, \ldots, \bm{\Sigma}_{1,n})$ be $N \times N$ block diagonal matrices with respective diagonal blocks $\bm{\Sigma}_i$ and $\bm{\Sigma}_{1,i}, i = 1, \ldots, n$.
As in \citet{Ning2020}, we first consider the most powerful Neyman-Pearson test $\phi_N = \mathbbm{1} \{ f_1 / f_0 \geq 1 \}$ for any density $f_1 \in \mathcal{F}_N$ such that $N^{-1}Q_{1/2}(f_0, f_1) \geq C_1 r_N^2$, where $C_1 > 0$ is the constant in \eqref{renyi-contraction}, and $f_1  = \prod_{i=1}^{n} f_{1,i}$ where $f_{1,i}$ is the density $\mathcal{N}_{n_i}( \bm{\xi}_{1,i}, \sigma_1^2 \bm{\Sigma}_{1,i})$. 
As shown in \cite{Jeong2019} and \cite{Ning2020}, if the average R\'{e}nyi divergence of order 1/2 between $f_0$ and $f_1$ is bigger than $C_1 r_N^2$, then
\begin{equation} \label{Neyman-Pearson-test-panel}
\begin{array}{l}
\mathbb{E}_{f_0} \phi_N \leq e^{- C_1 N r_N^2}, \\
\mathbb{E}_{f_1} (1 - \phi_N ) \leq e^{- C_1 N r_N^2}.
\end{array}
\end{equation}
For any $f \in \mathcal{F}_N$ where $f = \prod_{i=1}^{n} f_i$ and $f_i$ is the density $\mathcal{N}_{n_i}( \bm{\xi}_i, \sigma^2 \bm{\Sigma}_i ),$ we have by the Cauchy-Schwarz inequality that
\begin{equation} \label{upper-bound-type-ii-error}
\mathbb{E}_{f} (1-\phi_N ) \leq \{ \mathbb{E}_{f_1} (1-\phi_N) \}^{1/2} \{ \mathbb{E}_{f_1} (f / f_1)^2 \}^{1/2},
\end{equation}
and thus, following from \eqref{upper-bound-type-ii-error} and the second inequality in \eqref{Neyman-Pearson-test-panel}, we can control the probability of Type II error properly as in \eqref{testing-conditions-2} if $\mathbb{E}_{f_1} (f / f_1)^{2} \leq e^{7 C_1 N r_N^2 / 8}$. 

We first show that $\mathbb{E}_{f_1}(f/f_1)^2$ is bounded above by $e^{7 C_1 Nr_N^2/8}$ for every density $f_1$ with parameters $(\bbeta_1, \sigma_1^2, \rho_1)$ satisfying
\begin{equation} \label{densities-in-f1-panel}
\lVert \bm{\xi} - \bm{\xi}_1 \rVert_2^2 \leq \frac{C_1 N r_N^2}{16}~~ \text{ and }~~
\frac{1}{n} \lVert \sigma^2 \bm{\Sigma} - \sigma_1^2 \bm{\Sigma}_1 \rVert_F^2 \leq \frac{C_1^2 r_N^4}{4 n_{\max}^2},
\end{equation}
where $\bm{\xi}_1 = ( \bm{\xi}_{1,1}^\top, \ldots, \bm{\xi}_{1,n}^\top )^\top$ and $\bm{\xi}_{1,i}, i = 1, \ldots, n$, is the $n_i$-dimensional vector whose $t$th entry is $\xi_{1,it} = \beta_{1,0} (\bz_{it}) + \sum_{j=1}^{p} \beta_{1,j} (\bz_{it}) x_{itj}$.
	Let $\bm{\Sigma}_{1,i}^{\star} = (\sigma_1^2 / \sigma^2) \bm{\Sigma}_i^{-1/2} \bm{\Sigma}_{1,i} \bm{\Sigma}_i^{-1/2}$. We have by Lemma \ref{Lemma:1} and \eqref{densities-in-f1-panel} that
\begin{align*}
	\max_{1 \leq i \leq n} \lVert \bm{\Sigma}_{1,i}^{\star} - \bm{I}_{n_i} \rVert_{sp}^2 & \leq \max_{1 \leq i \leq n} \lVert \sigma^{-2} \bm{\Sigma}_i^{-1} \rVert_{sp}^2 \cdot \lVert \sigma^2 \bm{\Sigma}_i - \sigma_1^2 \bm{\Sigma}_{1,i} \rVert_{sp}^2 \\
	& \lesssim \frac{1}{n} \sum_{i=1}^{n} \lVert \sigma^2 \bm{\Sigma}_i - \sigma_1^2 \bm{\Sigma}_{1,i} \rVert_F^2 \\
	& \lesssim \frac{r_N^4}{4 n^2_{\max}}. 
\end{align*}
Furthermore, $\max_{1 \leq i \leq n} \lVert \bm{\Sigma}_{1,i}^{\star} - \bm{I}_{n_i} \rVert_{sp}$ is bounded below by $\max_{1 \leq i \leq n} | \text{eig}_k (\bm{\Sigma}_{1,i}^{\star}) - 1 |$ for every $k \leq n_i$, where $\text{eig}_k$ denotes the $k$th ordered eigenvalue of $\bm{\Sigma}_{1,i}^{\star}$. Thus, we have
\begin{equation} \label{densities-in-f1-panel-2}
	1 - \frac{r_N^2}{2 n_{\max}} \lesssim \min_{1 \leq i \leq n} \lambda_{\min} ( \bm{\Sigma}_{1,i}^{\star} ) \leq \min_{1 \leq i \leq n} \lambda_{\max} ( \bm{\Sigma}_{1,i}^{\star} ) \lesssim 1 + \frac{r_N^2}{2 n_{\max}}.
\end{equation} 
Since $r_N^2 / n_{\max} \rightarrow 0$, \eqref{densities-in-f1-panel-2} implies that $2 \bm{\Sigma}_{1,i}^{\star} - \bm{I}_{n_i}$ is nonsingular for every $1 \leq i \leq n$. Thus, for every density $f_1$ with parameters $(\bm{\beta}_1, \sigma_1^2, \rho_1)$ satisfying \eqref{densities-in-f1-panel}, we can explicitly calculate
\begin{align*} \label{densities-in-f1-panel-3}
	\mathbb{E}_{f_1} (f/f_1)^2 & \lesssim \prod_{i=1}^{n} \left\{ \text{det}(\bm{\Sigma}_{1,i}^{\star})^{1/2} \text{det} \left( 2 \bm{I}_{n_i} - \bm{\Sigma}_{1,i}^{\star -1} \right)^{-1/2} \right\} \\
	& \qquad \exp \left\{ \sum_{i=1}^{n} \lVert (2 \bm{\Sigma}_{1,i}^{\star} - \bm{I}_{n_i})^{-1/2} \sigma^{-1} \bm{\Sigma}_i^{-1/2} (\bm{\xi}_i - \bm{\xi}_{1,i}) \rVert_2^2 \right\}. \numbereqn
\end{align*}
Arguing as in Equations (29)-(30) in the proof of Lemma 2 of \cite{Jeong2019}, we have that
\begin{equation} \label{densities-in-f1-panel-4}
	\prod_{i=1}^{n} \left\{ \text{det}(\bm{\Sigma}_{1,i}^{\star})^{1/2} \text{det} ( 2 \bm{I}_{n_i} - \bm{\Sigma}_{1,i}^{\star -1} )^{-1/2} \right\} \leq e^{3 C_1 N r_N^2 / 4}.
\end{equation}
Further, for every density $f_1$ with parameters $(\bm{\beta}_1, \sigma_1^2, \rho_1)$ satisfying \eqref{densities-in-f1-panel}, we have that the exponent term in \eqref{densities-in-f1-panel-3} is bounded above by
\begin{equation} \label{densities-in-f1-panel-5}
	\left\{ \max_{1 \leq i \leq n} \lVert (2 \bm{\Sigma}_{1,i}^{\star} - \bm{I}_{n_i})^{-1} \rVert_{sp} \right\} \left\{ \max_{1 \leq i \leq n} \lVert \sigma^{-2} \bm{\Sigma}_i^{-1} \rVert_{sp} \right\} \lVert (\bm{\xi} - \bm{\xi}_1) \rVert_2^2 \leq \frac{C_1 N r_N^2}{8},
\end{equation}
since $\lVert \bm{\xi} - \bm{\xi}_1 \rVert_2^2 \leq C_1 N r_N^2 / 16$ by \eqref{densities-in-f1-panel}, $\max_{1 \leq \i \leq n} \lVert  2 \bm{\Sigma}_{1,i}^{\star} - \bm{I}_{n_i} \rVert_{sp} \leq 2$ for large $n$, and $\max_{1 \leq i \leq n} \lVert \sigma^{-2} \bm{\Sigma}_i \rVert_2 \lesssim 1$ by Lemma \ref{Lemma:1}. Combining \eqref{densities-in-f1-panel-3}-\eqref{densities-in-f1-panel-5} gives that $\mathbb{E}_{f_1} (f/f_1)^2$ in \eqref{upper-bound-type-ii-error} is bounded above by $e^{7 C_1 N r_N^2 /8}$ for every density $f_1$ with parameters $(\bm{\beta}_1, \sigma_1^2, \rho_1)$ satisfying \eqref{densities-in-f1-panel}. Combining this upper bound with the first inequality in \eqref{Neyman-Pearson-test-panel} finally gives that $\mathbb{E}_{f} (1- \phi_N)$ in \eqref{upper-bound-type-ii-error} can be upper bounded as 
\begin{align*}
	\mathbb{E}_f (1 - \phi_N ) \lesssim e^{- C_1 N r_N^2 / 16}.
\end{align*}
 for sufficiently large $N$. Coupled with the first inequality in \eqref{Neyman-Pearson-test-panel}, we deduce that on \emph{small subsets} of the sieve $\mathcal{F}_N$ in \eqref{sieve}, the desired exponentially powerful test satisfying \eqref{testing-conditions-2} can be obtained by taking the supremum of all tests $\phi_N$ constructed above.
 
Having established that small pieces of the alternative $H_1: N^{-1} Q_{1/2}(f, f_0) > C_1 r_N^2, f \in \mathcal{F}_N$, are locally testable with exponentially small errors, we now need to show that the metric entropy $\log N(r_N, \mathcal{F}_N, N^{-1} Q_{1/2})$, i.e. the logarithm of the minimum number of such pieces needed to cover $\mathcal{F}_N$, can be bounded above asymptotically by $N r_N^2$ (see Lemma D.3 of \cite{Ghosal2017}).  
 By Assumption (A2) and the Cauchy-Schwarz inequality, we have that $\frac{1}{N} \lVert \bm{\xi} - \bm{\xi}_1 \rVert_2^2 \leq D^2 (p+1) \sum_{j=0}^{p} \lVert \beta_j - \beta_{0j} \rVert_N^2$, and by Lemma \ref{Lemma:1}, the left-hand side of the second inequality in \eqref{densities-in-f1-panel} is bounded above by $n_{\max}^2 (\sigma^2 - \sigma_1^2)^2 + e^{4 C_4 N r_N^2 / \nu} n_{\max}^2 (\rho - \rho_1)^2$ on $\mathcal{F}_N$. Thus, for densities $f_1$ satisfying \eqref{densities-in-f1-panel}, the metric entropy can be bounded above by
\begin{align*} \label{upper-bound-metric-entropy-panel}
& \sum_{j=0}^{p} \log N \left( \frac{ C_1^{1/2} r_N}{4 D (p+1)^{1/2}}, \{ \beta_j \in \mathcal{B}_N^{j} : \lVert \beta_j - \beta_{0,j} \rVert_N < r_N \}, \lVert \cdot \rVert_N \right) \\
& \qquad + \log N \left( \frac{ C_1 r_N^2}{8^{1/2} n_{\max}^2}, \{ \sigma^2: 0 < \sigma^2 < e^{2 C_4 N r_N^2 / \nu} \}, | \cdot | \right) \\
& \qquad + \log N \left( \frac{C_1 r_N^2}{8^{1/2} n^2_{\max} e^{2 C_4 N r_N^2 / \nu}}, \{ \rho: 0 < \rho < 1 \}, | \cdot | \right). \numbereqn
\end{align*}
Given that $\sigma^2 \sim \lambda \nu \chi_2^{-2}$ and $\rho \sim \text{Uniform}(0,1)$, one can easily verify that the last two terms in \eqref{upper-bound-metric-entropy-panel} are upper bounded by a constant factor of $N r_N^2$. By modifying the proof of Theorem 7.1 in \citet{RockovaSaha2019} appropriately, we also have for some $A_2 > 0$ and small $\delta > 0$ that the first term in \eqref{upper-bound-metric-entropy-panel} can be upper bounded by
\begin{equation} \label{upper-bound-first-term-metric-entropy-panel}
\sum_{j=0}^{p} \left[ (l_N^j + 1) M \log (N R l_N^j) + A_2 M l_N^j \log \left( 972 D [ M C_1^{-1} (p+1) l_N^j]^{1/2} N^{1 + \delta/2} \right) \right],
\end{equation}
where $M$ is the number of trees in each ensemble $\mathcal{B}_N^{j}$. Recalling that $M$, $C_1^{-1}$, and $D$ are fixed positive constants, $l_N^j \asymp N ( r_N^j)^2 / \log N$ where $r_N^j = N^{-\alpha_j / (2 \alpha_j + R) } (\log  N)^{1/2}$, and Assumption (A3) that $p=O(1)$ and $R = O((\log N)^{1/2})$, it follows that \emph{each} summand in \eqref{upper-bound-first-term-metric-entropy-panel} can be upper bounded by $N(r_N^j)^2$. Thus, the total sum \eqref{upper-bound-first-term-metric-entropy-panel} is asymptotically bounded above by $N \sum_{j=0}^{p} (r_N^j)^2 = N r_N^2$. Altogether, from \eqref{upper-bound-metric-entropy-panel}-\eqref{upper-bound-first-term-metric-entropy-panel}, we have that for densities $f_1$ satisfying \eqref{densities-in-f1-panel},
\begin{align*}
\log N \left( r_N, \mathcal{F}_N, \frac{1}{N} Q_{1/2} \right) \lesssim N r_N^2.
\end{align*}
By Lemma D.3 of \cite{Ghosal2017}, this completes the proof of \eqref{testing-conditions-2}. Having established \eqref{prior-concentration-condition-2}-\eqref{testing-conditions-2}, it follows that we have posterior contraction with respect to average R\'{e}nyi divergence of order 1/2, i.e. \eqref{renyi-contraction}, holds.
\vspace{.5cm}

\noindent \textbf{Step 2: Posterior contraction in squared empirical $\ell_2$ norm}. Having proven that $\Pi ( N^{-1} Q_{1/2}(f, f_0) \lesssim r_N^2 \mid \bm{Y} ) \rightarrow 1$ in $\mathbb{P}_{f_0}^{(N)}$-probability as $N, R \rightarrow \infty$, where $r_N^2 = \log N \times \sum_{j=0}^{p} N^{-2 \alpha_j / (2 \alpha_j + R)}$, we can now convert this result to posterior contraction in terms of the empirical squared $\ell_2$ norm. 

Recall that $\bm{\Sigma}_i = \bm{\Sigma}_i(\rho)$ and $\bm{\Sigma}_{0,i} = \bm{\Sigma}_{0,i}(\rho_0)$ for short, while $\bm{\Sigma} = \text{bdiag}(\bm{\Sigma}_1, \ldots, \bm{\Sigma}_n)$ and $\bm{\Sigma}_0 = \text{bdiag}(\bm{\Sigma}_{01}, \ldots, \bm{\Sigma}_{0n})$ are $N \times N$ block diagonal matrices with respective diagonal blocks $\bm{\Sigma}_i$'s and $\bm{\Sigma}_{0,i}$'s.
	Following \cite{Jeong2019}, $N^{-1}Q_{1/2}(f, f_0)$ can be written explicitly as
\begin{align*}
	\frac{1}{N} Q_{1/2} (f, f_0) = & -\frac{1}{N} \left[ \sum_{i=1}^{n} \log \left\{ \frac{[\text{det}(\sigma^2 \bm{\Sigma}_i)]^{1/4} [\text{det}(\sigma_0^2 \bm{\Sigma}_{0,i})]^{1/4}}{[ \text{det} ( (\sigma^2 \bm{\Sigma}_i + \sigma_0^2 \bm{\Sigma}_{0,i}) / 2) ]^{1/2}} \right\} \right] \\
	& \qquad  + \frac{1}{4N} \lVert (\sigma^2\bm{\Sigma} + \sigma_0^2 \bm{\Sigma}_0)^{-1} (\bm{\xi} - \bm{\xi}_0) \rVert_2^2.
\end{align*}
Then $N^{-1} Q_{1/2}(f, f_0) \lesssim r_N^2$ implies that
\begin{equation} \label{Renyi-upper-bound-1}
	- \frac{1}{N} \sum_{i=1}^{n} \log \left\{ \frac{[\text{det}(\sigma^2 \bm{\Sigma}_i)]^{1/4} [\text{det}(\sigma_0^2 \bm{\Sigma}_{0,i})]^{1/4}}{[ \text{det} ( (\sigma^2 \bm{\Sigma}_i + \sigma_0^2 \bm{\Sigma}_{0,i}) / 2) ]^{1/2}} \right\} \lesssim r_N^2,
\end{equation}
and
\begin{equation} \label{Renyi-upper-bound-2}
	\frac{1}{4N} \lVert (\sigma^2\bm{\Sigma} + \sigma_0^2 \bm{\Sigma}_0)^{-1} (\bm{\xi} - \bm{\xi}_0) \rVert_2^2 \lesssim r_N^2.
\end{equation}
As in the proof of Theorem 3 of \citet{Jeong2019}, define the function $g$ as
\begin{align*}
	g^2 ( \sigma^2 \bm{\Sigma}_i, \sigma_0^2 \bm{\Sigma}_{0,i}) = 1 - \frac{[\text{det}(\sigma^2 \bm{\Sigma}_i)]^{1/4} [\text{det}(\sigma_0^2 \bm{\Sigma}_{0,i})]^{1/4}}{[ \text{det} ( (\sigma^2 \bm{\Sigma}_i + \sigma_0^2 \bm{\Sigma}_{0,i}) / 2) ]^{1/2}}.
\end{align*}
Then using the fact that $\log x \leq x - 1$, \eqref{Renyi-upper-bound-1} implies that
\begin{align*}
	r_N^2 \gtrsim - \frac{1}{N} \sum_{i=1}^{n} \log [ 1 - g^2 ( \sigma^2 \bm{\Sigma}_i, \sigma_0^2 \bm{\Sigma}_{0,i} ) ] \geq \frac{1}{N} \sum_{i=1}^{n} g^2 ( \sigma^2 \bm{\Sigma}_i, \sigma_0^2 \bm{\Sigma}_{0,i} ).
\end{align*}
By Lemma 10 of \cite{Jeong2019}, $g^2 ( \sigma^2 \bm{\Sigma}_i, \sigma_0^2, \bm{\Sigma}_{0,i} ) \gtrsim \lVert \sigma^2 \bm{\Sigma}_i - \sigma_0^2 \bm{\Sigma}_{0,i} \rVert_F^2$ when $r_N^2 \rightarrow 0$. Therefore, we can ultimately lower bound \eqref{Renyi-upper-bound-1} by
\begin{align} \label{Renyi-bound-3}
	r_N^2 \gtrsim \frac{1}{N} \lVert \sigma^2 \bm{\Sigma} - \sigma_0^2 \bm{\Sigma}_0 \rVert_F^2 \geq \max_{1 \leq i \leq n} \lVert \sigma^2 \bm{\Sigma}_i - \sigma_0^2 \bm{\Sigma}_{0,i} \rVert_F^2 \geq \max_{1 \leq i \leq n} \lVert \sigma^2 \bm{\Sigma}_i - \sigma_0^2 \bm{\Sigma}_{0,i} \rVert_{sp}^2.
\end{align}
Using Lemma \ref{Lemma:1} and \eqref{Renyi-bound-3}, we also have 
\begin{align} \label{Renyi-bound-4}
	\max_{1 \leq i \leq n} \lVert \sigma^2 \bm{\Sigma}_i + \sigma_0^2 \bm{\Sigma}_{0,i} \rVert_{sp}^2 & \leq 2 \max_{1 \leq i \leq n} \lVert \sigma^2 \bm{\Sigma}_i - \sigma_0^2 \bm{\Sigma}_{0,i} \rVert_{sp}^2 + 8 \max_{1 \leq i \leq n} \lVert \sigma_0^2 \bm{\Sigma}_{0,i} \rVert_{sp}^2 \lesssim r_N^2 + 1. \numbereqn
\end{align}
By \eqref{Renyi-bound-3}, the term on the left-hand side of \eqref{Renyi-upper-bound-1} is bounded below by something that is strictly nonnegative. Thus, a lower bound for $N^{-1} Q_{1/2} (f, f_0)$ is the left-hand side of \eqref{Renyi-upper-bound-2}. By \eqref{Renyi-bound-4}, we have that for sufficiently large $N$,
\begin{align*} \label{lower-bound-rN-panel}
r_N^2 & \gtrsim \frac{1}{4N ( r_N^2 + 1 ) } \lVert \bm{\xi} - \bm{\xi}_0 \rVert_2^2 \\
& = \frac{1}{4N (r_N^2 + 1)} \sum_{i=1}^{n} \sum_{t=1}^{n_i} \left[ \{ \beta_0(\bz_{it}) - \beta_{0,0} \bz_{it} \} + \sum_{j=1}^{p} \{ \beta_j(\bz_{it}) - \beta_{0,j} (\bz_{it}) \} x_{itj} \right]^2  \\
& \gtrsim \frac{1}{4N (r_N^2 + 1)} \sum_{i=1}^{n} \sum_{t=1}^{n_i} \sum_{j=0}^{p} [ \beta_j (\bz_{ij}) - \beta_{0,j} (\bz_{ij} ) ]^2  \\
& \asymp \frac{1}{N} \sum_{i=1}^{n} \sum_{t=1}^{n_i} \sum_{j=0}^{p} [ \beta_j (\bz_{ij}) - \beta_{0,j} (\bz_{ij} ) ]^2 = \lVert \bm{\beta} - \bm{\beta}_0 \rVert_N^2, \numbereqn
\end{align*}
where we used Assumption (A1) that $\lVert \beta_j \rVert_{\infty} < \infty$ for all $0 \leq j \leq p$ and Assumption (A2) that the covariates $x_{itj}, 1 \leq i \leq n, 1 \leq t \leq n_i, 1 \leq j \leq p$, are all uniformly bounded in the third line of the display. In the final line, we used the fact that $r_N^2 \rightarrow 0$ as $N \rightarrow \infty$, and thus, $4 N (r_N^2 + 1)$ is asymptotically the same order as $N$. By \eqref{lower-bound-rN-panel}, the VCBART posterior is asymptotically supported on the event $\{ \bm{\beta}: \lVert \bm{\beta} - \bm{\beta}_0 \rVert_N^2 \leq \widetilde{C} r_N^2 \}$ for sufficiently large $N$ and  some large constant $\widetilde{C} > 0$. This proves the theorem.
\end{proof}

%% file: hyperparameter_sensitivity.tex
\subsection{Convergence diagnostics \& number of MCMC iterations}
\label{sec:convergence_diagnostics}
\begin{figure}[h!]
\centering
\begin{subfigure}{0.48\textwidth}
\centering
\includegraphics[width = \textwidth]{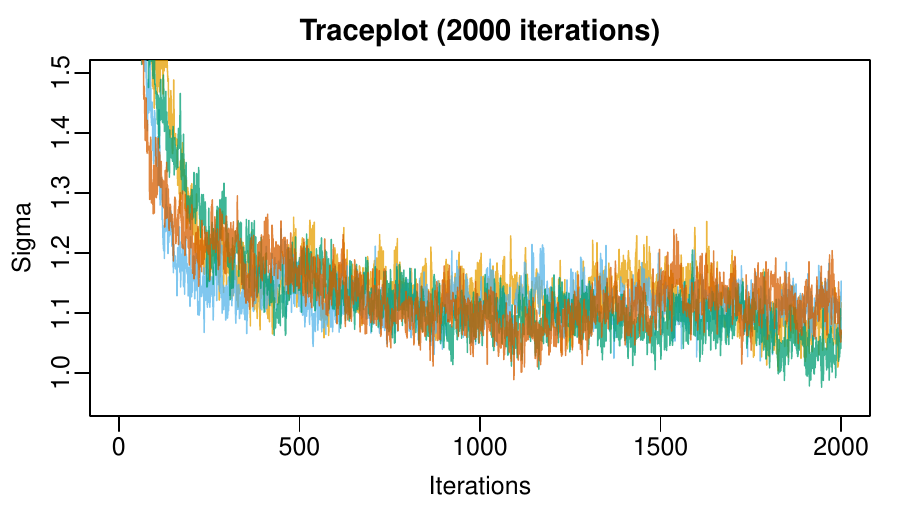}
\caption{}
\label{fig:mixing_traceplot2k}
\end{subfigure}
\begin{subfigure}{0.48\textwidth}
\centering
\includegraphics[width = \textwidth]{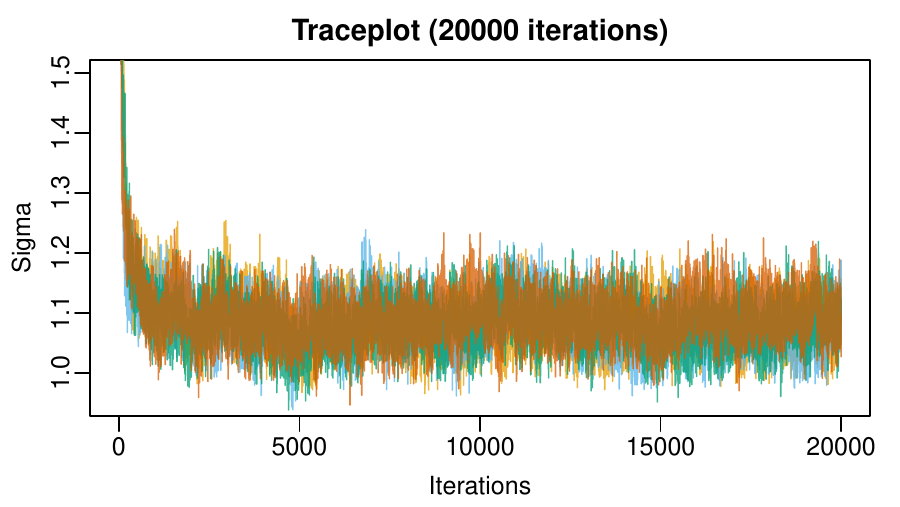}
\caption{}
\label{fig:mixing_traceplot20k}
\end{subfigure}
\caption{Trace plots of $\sigma$ samples obtained by running four chains for 2,000 (a) and 20,000 (b) iterations}
\label{fig:mixing_traceplot}
\end{figure}

In Section 4 of the main text, the reported VCBART results were based on simulating four Markov chains for 2,000 total iterations and then discarding the first 1,000 iterations of each chain as ``burn-in.''
Figure~\ref{fig:mixing_traceplot2k} shows the trace plots for $\sigma$ from one such run and Figure~\ref{fig:mixing_traceplot20k} shows the trace plot obtained by running each chain ten times longer (i.e., for 20,000 total iterations).
Although all chains appear to quickly navigate to the same region of $\sigma$-space, they do not appear to have mixed after only 2,000 iterations. 
In particular, after about 1,700 iterations, the green chain appears to explore somewhat smaller values of $\sigma$ than any of the other chains.

\begin{figure}[h!]
\begin{subfigure}{0.32\textwidth}
\centering
\includegraphics[width = \textwidth]{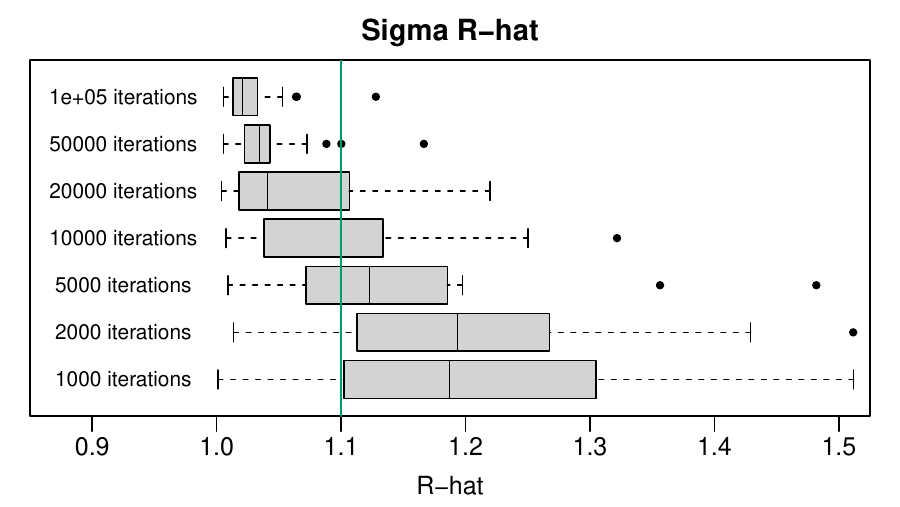}
\caption{}
\label{fig:mixing_rhat}
\end{subfigure}
\begin{subfigure}{0.32\textwidth}
\centering
\includegraphics[width = \textwidth]{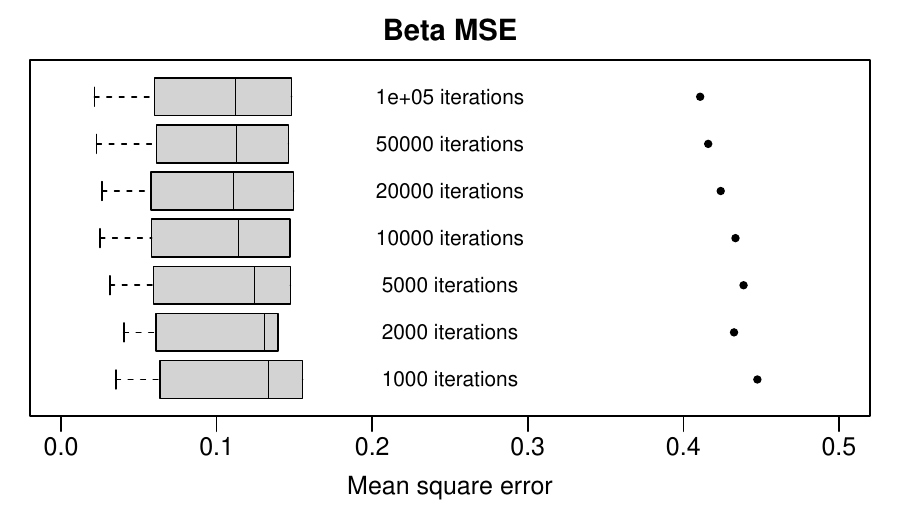}
\caption{}
\label{fig:mixing_beta_mse}
\end{subfigure}
\begin{subfigure}{0.32\textwidth}
\centering
\includegraphics[width = \textwidth]{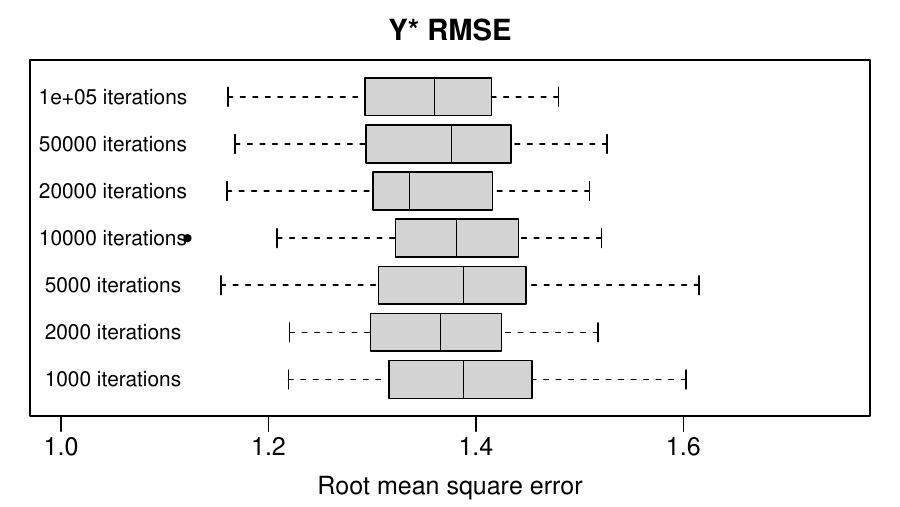}
\caption{}
\label{fig:mixing_ystar_rmse}
\end{subfigure}

\begin{subfigure}{0.32\textwidth}
\centering
\includegraphics[width = \textwidth]{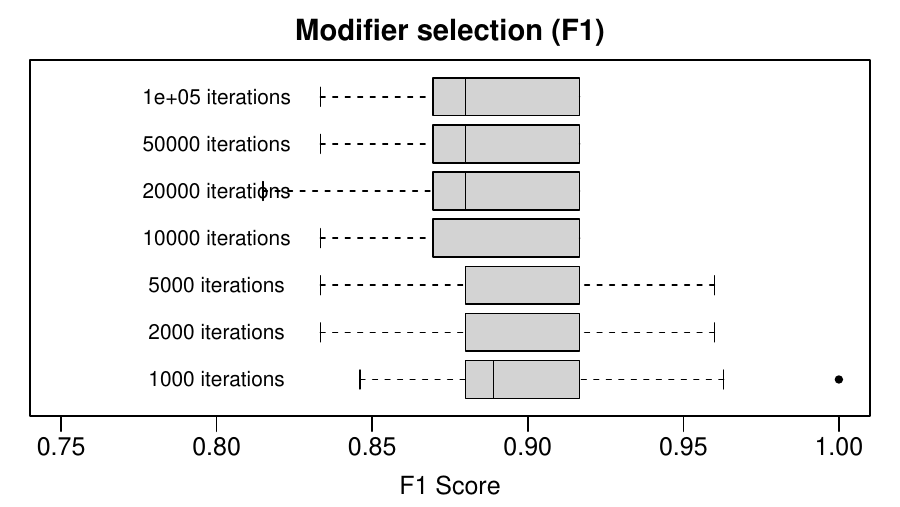}
\caption{}
\label{fig:mixing_f1}
\end{subfigure}
\begin{subfigure}{0.32\textwidth}
\centering
\includegraphics[width = \textwidth]{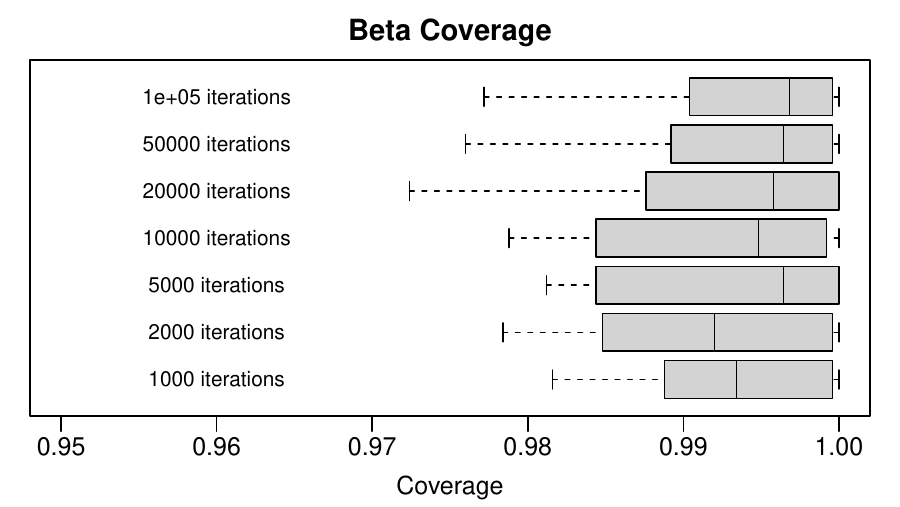}
\caption{}
\label{fig:mixing_beta_cov}
\end{subfigure}
\begin{subfigure}{0.32\textwidth}
\centering
\includegraphics[width = \textwidth]{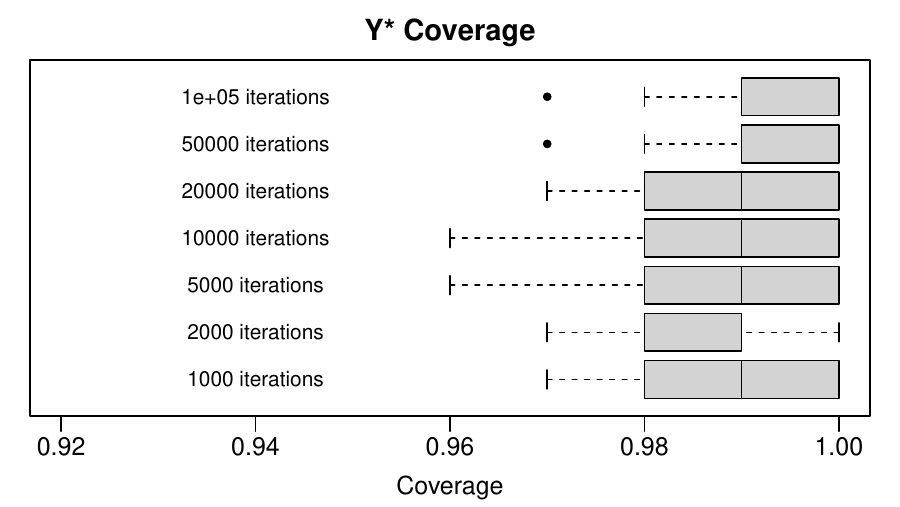}
\caption{}
\label{fig:mixing_ystar_cov}
\end{subfigure}
\caption{Boxplots of $\hat{R}$ values (a); total mean square errors for evaluating $\beta_{j}(\bz)$ out-of-sample (b); out-of-sample root mean square predictive error (c); $F_{1}$ score for recovering the support of the $\beta_{j}(\bZ)$'s (d); and uncertainty interval coverage for $\beta_{j}(\bz)$ and predictions (e) and (f) across 25 simulation replications. The heuristic $\hat{R}$ cut-off of 1.1 is indicated in green in (a).}
\label{fig:mixing_boxplots}
\end{figure}

To investigate this more formally, we conducted another experiment in which we ran 4 MCMC chains of length length $2 \times \texttt{nd}$ where $\texttt{nd} \in \{500, 1000, 2500, 5000, 10000, 25000, 50000 \}$ using the 25 synthetic datasets described in Section 4 of our manuscript.
We then computed the $\hat{R}$-statistic based on the last \texttt{nd} samples of $\sigma$.
Figure~\ref{fig:mixing_rhat} shows the distribution of the $\hat{R}$-values across 25 simulation replications. 

We see clearly that it was only after 10,000 (i.e., $\texttt{nd} = 5,000$) that we obtain $\hat{R} < 1.1$ in most simulation replications.
In fact, for these data, we only consistently obtained $\hat{R}$'s less than 1.1 when we set $\texttt{nd} = 25,000$ and ran each chain for 50,000 total iterations.
Interestingly, however, there was virtually no variation in the total mean square error for estimating $\beta_{j}(\bz)$'s out-of-sample (Figure~\ref{fig:mixing_beta_mse}); the root mean square error for predicting out-of-sample responses (Figure~\ref{fig:mixing_ystar_rmse}); the ability to identify relevant modifiers (Figure~\ref{fig:mixing_f1}); or the coverage of the uncertainty intervals for $\beta_{j}(\bz)$ and predictions (Figures~\ref{fig:mixing_beta_cov} and~\ref{fig:mixing_ystar_cov}).
So although 2,000 iterations may not be enough to mix formally, they were certainly adequate to obtain excellent predictions and reasonably well-calibrated uncertainty intervals.

\subsection{Number of trees $M$ \& jump variance $\tau_{j}$}
\label{sec:m_tau_sensitivity}
Recall from the main text that each ensemble contained $M = 50$ trees and that we set the prior jump variance $\tau_{j} = M^{-\frac{1}{2}}/2.$
This induced a marginal $\normaldist{0}{1/4}$ prior on each evaluation $\beta_{j}(\bz).$
Of course, if one has strong prior beliefs about the range of covariate effects, one can set $\tau_{j}$ in such a way that the implied marginal prior on $\beta_{j}(\bz)$ places substantial probability on this range.
In this section, we consider the sensitivity of VCBART's covariance effect recovery and predictive capabilities to different choices of $\tau_{j}.$
Specifically, we replicate the synthetic data experiment from Section 4 of the main text with $N = 1,000$ total observations and vary $M \in \{50, 100, 200\}$ and $\tau_{j} = \tau/\sqrt{M}$ for $\tau \in \{0.25, 0.5, 1, 2, 4\}.$

\begin{figure}[h!]
\centering
\begin{subfigure}[b]{0.32\textwidth}
\centering
\includegraphics[width = \textwidth]{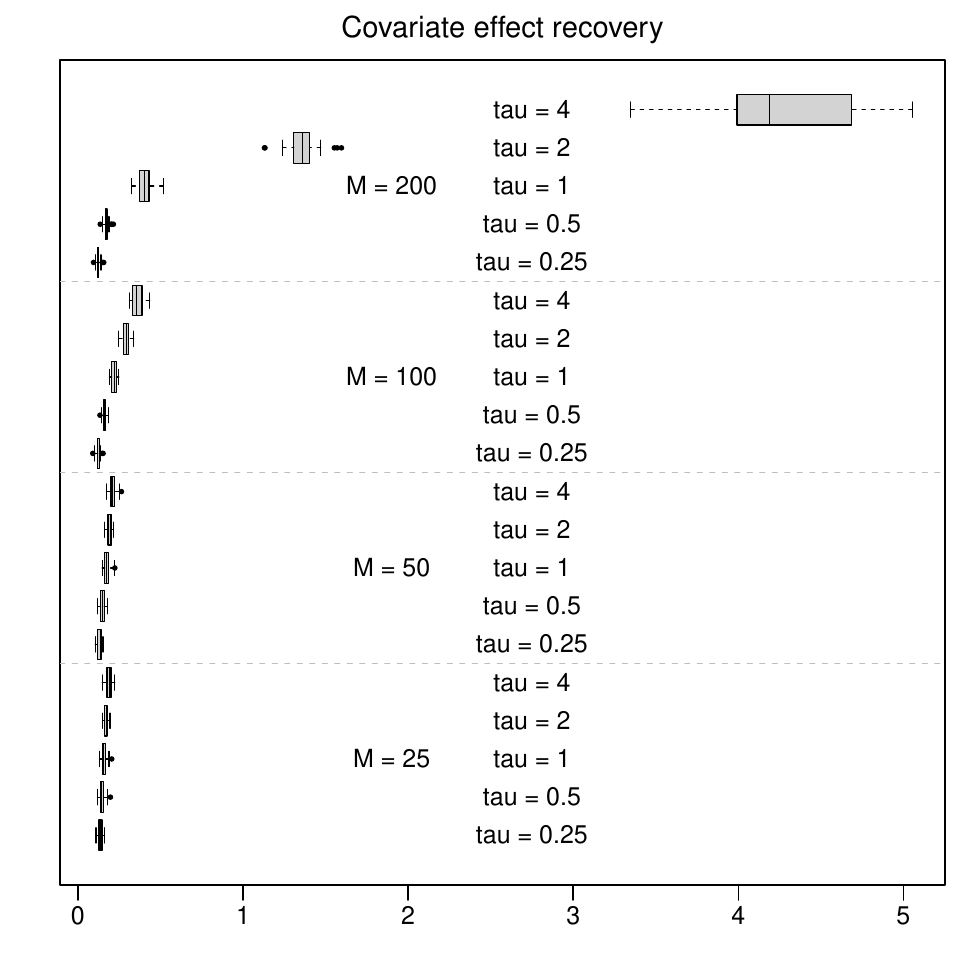}
\caption{}
\label{fig:hyperparam_beta_mse}
\end{subfigure}
\begin{subfigure}[b]{0.32\textwidth}
\centering
\includegraphics[width = \textwidth]{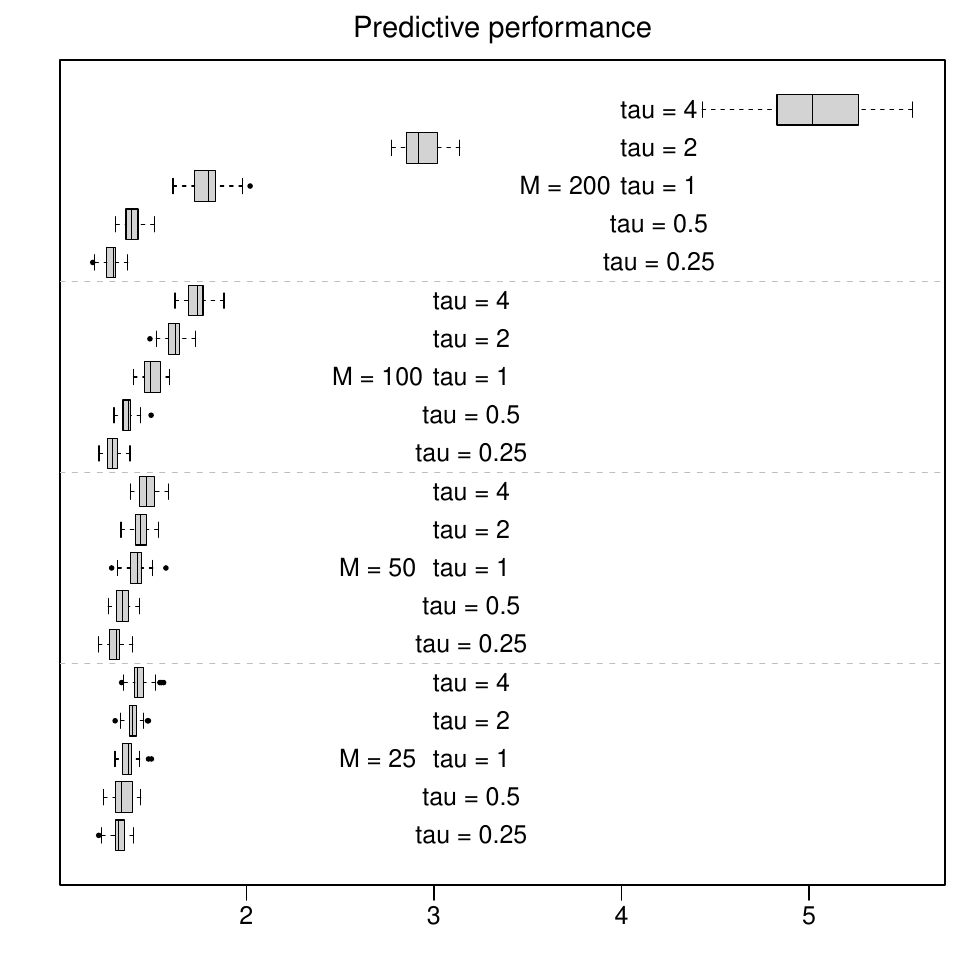}
\caption{}
\label{fig:hyperparam_ystar_rmse}
\end{subfigure}
\begin{subfigure}[b]{0.32\textwidth}
\centering
\includegraphics[width = \textwidth]{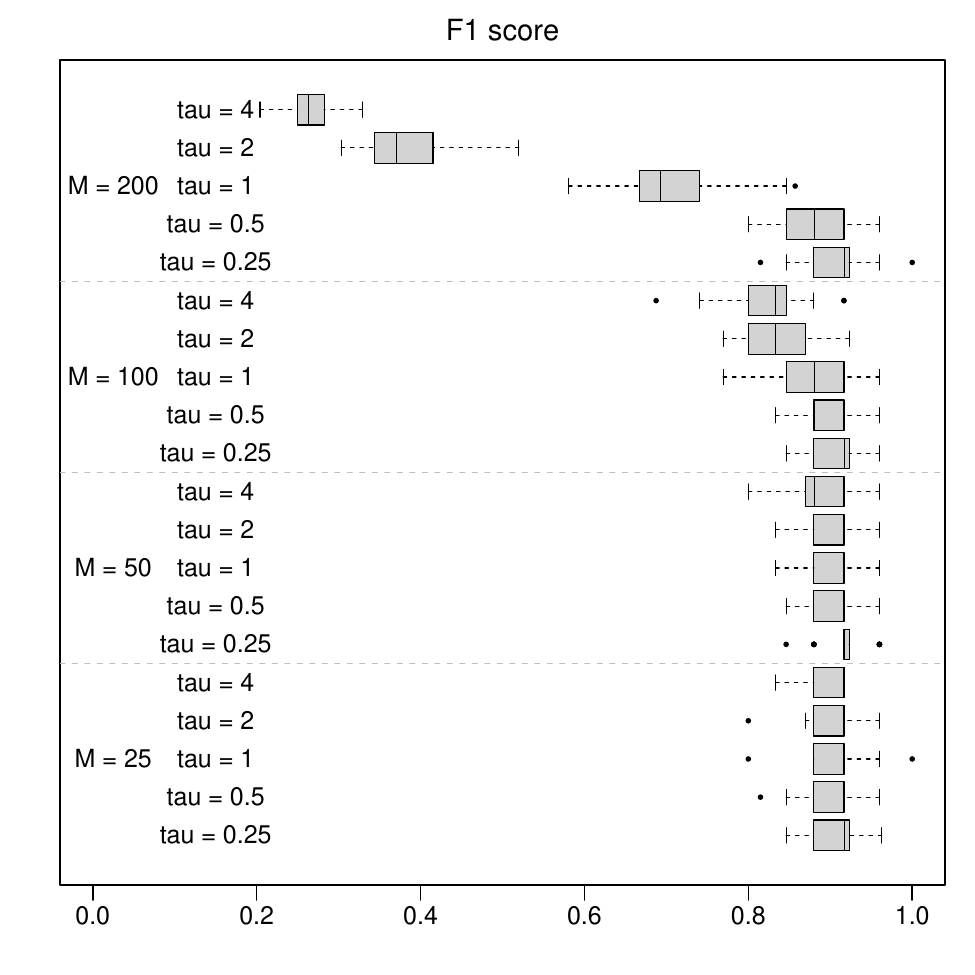}
\caption{}
\label{fig:hyperparam_varsel_f1}
\end{subfigure}
\caption{Covariate effect recovery, predictive, and modifier selection performance of VCBART run with different combinations of $(M, \tau_{j})$. (a) Total mean square error for estimating evaluations $\beta_{j}(\bz)$. (b) Predictive root mean square error. (c) F1 score of the median probability model. All measures reported over 25 testing datasets}
\label{fig:hyperparam}
\end{figure}

Figures~\ref{fig:hyperparam_beta_mse} and~\ref{fig:hyperparam_ystar_rmse} are the analogs of Figures 3a and 3c of the main text, comparing the covariate effect recovery and predictive performance of VCBART with different values of $(M,\tau).$
We observed that the coverages of posterior credible intervals for evaluations $\beta_{j}(\bz)$ and predictive intervals for testing outcomes were rather insensitive to both $M$ and $\tau$; in fact, for each combination of $M$ and $\tau$ the intervals all displayed much higher than nominal frequentist coverage.
Figure~\ref{fig:hyperparam_varsel_f1} shows the  F1 score for the median probability model for each combination of $M$ and $\tau.$

For each $M,$ we see that as the covariate effect recovery and predictive performance degrades as we increase (resp. decrease) $\tau$ (resp. the amount of regularization).
Interestingly, the degree to which performance degrades with increasing $\tau$ also increases with the total number of trees in the ensemble.
Put differently, at least for our synthetic data generating process, VCBART's performance is much more sensitive to $\tau$ when the number of trees $M$ is large.

We further observed that the modifier selection performance of VCBART was much worse for $M = 200$ than it was for smaller ensemble sizes.
On further inspection, we found that the estimated median probability model made too many false positive identifications in the supports of the $\beta_{j}(\bZ)$'s.
That is, the tree ensembles $\calE_{j}$ split on too many spurious modifiers $Z_{r}.$
We conjecture that the poor performance displayed in the top panel of Figure~\ref{fig:hyperparam_varsel_f1} is actually more general.
Specifically, the combination of a too large ensemble size $M$ and a Dirichlet prior on the vector of splitting probabilities will lead to a proliferation of false positives in the median probability model.

To develop some intuition about this conjecture, suppose that the function $\beta_{j}(\bZ)$ has been extremely well-approximated by the first $M' \ll M$ regression trees in an ensemble of size $M.$
The Gibbs sampling underlying VCBART (and indeed, virtually all BART extensions), updates regression trees based on a partial residual based on the approximation provided by the other $M-1$ trees.
Once the sampler reaches the $m$-th tree with $m > M'$, the relevant partial residual will have very little variance.
At this point, the tree is being used, essentially, to fit ``noise,'' with the first $M'$ trees in the ensemble accounting for virtually all the ``signal.''
In this situation, spurious decision rules --- that is, those based on modifiers that do not drive variation in $\beta_{j}(\bZ)$ --- are not heavily penalized in the Metropolis-Hastings update.
And once a split on a spurious variable has been accepted in a single MCMC iteration, the Dirichlet-Multinomial conjugacy underpinning our decision rule prior, will encourage further splitting on that same variable, leading to a proliferation of false positive identifications.
Introducing a dependence on $M$ into our decision rule prior might alleviate this behavior.
We leave such explorations for future work.

%% file: additional_sim_results.tex
\subsection{Implementation details}
\label{app:implementation_details}
VCBART is implemented in \textsf{C++} and interfaces with \Rlang \citep{R_citation} through \textbf{Rcpp} \citep{Eddelbuettel2011} and \textbf{RcppArmadillo} \cite{Eddelbuettel2014}.
Our implementation uses the basic class structures of \textbf{flexBART} \citep{Deshpande2022_flexBART}.
We have created an \Rlang package \textbf{VCBART}, which is available online at \url{https://github.com/skdeshpande91/VCBART}. 
We performed all of the experiments reported in Sections 4 and 5 of the main text on a high throughput computing cluster \citep{chtc}.
All experiments were run in \Rlang version 4.1.3 on compute nodes.

Where possible, we ran the competing methods in our simulations with package defaults and did not implement any additional tuning procedures.
For instance, the default implementation of \citet{ZhouHooker2019}'s boosted tree procedure (referred to as \texttt{BTVCM} in Section 4 of the main text) does not automatically perform cross-validation to set the learning rate or number of boosting iterations.
Instead, we ran \texttt{BTVCM} for 200 iterations and with a learning rate of 0.5, which are the same settings as the example provided by the authors at \url{https://github.com/siriuz42/treeboostVCM}.
Similarly the implementation of extremely randomized trees in the \textbf{ranger} package \citep{Wright2017} does not automatically perform cross-validation to select the number of possible splitting variables at each node (i.e.\ the parameter \texttt{mtry}).
In our experiments, we used the package default, setting \texttt{mtry} equal to the square root of the number of inputs.

\subsection{Function-by-function covariate recovery performance}
\label{app:p5R20_function_by_function}
\begin{figure}[h!]
\centering
\begin{subfigure}[b]{0.32\textwidth}
\centering
\includegraphics[width = \textwidth]{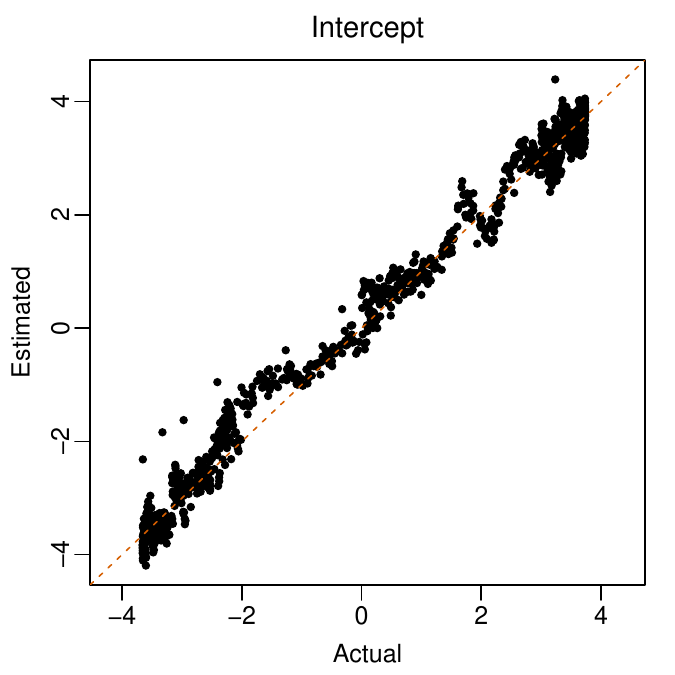}
\caption{}
\label{fig:p5R20_demo_beta0}
\end{subfigure}
\begin{subfigure}[b]{0.32\textwidth}
\centering
\includegraphics[width = \textwidth]{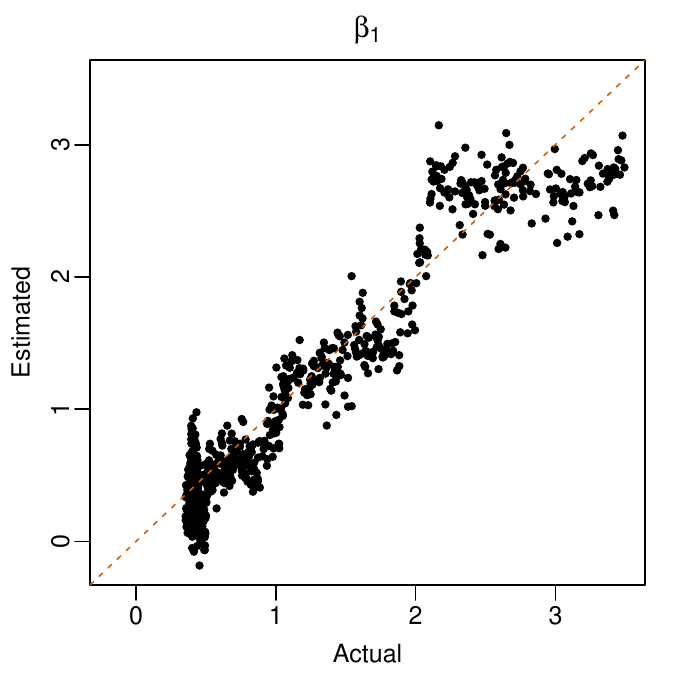}
\caption{}
\label{fig:p5R20_demo_beta1}
\end{subfigure}
\begin{subfigure}[b]{0.32\textwidth}
\centering
\includegraphics[width = \textwidth]{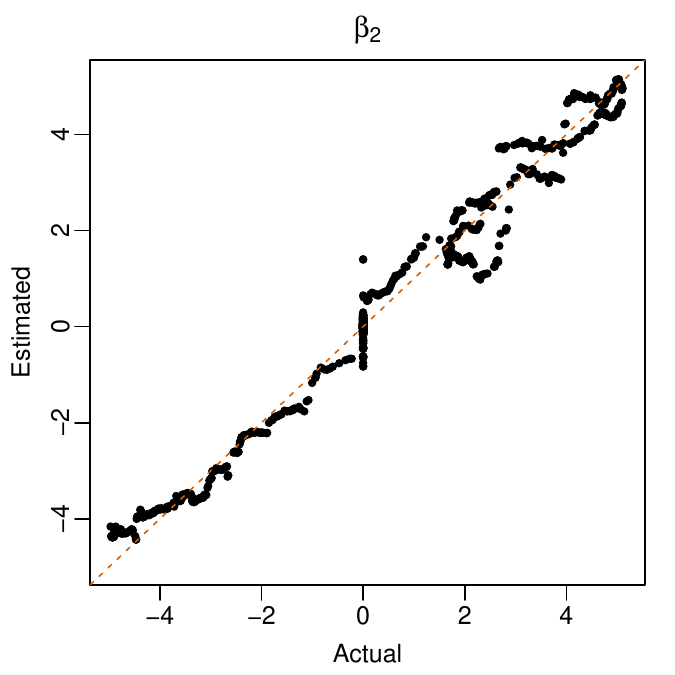}
\caption{}
\label{fig:p5R20_demo_beta2}
\end{subfigure}

\begin{subfigure}[b]{0.32\textwidth}
\centering
\includegraphics[width = \textwidth]{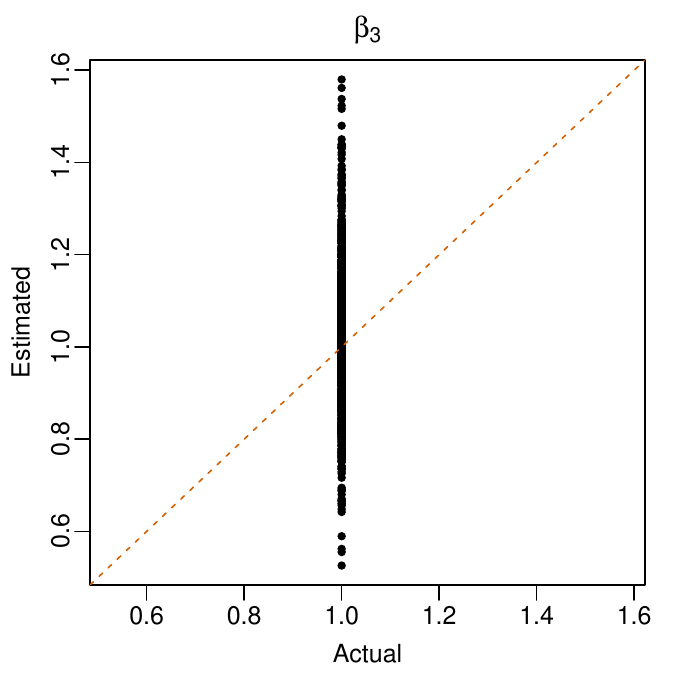}
\caption{}
\label{fig:p5R20_demo_beta3}
\end{subfigure}
\begin{subfigure}[b]{0.32\textwidth}
\centering
\includegraphics[width = \textwidth]{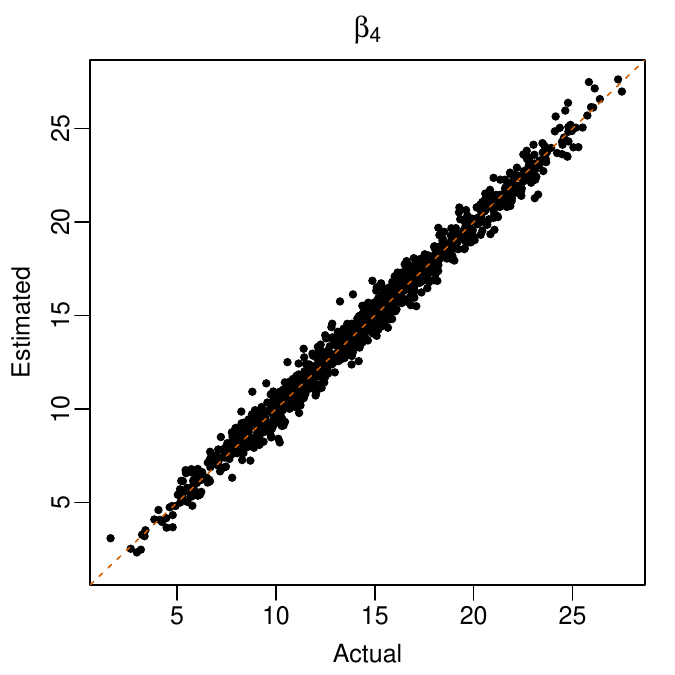}
\caption{}
\label{fig:p5R20_demo_beta4}
\end{subfigure}
\begin{subfigure}[b]{0.32\textwidth}
\centering
\includegraphics[width = \textwidth]{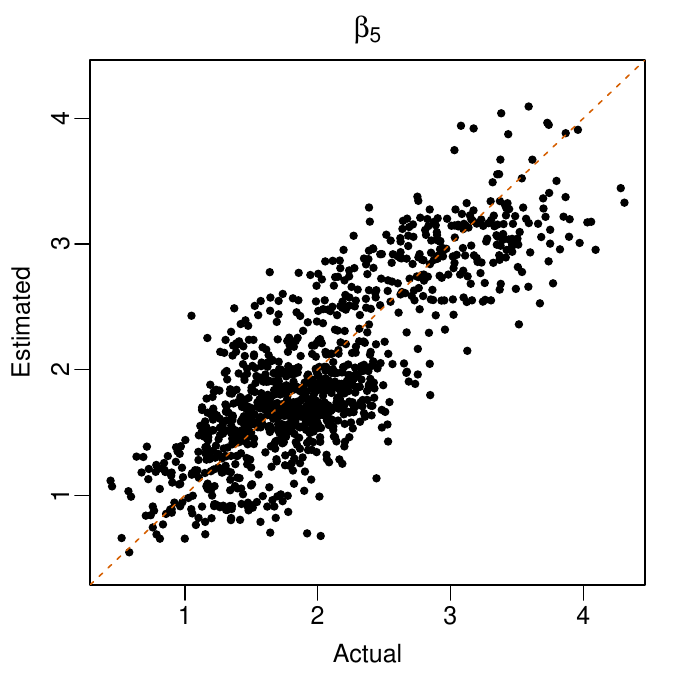}
\caption{}
\label{fig:p5R20_demo_beta5}
\end{subfigure}
\caption{Actual and VCBART estimates of the $\beta_{j}(\bz)$ used to generate the data underpinning Figure 2 of the main text. For the non-constant functions $\beta_{j}(\bz)$ for $j \neq 3,$ we see that the estimates are quite close to the actual values}
\label{fig:p5R20_demo_actual_estimated}
\end{figure}

Figure 2 of the main text plotted estimates of $\beta_{0}(\bz), \beta_{1}(\bz), \beta_{2}(\bz)$ and $\beta_{3}(\bz)$ computed after fitting a VCBART model to $N = 1000$ total observations.
Figure~\ref{fig:p5R20_demo_actual_estimated} plots the posterior means of each $\beta_{j}(\bz)$ against the actual value.
With the exception of Figure~\ref{fig:p5R20_demo_beta3}, we see that the points fall close to the 45-degree line, indicating that our posterior mean estimates of $\beta_{j}(\bz)$ are close to the actual values used to generate the data.
Recall that we generated the data with a constant $\beta_{3}(\bz) = 1$ and with a residual standard deviation of $\sigma = 1.$
On further inspection, we found that the vast majority of the estimated $\beta_{3}(\bz)$'s were within half a standard deviation of the true value $\beta_{3}(\bz) = 1.$
So although the VCBART estimates of the constant function were not constant, the estimated values were still quite close to the true function value. 

\begin{figure}[h!]
\centering
\begin{subfigure}[b]{0.32\textwidth}
\centering
\includegraphics[width = \textwidth]{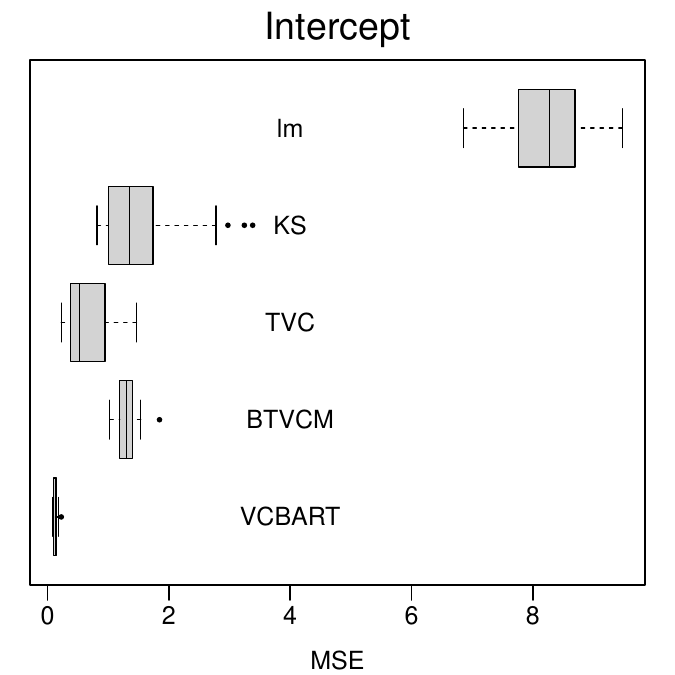}
\caption{}
\label{fig:p5R20_beta0_rmse}
\end{subfigure}
\begin{subfigure}[b]{0.32\textwidth}
\centering
\includegraphics[width = \textwidth]{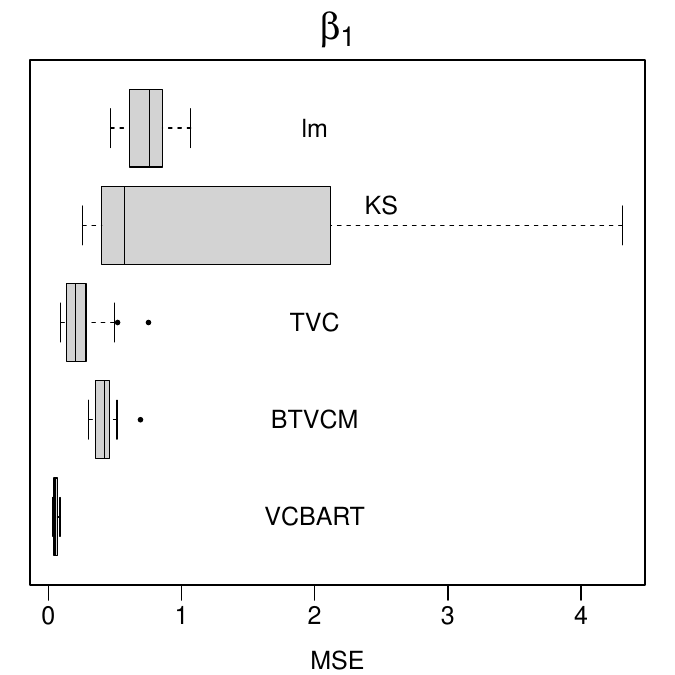}
\caption{}
\label{fig:p5R20_beta1_rmse}
\end{subfigure}
\begin{subfigure}[b]{0.32\textwidth}
\centering
\includegraphics[width = \textwidth]{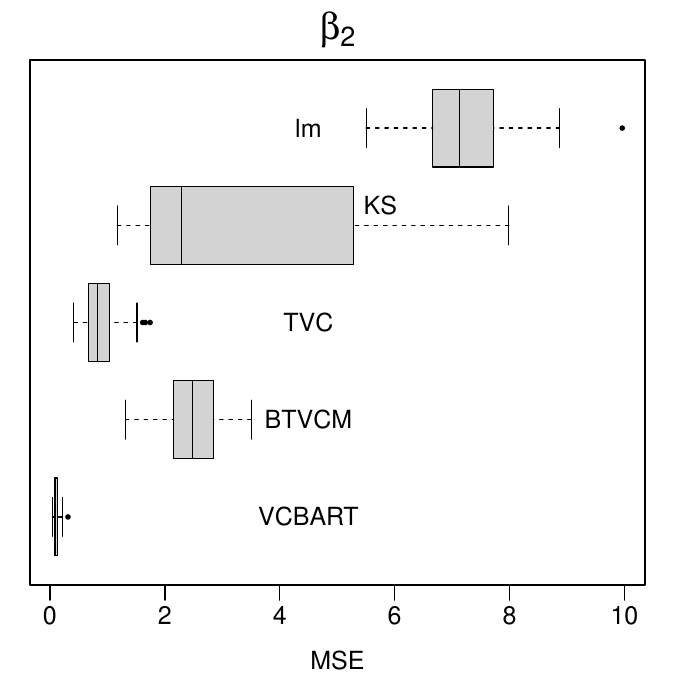}
\caption{}
\label{fig:p5R20_beta2_rmse}
\end{subfigure}

\begin{subfigure}[b]{0.32\textwidth}
\centering
\includegraphics[width = \textwidth]{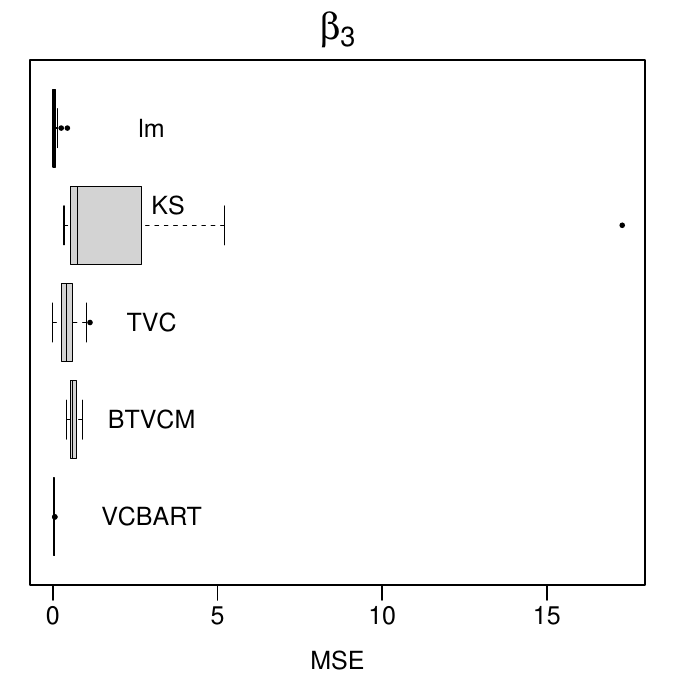}
\caption{}
\label{fig:p5R20_beta3_rmse}
\end{subfigure}
\begin{subfigure}[b]{0.32\textwidth}
\centering
\includegraphics[width = \textwidth]{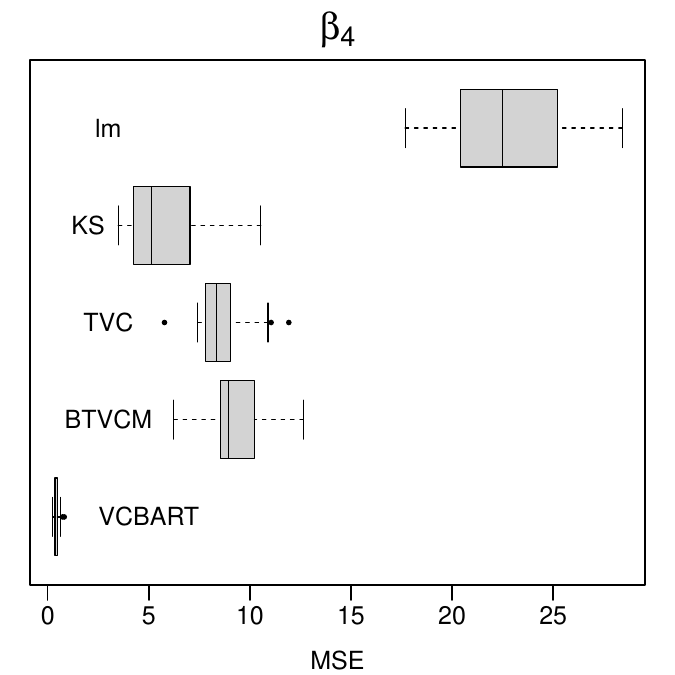}
\caption{}
\label{fig:p5R20_beta4_rmse}
\end{subfigure}
\begin{subfigure}[b]{1.66in}
\centering
\includegraphics[width = \textwidth]{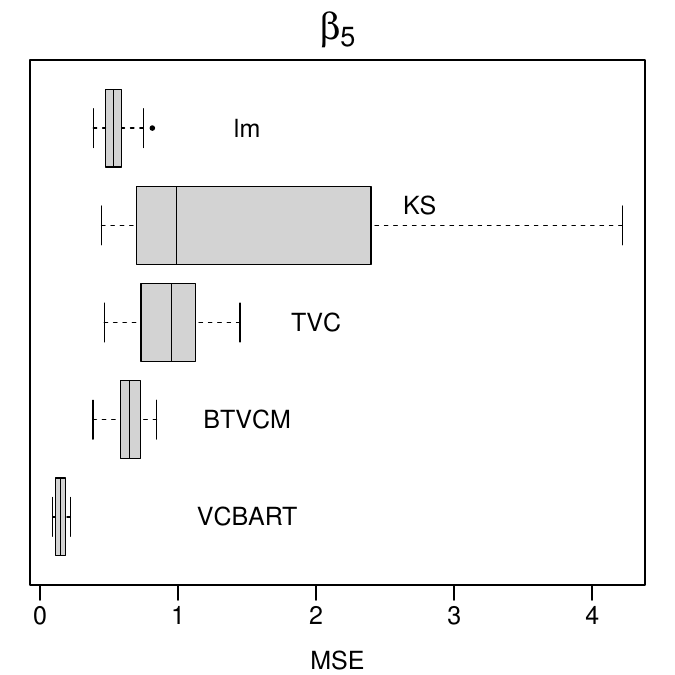}
\caption{}
\label{fig:p5R20_beta5_mse}
\end{subfigure}
\caption{Analog of Figure 3a of the main text showing the out-of-sample mean square error for estimating each $\beta_{j}(\bz)$ in our main experiment function-by-function.}
\label{fig:beta_mse_supp}
\end{figure}

\begin{figure}[h!]
\centering
\begin{subfigure}[b]{0.32\textwidth}
\centering
\includegraphics[width = \textwidth]{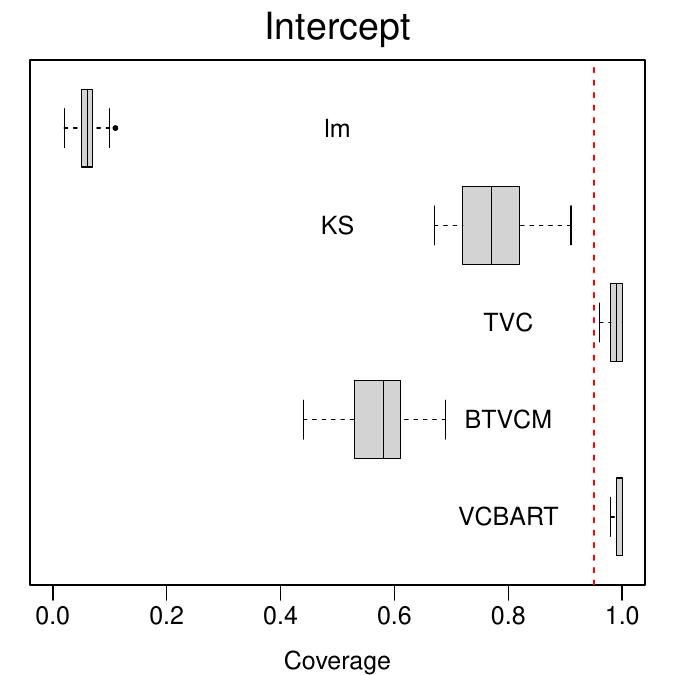}
\caption{}
\label{fig:p5R20_beta0_cov}
\end{subfigure}
\begin{subfigure}[b]{0.32\textwidth}
\centering
\includegraphics[width = \textwidth]{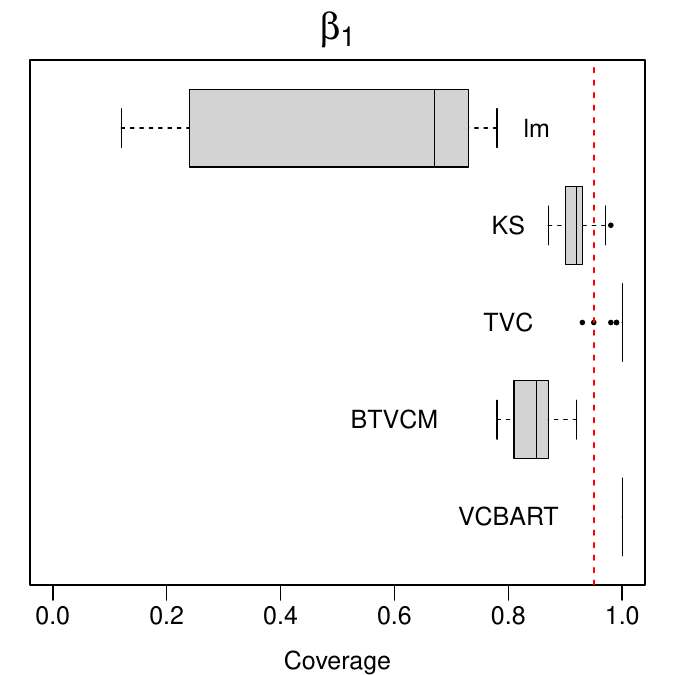}
\caption{}
\label{fig:p5R20_beta1_cov}
\end{subfigure}
\begin{subfigure}[b]{0.32\textwidth}
\centering
\includegraphics[width = \textwidth]{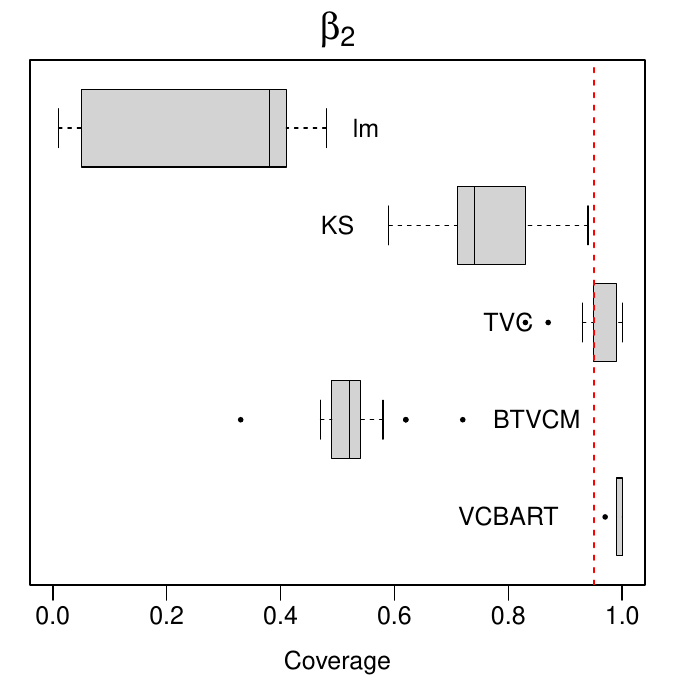}
\caption{}
\label{fig:p5R20_beta2_cov}
\end{subfigure}

\begin{subfigure}[b]{0.32\textwidth}
\centering
\includegraphics[width = \textwidth]{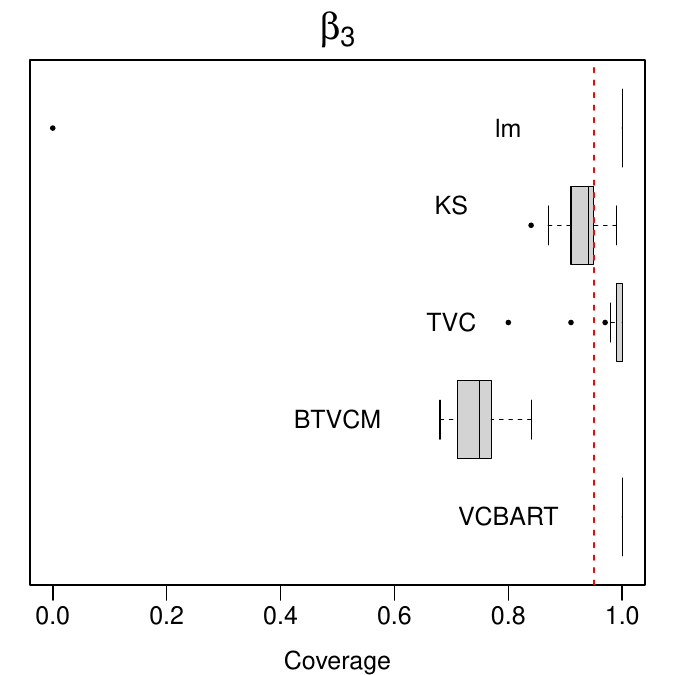}
\caption{}
\label{fig:p5R20_beta3_cov}
\end{subfigure}
\begin{subfigure}[b]{0.32\textwidth}
\centering
\includegraphics[width = \textwidth]{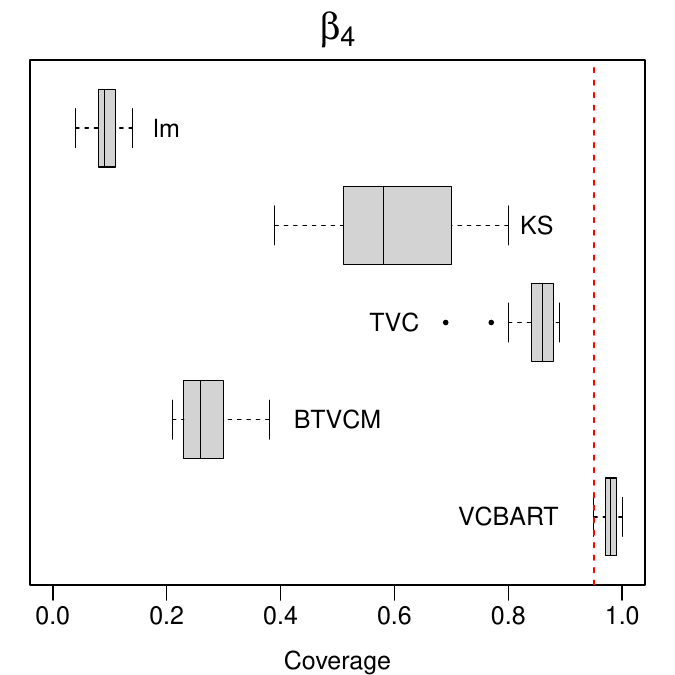}
\caption{}
\label{fig:p5R20_beta4_cov}
\end{subfigure}
\begin{subfigure}[b]{0.32\textwidth}
\centering
\includegraphics[width = \textwidth]{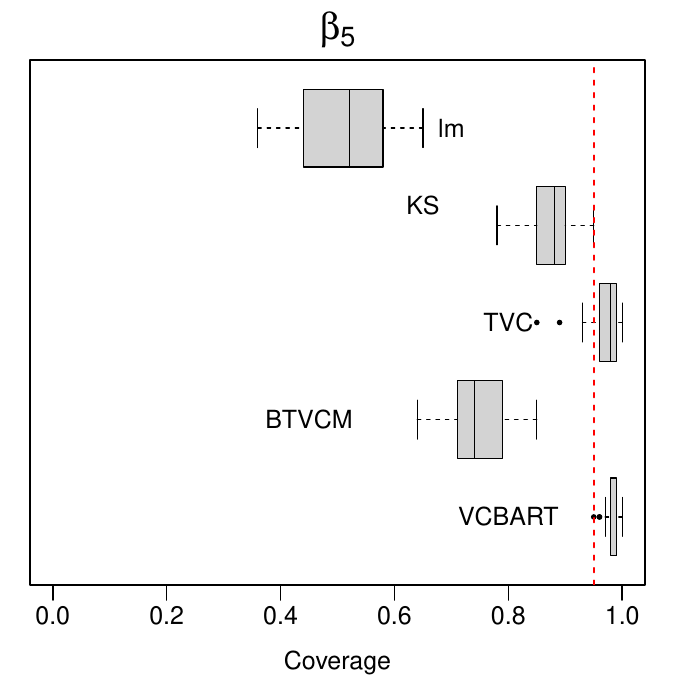}
\caption{}
\label{fig:p5R20_beta5_cov}
\end{subfigure}
\caption{Analog of Figure 3b of the main text showing the coverage of the 95\% credible intervals for estimating each $\beta_{j}(\bz)$ out-of-sample in our main experiment function-by-function.}
\label{fig:beta_cov_supp}
\end{figure}

Across the 25 training--testing splits considered in Section 4 of the main text, \texttt{VCBART} produced consistently more accurate estimates of each individual function $\beta_{j}.$
Figures 3a and 3b in the main text compared the mean square error and uncertainty interval coverage of VCBART and several competitors averaged across all functions and test-set inputs.
Figures~\ref{fig:beta_mse_supp} and~\ref{fig:beta_cov_supp} are analogs of those two figures that compare the function-by-function mean square and uncertainty interval coverage.
We see that VCBART clearly outperforms the other competitors in terms of recovering each function. 

\subsection{Scalability and stability}
\label{sec:scalability}

\begin{figure}[h!]
\centering
\begin{subfigure}[b]{0.24\textwidth}
\centering
\includegraphics[width = \textwidth]{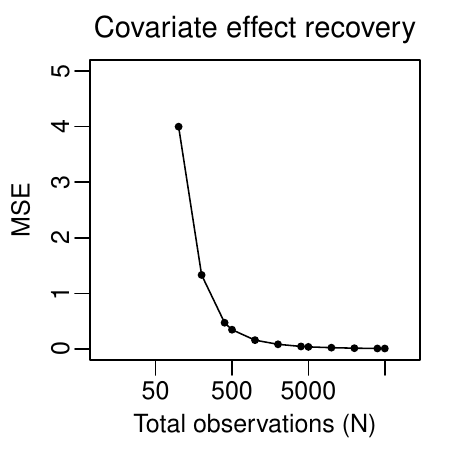}
\caption{}
\end{subfigure}
\begin{subfigure}[b]{0.24\textwidth}
\centering
\includegraphics[width = \textwidth]{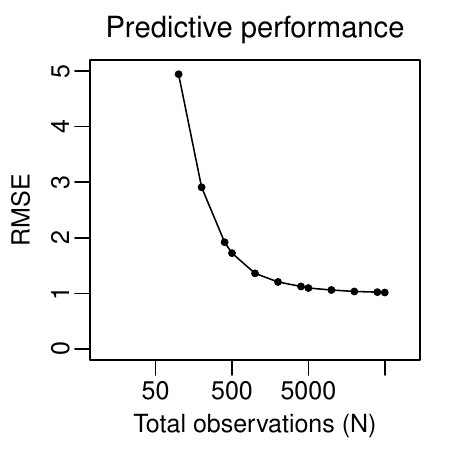}
\caption{}
\end{subfigure}
\begin{subfigure}[b]{0.24\textwidth}
\centering
\includegraphics[width = \textwidth]{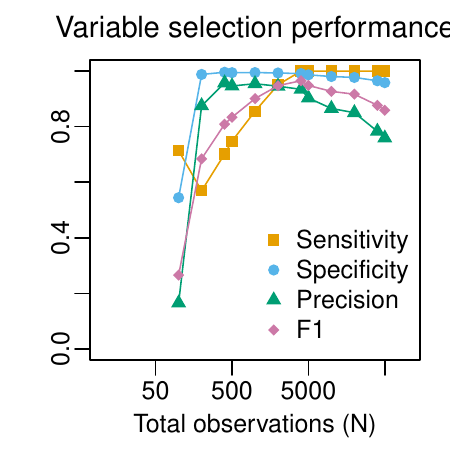}
\caption{}
\end{subfigure}
\begin{subfigure}[b]{0.24\textwidth}
\centering
\includegraphics[width = \textwidth]{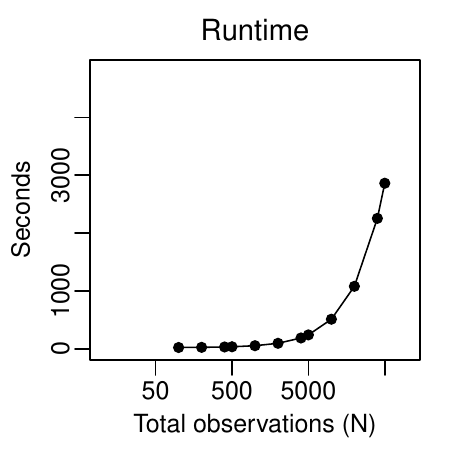}
\caption{}
\end{subfigure}
\caption{VCBART's estimation, prediction, and variable selection performance as a function of the total number of observations $N$.Average MSE for evaluating $\beta_{j}$'s (a) RMSE for predicting responses $y$ (b), $F1$ score, precision, and sensitivity (c), and total time in seconds to run four chains sequentially (d). Note the log-scale on the horizontal axis}
\label{fig:scalability}
\end{figure}

Recall from Section 4 of the main text that we performed a second synthetic experiment to study VCBART's ability to scale to larger datasets.
Like the main simulation study, we generated data from a varying coefficient model in which each of $n$ subject contributed four observations.
By varying $N$ between 25 and 12,500, we were able to see how VCBART performed when trained on as few as 100 and as many 50,000 total observations.
Like our main experiment, we simulated 4 MCMC chains for 2,000 iterations, discarding the first half of each chains as burn-in.
Figure~\ref{fig:scalability} plots the covariate effect estimation, uncertainty interval coverage, modifier selection, and runtime as functions of the total number of observations $N.$

\subsection{Increasing $R$ and correlated modifiers}
\label{sec:largeR_correlatedZ}

\begin{figure}[h!]
\centering
\begin{subfigure}[b]{0.32\textwidth}
\centering
\includegraphics[width = \textwidth]{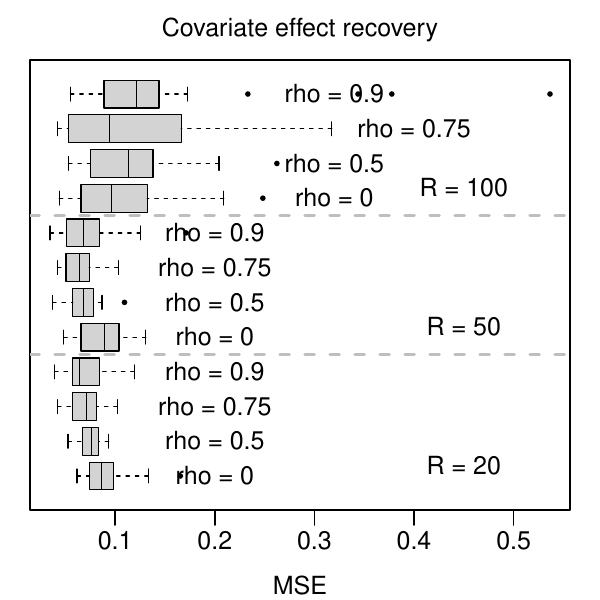}
\caption{}
\label{fig:zvar_beta_mse}
\end{subfigure}
\begin{subfigure}[b]{0.32\textwidth}
\centering
\includegraphics[width = \textwidth]{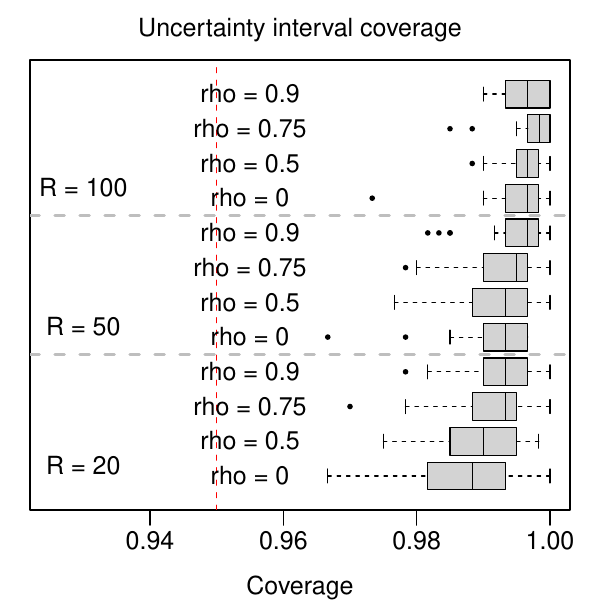}
\caption{}
\label{fig:zvar_beta_cov}
\end{subfigure}
\begin{subfigure}[b]{0.32\textwidth}
\centering
\includegraphics[width = \textwidth]{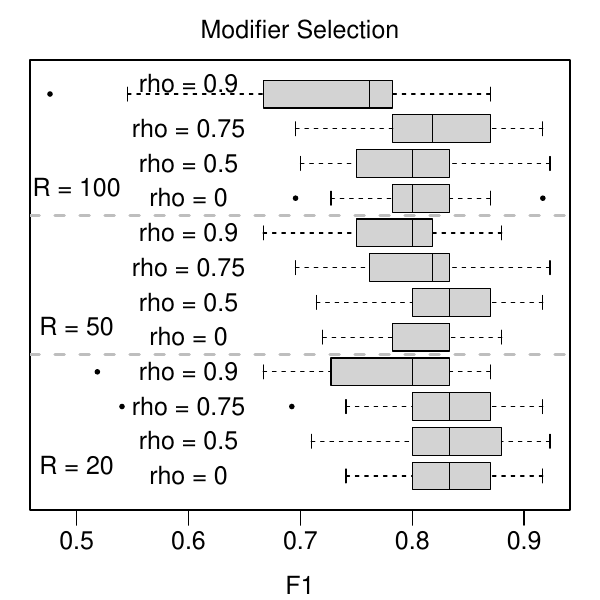}
\caption{}
\label{fig:zvar_f1}
\end{subfigure}
\caption{VCBART's estimation performance for different numbers of modifiers $R$ and correlation between modifiers. (a) Average mean square error for estimating evaluations $\beta_{j}(\bz)$. (b) Average coverage of 95\% uncertainty intervals for evaluations $\beta_{j}(\bz).$ (c) $F_{1}$ scores for modifier selection. All measures reported over 25 testing datasets.}
\label{fig:zvar_beta_performance}
\end{figure}

\begin{figure}[h!]
\begin{subfigure}[b]{0.32\textwidth}
\centering
\includegraphics[width = \textwidth]{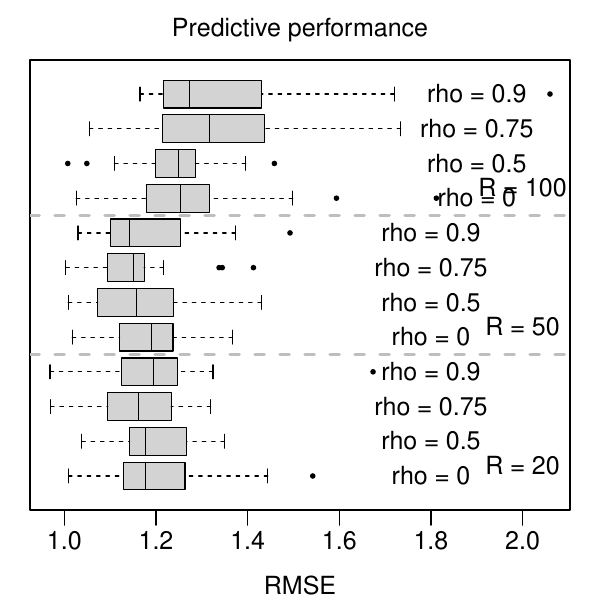}
\caption{}
\label{fig:zvar_ystar_rmse}
\end{subfigure}
\begin{subfigure}[b]{0.32\textwidth}
\centering
\includegraphics[width = \textwidth]{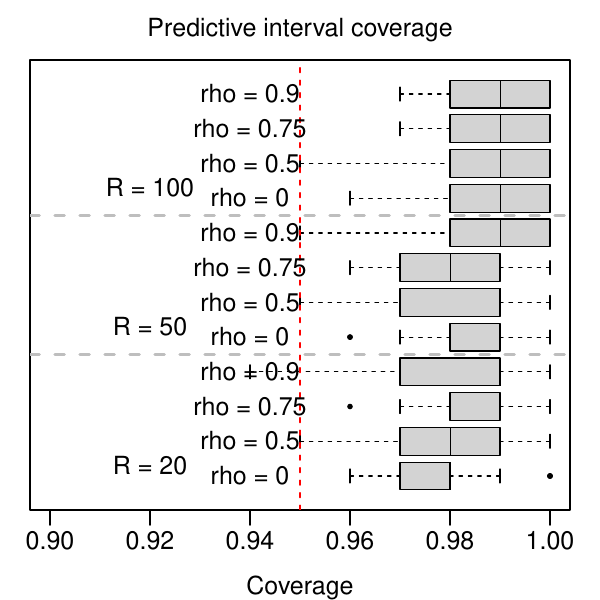}
\caption{}
\label{fig:zvar_ystar_cov}
\end{subfigure}
\begin{subfigure}[b]{0.32\textwidth}
\centering
\includegraphics[width = \textwidth]{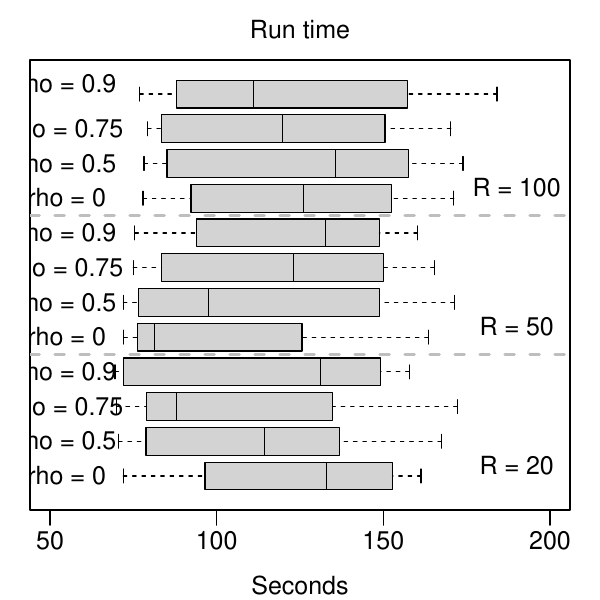}
\caption{}
\label{fig:zvar_timing}
\end{subfigure}
\caption{VCBART's predictive performance for different numbers of modifiers $R$ and correlation between modifiers.(a) Predictive root mean square error. (b) Coverage of 95\% prediction intervals. (c) Runtime in seconds. All measures reported over 25 testing datasets.}
\label{fig:zvar_ystar_performance}
\end{figure}

In our main experiment (Section 4 of the main text), we drew the effect modifiers independently and uniformly from the unit interval $[0,1].$
We conducted an additional experiment to assess how VCBART performs when the number of modifiers $R$ increases and when the effect modifiers are correlated.
We fixed $p = 5$ and drew each $\bx \sim \mathcal{N}_{p}(\mathbf{0}_{p}, \Sigma_{X})$ where the $(j,j')$ element of $\Sigma_{X}$ was equal to $0.5^{\lvert j - j' \rvert}.$
We used the same covariate effect functions as in our main experiment.
To generate correlated effect modifiers, we first drew vectors $\bz \sim \mathcal{N}_{R}(\mathbf{0}_{R}, \Sigma_{Z}),$ where the $(r,r')$ entry of $\Sigma_{Z}$ was equal to $\rho_{z}^{\lvert r -r' \rvert}.$
We then re-scaled each coordinate to the range $[0,1]$ and then generated data from a varying coefficient model using the same $\beta_{j}(\bZ)$'s as in our original experiment.
We generated 25 synthetic datasets for each combination of $R \in \{20, 50, 100\}$ and $\rho_{z} \in \{0, 0.5, 0.75, 0.9\}.$
Figure~\ref{fig:zvar_beta_performance} compares VCBART's ability to estimate $\beta_{j}(\bZ)$ and its support while Figure~\ref{fig:zvar_ystar_performance} compares VCBART's ability to predict new observations and its runtime for each combination of $(R, \rho_{z}).$. 

For $R = 20$ and $R = 50,$ VCBART's ability to estimate $\beta_{j}(\bz)$'s improved slightly as $\rho_{z}$ increased but was somewhat insensitive to $\rho_{z}$ when $R = 100.$
We additionally observe that for each $R,$ as $\rho_{z}$ increased, the $F_{1}$ score for the median probability model tended to decrease slightly.
However, Figure~\ref{fig:zvar_f1} suggests that VCBART's ability to recover the support of the $\beta_{j}(\bZ)$'s was fairly stable except for the case of $R = 100$ highly correlated modifiers (i.e., $\rho_{z} = 0.9$).
Turning to prediction, we see almost no differences in the predictive RMSE or predictive interval coverages across all values of $\rho_{z}$ when $R = 20$ or $R = 50.$
However, we do see some slightly higher predictive RMSEs when $R = 100.$
We further observe that the run time is wholly unaffected by $R$ (Figure~\ref{fig:zvar_timing}).

\subsection{Correlated covariates}
\label{sec:correlatedX}

Recall from Remark 1 that the vector of covariate effect function evaluations $\bbeta(\bz) = (\beta_{0}(\bz), \ldots, \beta_{p}(\bz))$ is identified so long as the matrix $\E[\bx\bx^{\top} \vert \bZ = \bz]$ is positive definite.
In Section 4 of the main text, we saw that VCBART was able to estimate these evaluations very well even when there was moderate correlation among the covariates.
We ran a similar experiment to determine how well VCBART performs in the presence of larger between-covariate correlation.
We generated data from the model in Equation (3.1) with $R = 20$ potential effect modifiers drawn independently from the unit interval and $p = 5$ correlated covariates. 
We used exactly the same covariate effect functions as in that simulation study and generated responses with $\sigma = 1$ and independent within-subject errors.
For this experiment, we drew the covariate vectors from mean-zero multivariate normal distributions $\bx \sim \mathcal{N}_{p}(\mathbf{0}_{p}, \Sigma_{X}).$
We considered two different classes of covariance structure $\Sigma_{X}$: banded structure, in which the $(j,j')$ entry of $\Sigma_{X}$ is equal to $\rho^{\lvert j - j' \rvert},$ and constant, in which the $(j,j')$ entry of $\Sigma_{X}$ is equal to $\rho$ if $j \neq j'$ and 1 otherwise.
For each covariance structure, we considered six different values of $\rho \in \{0, 0.25, 0.5, 0.75, 0.9, 0.99\}$ and generated 25 synthetic data sets for each $\Sigma_{X}.$

\begin{figure}[h!]
\centering
\begin{subfigure}[b]{0.32\textwidth}
\centering
\includegraphics[width = \textwidth]{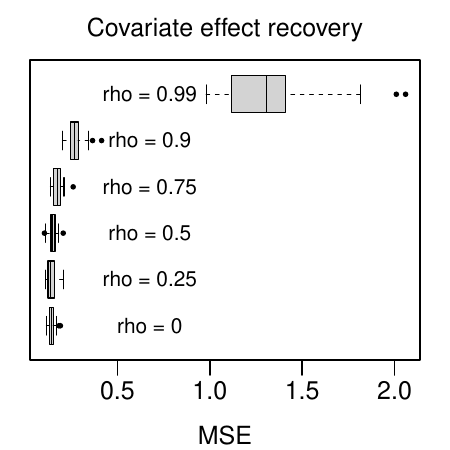}
\caption{}
\label{fig:banded_beta_mse}
\end{subfigure}
\begin{subfigure}[b]{0.32\textwidth}
\centering
\includegraphics[width = \textwidth]{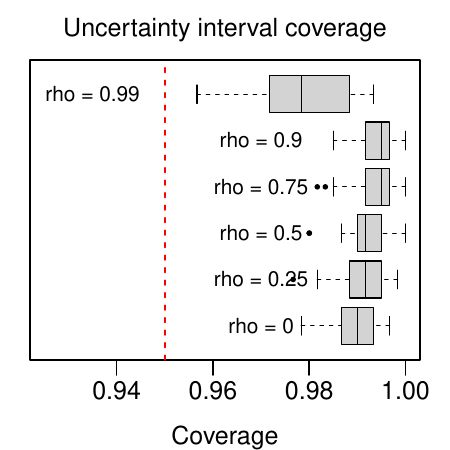}
\caption{}
\label{fig:banded_beta_cov}
\end{subfigure}
\begin{subfigure}[b]{0.32\textwidth}
\centering
\includegraphics[width = \textwidth]{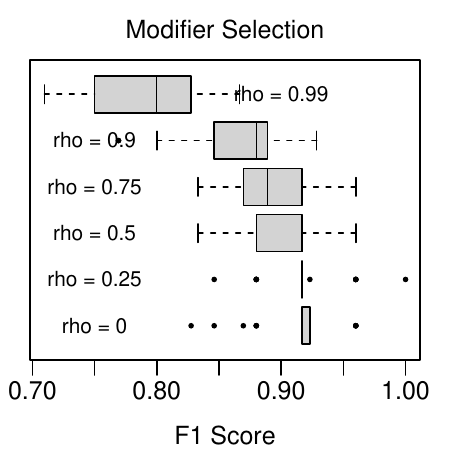}
\caption{}
\label{fig:banded_f1}
\end{subfigure}

\begin{subfigure}[b]{0.32\textwidth}
\centering
\includegraphics[width = \textwidth]{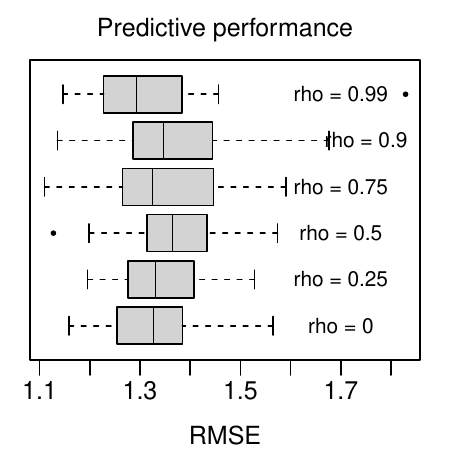}
\caption{}
\label{fig:banded_ystar_rmse}
\end{subfigure}
\begin{subfigure}[b]{0.32\textwidth}
\centering
\includegraphics[width = \textwidth]{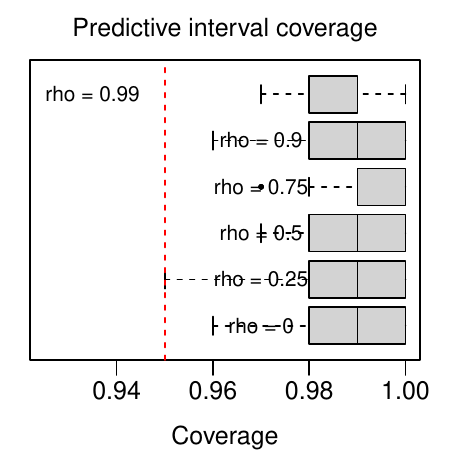}
\caption{}
\label{fig:banded_ystar_cov}
\end{subfigure}
\begin{subfigure}[b]{0.32\textwidth}
\centering
\includegraphics[width = \textwidth]{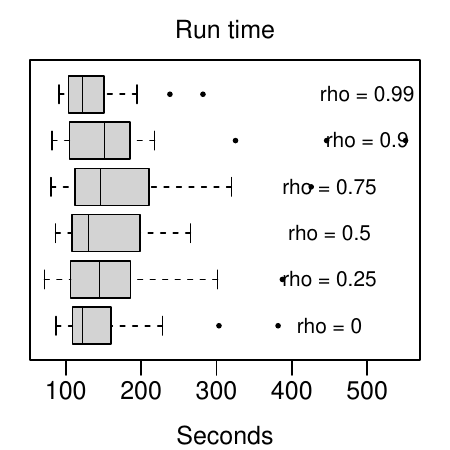}
\caption{}
\label{fig:banded_timing}
\end{subfigure}

\caption{Performance of VCBART when covariates have banded correlation structure. (a) Average mean square error for estimating evaluations $\beta_{j}(\bz)$. (b) Average coverage of 95\% uncertainty intervals for evaluations $\beta_{j}(\bz).$ (c) $F_{1}$ scores for modifier selection. (d) Predictive root mean square error. (e) Coverage of 95\% prediction intervals. (f) Runtime in seconds.  All measures reported over 25 testing datasets.}
\label{fig:banded_performance}
\end{figure}

\begin{figure}[h!]
\centering
\begin{subfigure}[b]{0.32\textwidth}
\centering
\includegraphics[width = \textwidth]{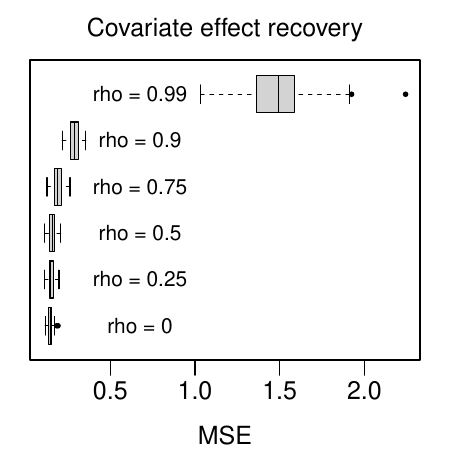}
\caption{}
\label{fig:const_beta_mse}
\end{subfigure}
\begin{subfigure}[b]{0.32\textwidth}
\centering
\includegraphics[width = \textwidth]{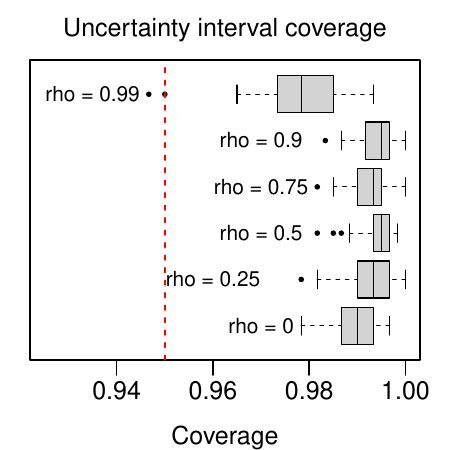}
\caption{}
\label{fig:const_beta_cov}
\end{subfigure}
\begin{subfigure}[b]{0.32\textwidth}
\centering
\includegraphics[width = \textwidth]{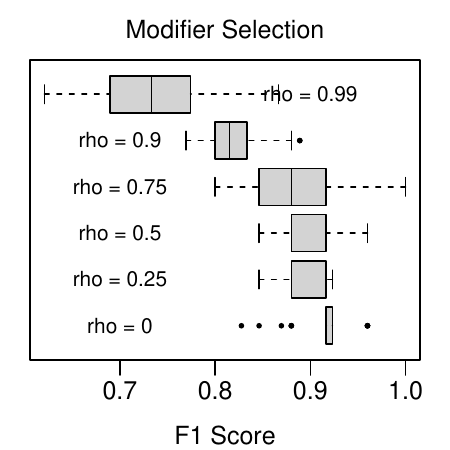}
\caption{}
\label{fig:const_f1}
\end{subfigure}

\begin{subfigure}[b]{0.32\textwidth}
\centering
\includegraphics[width = \textwidth]{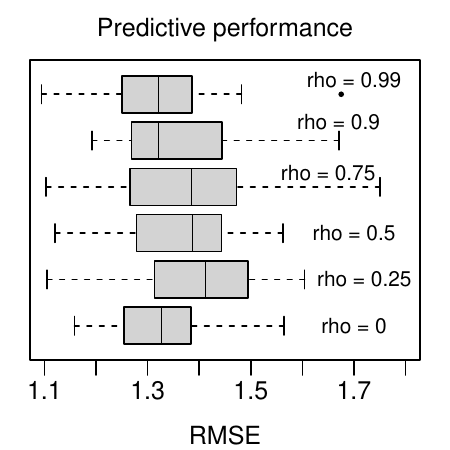}
\caption{}
\label{fig:const_ystar_rmse}
\end{subfigure}
\begin{subfigure}[b]{0.32\textwidth}
\centering
\includegraphics[width = \textwidth]{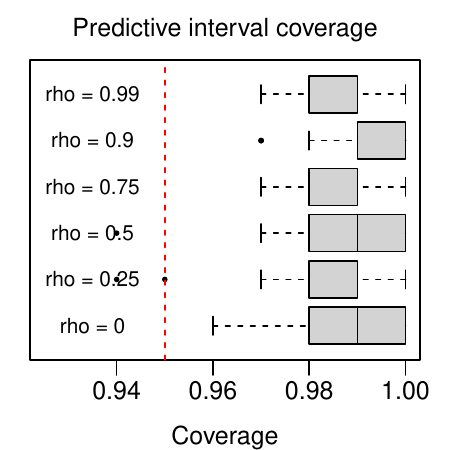}
\caption{}
\label{fig:const_ystar_cov}
\end{subfigure}
\begin{subfigure}[b]{0.32\textwidth}
\centering
\includegraphics[width = \textwidth]{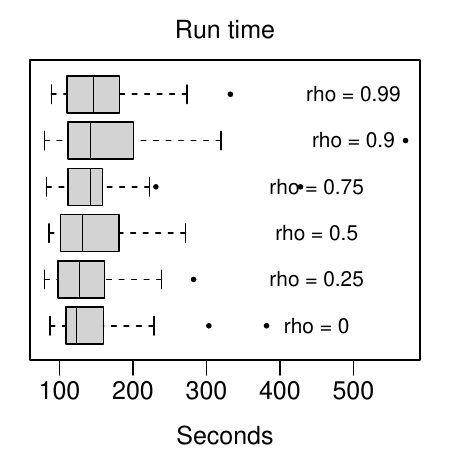}
\caption{}
\label{fig:const_timing}
\end{subfigure}
\caption{Performance of VCBART when covariates have constant correlation structure. (a) Average mean square error for estimating evaluations $\beta_{j}(\bz)$. (b) Average coverage of 95\% uncertainty intervals for evaluations $\beta_{j}(\bz).$ (c) $F_{1}$ scores for modifier selection. (d) Predictive root mean square error. (e) Coverage of 95\% prediction intervals. (f) Runtime in seconds.  All measures reported over 25 testing datasets.}
\label{fig:const_performance}
\end{figure}

Figure~\ref{fig:banded_performance} (resp. Figure~\ref{fig:const_performance}) shows the prediction, uncertainty quantification, and modifier selection performance for VCBART for banded (reps. constant) covariance structures. 
As we might expect, in light of Remark 1 and the identifiability of the $\bbeta(\bz)$'s, as we increased $\rho,$ VCBART's ability to estimate $\bbeta(\bz)$ out-of-sample and to select the relevant modifiers decayed.
That said, the degradation was not especially severe until $\rho = 0.99,$ which corresponds to the situation where there is extremely high correlation between covariates.
Across both covariance structures, we did not observe substantial degradation in the coverage of the VCBART posterior credible intervals for evaluations $\beta_{j}(\bz)$ nor did we observe changes in the predictive performance, posterior predictive interval coverage, or runtime.

%% file: gibbs_sampler.tex
Recall that we have $n_{i}$ observations for each subject $i = 1, \ldots, n.$
We model for each $i = 1, \ldots, n,$ and $t = 1, \ldots, n_{i}$
\begin{equation}
\label{eq:panel_model_supp}
y_{it} = \beta_{0}(\bz_{it}) + \sum_{j = 1}^{p}{\beta_{j}(\bz_{it})x_{itj}} + \sigma \varepsilon_{it},
\end{equation}
where $\boldsymbol{\varepsilon}_{i} \sim \mathcal{N}(\mathbf{0}_{n_{i}}, \bSigma_{i}(\rho))$ and the diagonal elements of $\bSigma_{i}(\rho)$ are equal to one.

Let $\bTheta = \{\btheta_{0}, \ldots, \btheta_{p}\}$ be the collection of splitting index probabilities.
We further model each $\btheta_{j} \vert \eta_{j} \sim \text{Dirichlet}(\eta_{j}/R, \ldots, \eta_{j}/R)$ and $\eta_{j}/(\eta_{j} +R) \sim \text{Beta}(0.5, 1).$
We complete our prior specification with $\sigma^{2} \sim \lambda \nu/\chi^{2}_{\nu}$ and $\rho \sim \text{Uniform}[0,1].$

We use a Gibbs sampling strategy to simulate draws from the joint posterior of $(\bEcal, \sigma^{2}, \bTheta, \boldsymbol{\eta}, \rho) \vert \bY,$ where $\bEcal$ denotes the collection of all regression trees.
We divide our derivation and description of each step into two parts: updating the tree ensemble $\bEcal$ (Section~\ref{app:tree_updates}); and updating $\bTheta, \boldsymbol{\eta}, \sigma^{2}$ and $\rho$ (Section~\ref{app:theta_eta_updates}).
Before describing these updates, we must introduce some additional notation.

Given a regression tree $(T, \bmu)$ and a vector of modifiers $\bz \in \calZ$  let $\ell(\bz; T)$ be the leaf reached by the decision-following path of $\bz$.
Let $g(\bz; T, \bmu)$ be the evaluation function that returns the jump associated with leaf $\ell(\bz;T);$ that is
$$
g(\bz; T, \bmu) = \mu_{\ell(\bz;T)}.
$$

For each $j = 0, \ldots, p,$ let $\calE^{(j)} = \left\{(T^{(j)}_{m}, \bmu_{m}^{(j)})\right\}_{m = 1}^{M}$ denote the collection of $M$ regression trees used to express the function $\beta_{j}(\bZ)$ in Equation~\eqref{eq:panel_model_supp}.
Given $\calE^{(j)},$ we have
$$
\beta_{j}(\bz) = \sum_{m = 1}^{M}{g(\bz; T_{m}^{(j)}, \bmu^{(j)}_{m})}.
$$
Additionally, let $\bcalE = \left\{\calE^{(0)}, \ldots, \calE^{(p)}\right\}$ denote the collection of all regression trees in our model.
The likelihood function implied by Equation~\eqref{eq:panel_model_supp} is therefore
\begin{equation}
\label{eq:likelihood}
p(\bY \vert \bcalE, \sigma^{2}, \rho) = \prod_{i = 1}^{n}{\sigma^{-n_{i}}\lvert \bOmega_{i}(\rho)\rvert^{\frac{1}{2}}\exp\left\{-\frac{\bR_{i}^{\top}\bOmega_{i}(\rho)\bR_{i}}{2\sigma^{2}}\right\}}
\end{equation}
where for each subject $i = 1, \ldots, n,$ $\bOmega_{i}(\rho) = \bSigma_{i}^{-1}(\rho)$ and $\bR_{i} = (R_{i1}, \ldots, R_{in_{i}})^{\top}$ is the vector of \textit{full residuals}
\begin{equation}
\label{eq:full_residual}
R_{it} = y_{it} - \sum_{m = 1}^{M}{g(\bz_{it}; T_{m}^{(0)}, \bmu_{m}^{(0)})} - \sum_{j = 1}^{p}{\sum_{m = 1}^{M}{x_{itj}g(\bz_{i}; T_{m}^{(j)}, \bmu_{m}^{(j)})}}.
\end{equation}

The joint posterior density of $(\bcalE, \sigma^{2}, \bTheta, \boldsymbol{\eta}, \rho)$ is given by
\begin{equation}
\begin{split}
\label{eq:full_posterior_panel_supp}
p(\bEcal, \sigma^{2}, \bTheta, \boldsymbol{\eta}, \rho \mid \bY) &\propto \prod_{i = 1}^{n}{\sigma^{-n_{i}}\lvert \bOmega_{i}(\rho)\rvert^{\frac{1}{2}}\exp\left\{-\frac{\bR_{i}^{\top}\bOmega_{i}(\rho)\bR_{i}}{2\sigma^{2}}\right\}} \\
&\times \prod_{j = 0}^{p}{p(\eta_{j})p(\btheta_{j} \vert \eta_{j})\prod_{m = 1}^{M}{p(T^{(j)}_{m} \vert \btheta_{j})\tau^{-L_{m}^{(j)}}_{j}\exp\left\{-\frac{\lVert \mu_{m}^{(j)} \rVert_{2}^{2}}{2\tau^{2}_{j}}\right\}}} \\
&\times \left(\sigma^{2}\right)^{-1 - \frac{\nu}{2}}\exp\left\{-\frac{\lambda\nu}{2\sigma^{2}}\right\}
\end{split}
\end{equation}


\subsection{Conditional regression tree updates}
\label{app:tree_updates}

We now describe the update of the $m^{\text{th}}$ regression tree $(T_{m}^{(j)}, \bmu^{(j)}_{m})$ in the ensemble $\mathcal{E}_{j}$ used to approximate $\beta_{j}.$
Let $\bEcal^{-}$ denote the collection of all remaining $M(p+1) - 1$ regression trees, where, for notational compactness, we have suppressed the dependence of this collection on the indices $j$ and $m.$
Additionally, let $\br_{i} = (r_{i1}, \ldots, r_{in_{i}})^{\top}$ be the vector of subject $i$'s \textit{partial residuals} where
\begin{equation}
\label{eq:partial_residual}
r_{it} = R_{it} + x_{itj}g(\bz_{it}; T_{m}^{(j)}, \bmu_{m}^{(j)}),
\end{equation}
where we have again suppressed the dependence of $r_{it}$ on $j$ and $m$ for brevity.

Before describing the update of the tree, we require some additional notation.
For an arbitrary decision tree $T$ with $L$ leaves, let $I_{i}(\ell)$ be the set of indices $t$ such that $\bz_{it}$ is contained in leaf $\ell$ of tree $t.$
That is, $I_{i}(\ell) = \{t: \ell(\bz_{it}) = \ell\}.$
Further, let $\bX_{i}(T)$ be the $n_{i} \times L$ matrix whose $(t,\ell)$ entry is equal to $x_{itj}$ if $t \in I_{i}(\ell,T)$ and is zero otherwise.
With this additional notation, we have $\br_{i} = \bR_{i} + \bX_{i}(T^{(j)}_{m})\bmu^{(j)}_{m}.$

Suppressing the conditioning variables from our notation, the conditional posterior density of a single regression tree $(T, \bmu)$ is given by
\begin{equation}
\label{eq:T_mu_panel}
\pi(T, \bmu \mid \ldots) \propto \pi(T \vert \btheta_{j})\tau^{-L(T)}\exp\left\{-\frac{1}{2}\left[\bmu^{\top}\Lambda(T)^{-1}\bmu - 2\bmu^{\top}\Theta(T)\right]\right\}
\end{equation}
where
\begin{align*}
\Lambda(T) &= \left[ \tau^{-2}\text{I}_{L(T)} + \sum_{i = 1}^{n}{\bX_{i}(T)^{\top}\bOmega_{i}(\rho)\bX_{i}(T)}\right]^{-1} \\
\Theta(T) &= \sum_{i = 1}^{n}{\bX_{i}(T)^{\top}\bOmega_{i}(\rho)\br_{i}}.
\end{align*}

In order to update the tree $(T_{m}^{(j)}, \bmu_{m}^{(j)}),$ we use the same strategy as \citet{Chipman2010}: we first draw a new decision tree from the marginal distribution $\pi(T \vert \bY, \bEcal^{-}, \bTheta, \boldsymbol{\eta}, \sigma, \rho)$ with a MH step and then draw a new set of jump conditionally on the new decision tree.
To this end, by integrating $\bmu$ out of~\eqref{eq:T_mu_panel}, the marginal posterior mass function of $T$ is given by
\begin{equation}
\label{eq:T_posterior}
\pi(T \mid \bY, \bEcal^{-}, \bTheta, \boldsymbol{\eta}, \sigma, \rho) \propto \lvert \Lambda(T) \rvert^{\frac{1}{2}}\tau^{-L(T)}\exp\left\{\frac{1}{2}\Theta(T)^{\top}\Lambda(T)\Theta(T)\right\}.
\end{equation}
In our MH step, we propose a new tree $\tilde{T}$ by growing or pruning the current tree $T_{m}^{(j)}.$

We can also read off the conditional density of $\bmu \mid T$ directly from~\eqref{eq:T_mu_panel}
\begin{equation}
\label{eq:mu_posterior}
\pi(\bmu \mid T, \bY, \bEcal^{-}, \bTheta, \boldsymbol{\eta}, \sigma, \rho) \propto \exp\left\{-\frac{1}{2}\left[\bmu^{\top}\Lambda^{-1}(T)\bmu - 2\bmu^{\top}\Theta(T)\right]\right\},
\end{equation}
which is proportional to the density of a $\mathcal{N}(\Lambda(T)\Theta(T), \Lambda(T))$ distribution.

\subsection{Remaining updates}
\label{app:theta_eta_updates}
When we update each decision tree $T_{m}^{(j)}$ in the ensemble $\mathcal{E}_{j},$ we keep track the number of times that each modifier is used in 
This facilitates a simple Dirichlet-Multinomial conjugate update for $\btheta_{j}$ \citep[Equation 6;][]{Linero2018}.
We then update the quantity $u_{j} = \eta_{j}/(\eta_{j} + R)$ with an independence Metropolis step that draws new proposals from the $\text{Beta}(0.5,1)$ prior.
The conditional log-posterior density of $u$ is, up to additive constants, given by
\begin{equation}
\label{eq:u_update}
\log p(u_{j} \vert \ldots) = \frac{u}{1-u} \times \sum_{r}{\log(\theta_{jr})} + \log\Gamma\left(R \times \frac{u_{j}}{1-u_{j}}\right) - R\log\left(\frac{u_{j}}{1-u_{j}}\right)
\end{equation}

From Equation~\eqref{eq:full_posterior_panel_supp}, we compute
\begin{align}
\begin{split}
\label{eq:sigma2_conditional}
p(\sigma^{2} \vert \bY, \bcalE, \bTheta, \boldsymbol{\eta}, \rho) &\propto \prod_{i = 1}^{n}{\left(\sigma^{2}\right)^{-\frac{n_{i}}{2}} \times \exp\left\{-\frac{\bR_{i}^{\top}\bOmega_{i}(\rho)\bR_{i}}{2\sigma^{2}}\right\}} \\
&\times \left(\sigma^{2}\right)^{-1 - \frac{\nu}{2}}\exp\left\{-\frac{\lambda\nu}{2\sigma^{2}}\right\}.
\end{split}
\end{align}
We can therefore draw $\sigma^{2}$ conditionally given all other parameters from an inverse gamma distribution.

We further compute
\begin{align*}
p(\rho \vert \bY, \bcalE, \bTheta, \boldsymbol{\eta}) \propto \prod_{i = 1}^{n}{\lvert \bOmega_{i}(\rho)\rvert^{\frac{1}{2}}\exp\left\{-\frac{\bR_{i}^{\top}\bOmega_{i}(\rho)\bR_{i}}{2\sigma^{2}}\right\}}
\end{align*}
Instead of updating $\rho$ directly, we update the unconstrained quantity $\log\left(\frac{\rho}{1-\rho}\right)$ with a random-walk Metropolis step in which proposals are drawn from a normal distribution centered at the current value with variance $0.01.$

%% file: additional_figures.tex
\begin{figure}[H]
\centering
\begin{subfigure}[b]{0.32\textwidth}
\centering
\includegraphics[width = \textwidth]{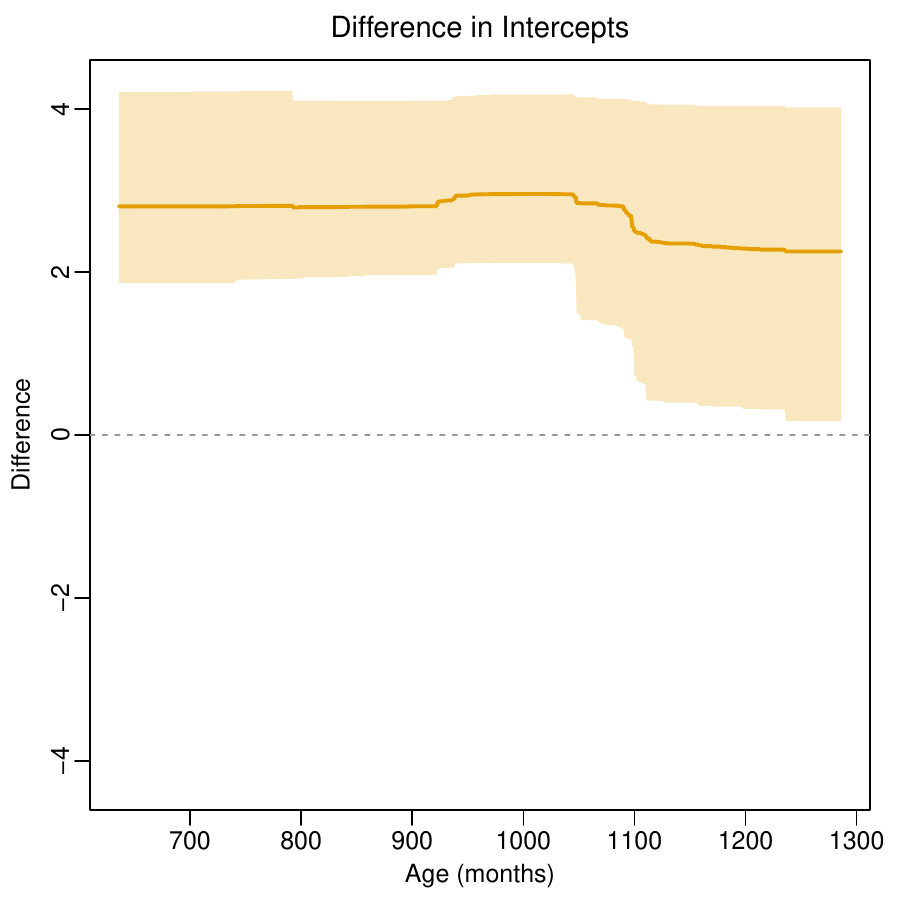}
\caption{}
\label{fig:hrs_intercept}
\end{subfigure}
\begin{subfigure}[b]{0.32\textwidth}
\centering
\includegraphics[width = \textwidth]{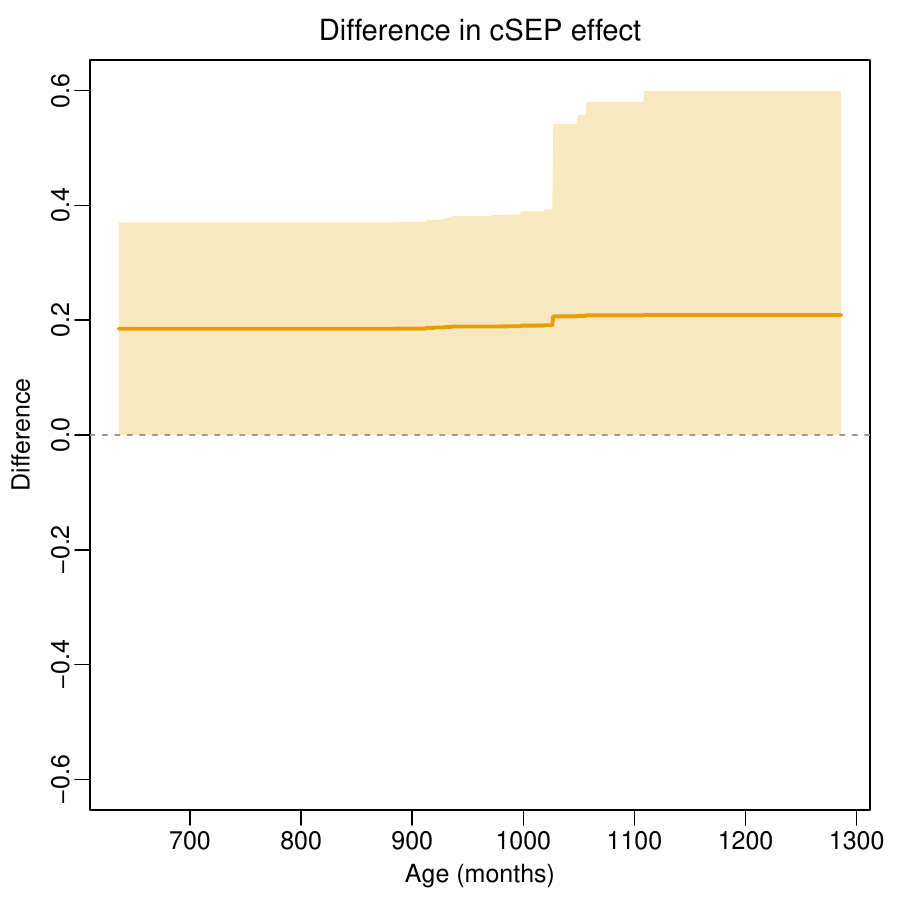}
\caption{}
\label{fig:hrs_cSEP}
\end{subfigure}
\begin{subfigure}[b]{0.32\textwidth}
\centering
\includegraphics[width = \textwidth]{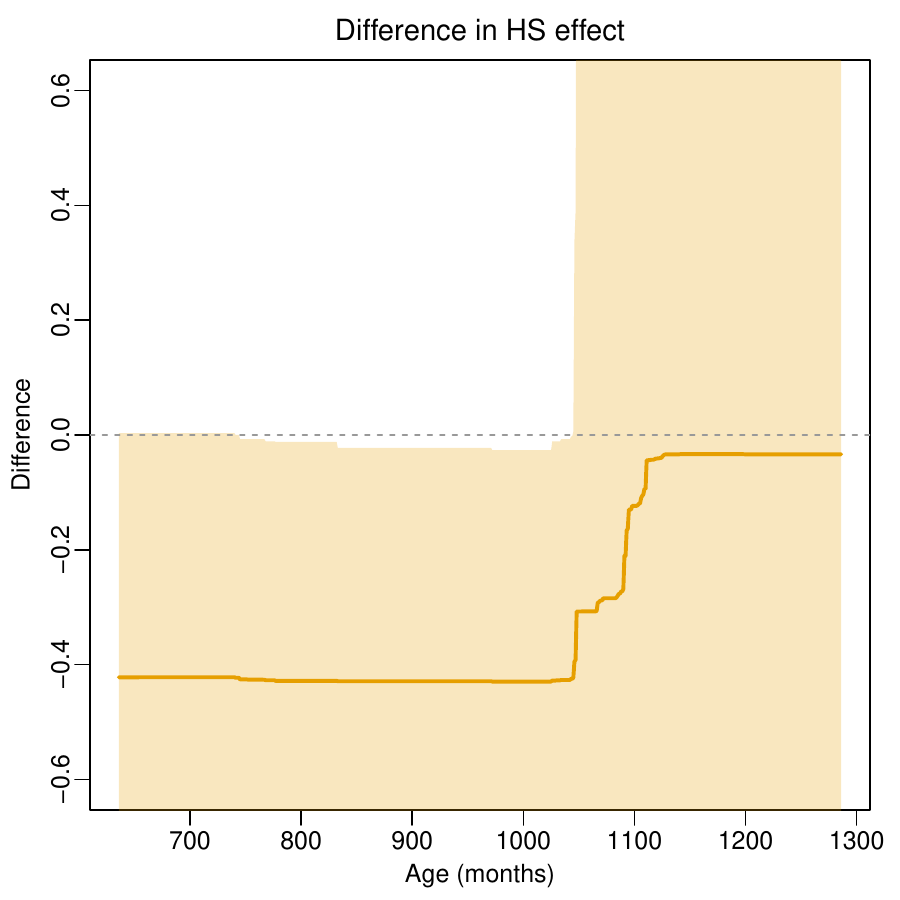}
\caption{}
\label{fig:hrs_hs}
\end{subfigure}
\caption{Posterior mean and pointwise 95\% credible intervals of the difference in intercept (a) and partial effects of childhood SEP (b) and high school completion (c) on later-life cognitive score as functions of age between the two individuals considered in the main text.}
\label{fig:hrs_diff}
\end{figure}

Figure~\ref{fig:hrs_diff} shows the posterior mean and pointwise 95\% credible intervals for the difference between the curves in Figure 4 of the main text.

\begin{figure}[H]
\centering
\includegraphics[width = 0.5\textwidth]{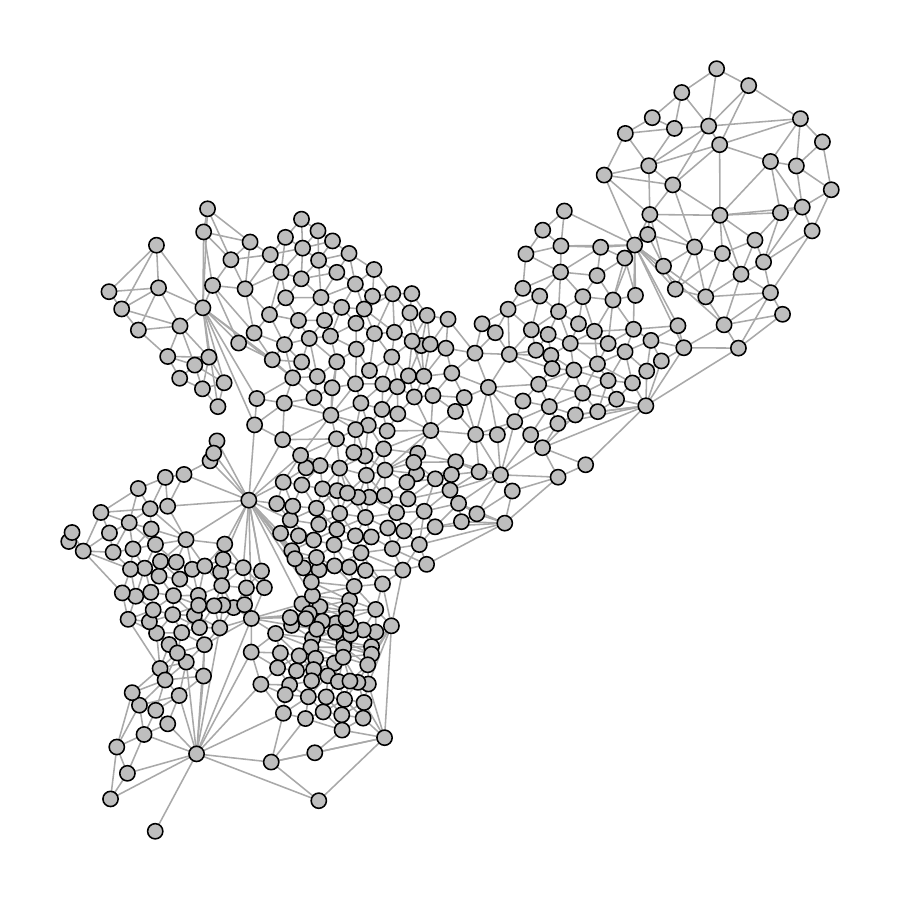}
\caption{Network vertices correspond to census tracts and edges are drawn between vertices whose corresponding tracts are spatially adjacent}
\label{fig:philly_network}
\end{figure}

Figure~\ref{fig:philly_network} shows a network encoding the spatial adjacency relationships between the census tracts in Philadelphia.

